\documentclass[a4paper,UKenglish,cleveref,autoref,thm-restate]{lipics-v2021}
\pdfoutput=1 

\hideLIPIcs  

\usepackage{standalone}
\usepackage{xcolor}
\usepackage{stmaryrd}
\usepackage{tikz}
\usetikzlibrary{automata, positioning, arrows, petri, backgrounds, calc, shapes}
\pgfdeclarelayer{farbackground}
\pgfsetlayers{farbackground,background,main}

\usepackage{amsmath}

\newcommand{\N}{\mathbb{N}}

\newcommand{\Z}{\mathbb{Z}}

\newcommand{\Q}{\mathbb{Q}}
\newcommand{\R}{\mathbb{R}}

\def\O{\mathcal{O}}

\newcommand{\Set}[1]{\left\{#1\right\}}

\newcommand{\Support}[1]{\operatorname{supp}(#1)}

\renewcommand{\dots}{...}
\renewcommand{\ldots}{...}
\renewcommand{\colon}{:}

\definecolor{nicebg}{HTML}{f6f0e4}
\definecolor{nicered}{HTML}{7f0a13}
\definecolor{nicebgred}{HTML}{f2e7e8}
\definecolor{niceblue}{HTML}{104354}
\definecolor{nicebgblue}{HTML}{e8edee}
\definecolor{nicegreen}{HTML}{217516}
\definecolor{nicebggreen}{HTML}{e9f1e8}
\definecolor{nicepurple}{HTML}{884bab}
\definecolor{nicebgpurple}{HTML}{f3edf7}
\definecolor{niceorange}{HTML}{d27c11}
\definecolor{nicebgorange}{HTML}{fbf2e8}
\definecolor{nicepink}{HTML}{e95f9f}
\definecolor{nicebgpink}{HTML}{fdeff6}
\definecolor{niceredlight}{HTML}{c9888d}
\definecolor{nicebluelight}{HTML}{78a4b8}
\definecolor{nicegreenlight}{HTML}{76de68}
\definecolor{nicepurplelight}{HTML}{bc87db}
\definecolor{niceredbright}{HTML}{bd0310}
\definecolor{nicebgredbright}{HTML}{f9e6e8}
\definecolor{nicebluebright}{HTML}{197b9b}
\definecolor{nicebgbluebright}{HTML}{e8f2f5}

\def \ifemptycheck#1{\def\temp{#1} \ifx\temp\empty }
\newcommand{\emptymultiset}{\emptyset}
\def \multiset#1{ \ifemptycheck{#1} \emptymultiset \else \Lbag#1\Rbag \fi }

\newcommand{\Prot}{\mathcal{P}} 
\newcommand{\Npred}{s} 
\newcommand{\Nvar}{v} 
\newcommand{\h}{d} 	
\newcommand{\Modulus}{\theta}
\newcommand{\splitsize}{r} 	
\newcommand{\StateInProt}[2]{\left(#1\right)_{#2}} 

\DeclareMathOperator{\size}{size}

\newcommand{\pow}{\textit{Pow}}
\newcommand{\powp}{\textit{Pow}^+}
\newcommand{\multisum}[1]{\operatorname{sum}(#1)}

\newcommand{\Bin}[1]{\operatorname{bin}(#1)}
\newcommand{\sign}[1]{\operatorname{sign}(#1)}

\newcommand{\State}[1]{\mathsf{#1}} 

\newcommand{\TraNs}{tra}
\newcommand{\TraName}[2][]{ \ifemptycheck{#1} \tag*{⟨\textsf{#2}⟩}\label{\TraNs:#2} \else \tag*{⟨\textsf{#2}⟩}\label{\TraNs:#1} \fi }
\newcommand{\TraRef}[1]{\text{\ref{\TraNs:#1}}}
\newcommand{\Cterm}{C_{\mathrm{term}}}
\newcommand{\QFPA}{\mathit{QFPA}}

\newcommand{\E}{\mathbb{E}}
\newcommand{\Abs}[1]{\mathopen|#1\mathclose|}
\newcommand{\Absbig}[1]{\mathopen\big|#1\mathclose\big|}

\definecolor{notquitegray}{HTML}{bab4a2}

\newcommand{\Double}{\operatorname{double}}
\newcommand{\Poly}{\operatorname{poly}}
\newcommand{\Polylog}{\operatorname{polylog}}

\newcommand{\Qorig}{Q_\mathrm{orig}}
\newcommand{\Qsupp}{Q_\mathrm{supp}}
\newcommand{\Qgate}{Q_\mathrm{gate}}
\newcommand{\Qreset}{Q_\mathrm{reset}}

\newcommand{\Nand}{\operatorname{\mathsf{nand}}}

\DeclareMathOperator{\preprocessconv}{Preprocess}
\DeclareMathOperator{\binarise}{Binarise}
\DeclareMathOperator{\focalise}{Focalise}
\DeclareMathOperator{\autarkify}{Autarkify}
\DeclareMathOperator{\distribute}{Distribute}
\DeclareMathOperator{\preprocess}{Preprocess}

\usepackage{tcolorbox}

\newcounter{specificationcount}
\newenvironment{specification}[1]{%
    \refstepcounter{specificationcount}
    \begin{tcolorbox}[colframe=niceblue,colback=nicebgblue]
    {\Large\bfseries\sffamily Specification: $#1$}\\[3mm]
    \begin{tabular}{@{}lp{11cm}} 
}{%
    \end{tabular}
    \end{tcolorbox}
}

\title{Fast and Succinct Population Protocols for Presburger Arithmetic}

\newcommand{\Affil}{Department of Computer Science, Technical University of Munich, Germany}
\author{Philipp Czerner}{\Affil\and \url{https://nicze.de/philipp}}{czerner@in.tum.de}{https://orcid.org/0000-0002-1786-9592}{}
\author{Roland Guttenberg}{\Affil\and \url{https://rolandguttenberg.de}}{guttenbe@in.tum.de}{https://orcid.org/0000-0001-6140-6707}{}
\author{Martin Helfrich}{\Affil\and \url{https://martinhelfrich.de}}{helfrich@in.tum.de}{https://orcid.org/0000-0002-3191-8098}{}
\author{Javier Esparza}{\Affil\and \url{https://www7.in.tum.de/~esparza}}{esparza@in.tum.de}{https://orcid.org/0000-0001-9862-4919}{}

\authorrunning{P. Czerner, R. Guttenberg, M. Helfrich and J. Esparza} 

\Copyright{Philipp Czerner, Roland Guttenberg, Martin Helfrich and Javier Esparza} 



\category{} 

\funding{This work was supported by an ERC Advanced Grant (787367: PaVeS) and by the Research Training Network of the Deutsche Forschungsgemeinschaft (DFG) (378803395: ConVeY).}

\nolinenumbers 

\EventEditors{James Aspnes and Othon Michail}
\EventNoEds{2}
\EventLongTitle{1st Symposium on Algorithmic Foundations of Dynamic Networks (SAND 2022)}
\EventShortTitle{SAND 2022}
\EventAcronym{SAND}
\EventYear{2022}
\EventDate{March 28--30, 2022}
\EventLocation{Virtual Conference}
\EventLogo{}
\SeriesVolume{221}
\ArticleNo{16}

\begin{document}

\maketitle

\begin{abstract}
In their 2006 seminal paper in Distributed Computing,  Angluin et al.\ present a construction that, given any Presburger predicate as input, outputs a leaderless population protocol that decides the predicate. 
The protocol for a predicate of size $m$ (when expressed as a boolean combination of threshold and remainder predicates with coefficients in binary) runs in $\O(m \cdot n^2 \log n)$ expected number of interactions, which is almost optimal in $n$, the number of interacting agents. However, the number of states of the protocol is exponential in $m$. This is a problem for natural computing applications, where a state corresponds to a chemical species and it is difficult to implement protocols with many states. Blondin et al.\ presented at STACS 2020 another construction that produces protocols with a polynomial number of states, but exponential expected number of interactions. We present a construction that produces protocols with $\O(m)$ states that run in expected $\O(m^{7} \cdot n^2)$ interactions, optimal in $n$, for all inputs of size $\Omega(m)$. For this, we introduce population computers, a carefully crafted generalization of population protocols easier to program, and show that our computers for Presburger predicates can be translated into fast and succinct population protocols.

\end{abstract}


\section{Introduction} \label{sec:intro}

\newcommand{\etal}{et al.}

Population protocols are a model of computation in which indistinguishable, mobile finite-state agents, randomly interact in pairs to decide whether their initial configuration satisfies a given property, modelled as a predicate on the set of all configurations~\cite{AADFP06}. The decision is taken by \emph{stable consensus}; eventually all agents agree on whether the property holds or not, and never change their mind again. 
Population protocols are very close to chemical reaction networks, a model in which agents are molecules and interactions are chemical reactions.

In a seminal paper, Angluin~\etal\ proved that population protocols
decide exactly the predicates definable in Presburger arithmetic (PA)~\cite{AAER07}. 
One direction of the result is proved in~\cite{AADFP06} by means of a construction that takes as input a Presburger predicate and outputs a protocol that decides it. 
The construction uses the quantifier elimination procedure for PA: every Presburger formula $\varphi$ can be transformed into an equivalent boolean combination of \emph{threshold} predicates of the form $ \vec{a} \cdot \vec{x} \ge c$  and \emph{remainder} predicates of the form $\vec{a} \cdot \vec{x} \equiv_{\Modulus} c$,  where $\vec{a}$ is an integer vector, $c$ and $\Modulus$ are integers, and $\equiv_{\Modulus}$ denotes congruence modulo $\Modulus$~\cite{Haase18}. Slightly abusing language, we call the set of these boolean combinations \textit{quantifier-free Presburger arithmetic} (QFPA).%
\footnote{Remainder predicates cannot be directly expressed in Presburger arithmetic without quantifiers.}
Using that PA and QFPA have the same expressive power, Angluin~\etal\ first construct protocols for all threshold and remainder predicates, and then show that the predicates computed by protocols are closed under negation and conjunction.  

Two fundamental parameters of a protocol are the expected number of interactions until a stable consensus is reached (that is, a consensus that cannot be broken by any possible continuation of the execution that leads to it),
and the number of states of each agent. The expected number of interactions divided by the number of agents, also called the parallel stabilisation time, is an adequate measure of the runtime of a protocol when interactions occur in parallel according to a Poisson process~\cite{AAE08}. The number of states measures the complexity of an agent. In many natural computing applications, where a state corresponds to a chemical species, it is difficult to implement protocols with many states (see \cite{CardelliN12} for a particularly simple example and, in general, the literature on programming chemical reaction networks, a model very close to population protocols \cite{SoloveichikCWB08}).

Given a formula $\varphi$ of QFPA, let $m$ be the number of bits needed to write $\varphi$ with coefficients in binary, and let \(n\) be the number of agents participating in the protocol. The construction of~\cite{AADFP06} yields a protocol with $\O(m \cdot n^2 \log n)$ expected interactions. Observe that the protocol does not have a leader (an auxiliary agent helping the other agents to coordinate), and agents have a fixed number of states, independent of the size of the population. Under these assumptions, which are also the assumptions of this paper, every protocol for the majority predicate needs $\Omega(n^2)$ expected interactions~\cite{AlistarhAEGR17}, and so the construction is nearly optimal\footnote{See the related work section for other results when these assumptions are given up.}.%
 However, the number of states is $\Omega(2^{m})$. This is well beyond the only known lower bound, showing that for every construction there exists an infinite subset of predicates $\varphi$ for which the construction produces protocols with $\Omega(m^{1/4})$ states~\cite{BlondinEJ18}. So the construction of~\cite{AADFP06} produces \emph{fast} but \emph{very large} protocols. The same happens with the constructions of~\cite{AAE08}, where protocols simulate register machines and the state of an agent stores (among other information) one bit for every register, and with those of~\cite{KosowskiU18}, which embed the other two.

In~\cite{BlondinEJ18,BlondinEGHJ20} Blondin~\etal\ exhibit a construction that produces \emph{succinct} protocols, that is, protocols with $\O(\Poly(m))$ states. However, they do not analyse their stabilisation time. We demonstrate that they run in $\Omega(2^n)$ expected interactions. Loosely speaking, the reason is the use of transitions that ``revert'' the effect of other transitions. This allows the protocol to ``try out'' different distributions of agents, retracing its steps until it hits the right one, but also makes it very slow. So~\cite{BlondinEJ18,BlondinEGHJ20} produce \emph{succinct} but \emph{very slow} protocols. 

Is it possible to produce protocols that are both \emph{fast} and \emph{succinct}? We give an affirmative answer. We present a construction that yields for every formula $\varphi$ of QFPA of size $m$ a protocol with $\O(\Poly(m))$ states and stabilizing after
$\O(m^7 \cdot n^2)$ expected interactions. So our construction achieves optimal parallel stabilisation time in $n$, and, at the same time, yields protocols that are as succinct as the construction of~\cite{BlondinEGHJ20}. Moreover, for inputs of size $\Omega(m)$ (a very mild constraint when agents are molecules), the protocols have $\O(m)$ states. 

Our construction relies on  \emph{population computers}, a carefully crafted generalization of the  population protocol model of~\cite{AADFP06}. Population computers extend population protocols in three ways. First, they have \emph{$k$-way interactions} between more than two agents (but these are limited to involve at most two types of agents).
Second, they have a more flexible \emph{output condition}, defined by an arbitrary function that assigns an output to every subset of states, instead of to every state%
\footnote{Other output conventions for population protocols have been considered, see e.g.~\cite{brijder2018democratic}.}.
Finally, population computers can use  \emph{helpers}: auxiliary agents that, like leaders, help regular agents to coordinate themselves but whose number, contrary to leaders, is not known \emph{a priori}. 
The construction proceeds in three steps. First, we exhibit succinct population computers for all Presburger predicates in which every run is finite, that is, every run reaches a configuration at which no transition can occur.
We call these computers \emph{bounded}.
In a second step we prove a very general conversion theorem: \emph{Any} bounded computer of size $m$ can be translated into a population protocol with $\O(m^2)$ states and stabilizing in $2^{\O(m^2 \log m)} \cdot n^3$ expected interactions.
  
An important ingredient of the proof is a novel simulation of  interactions between an arbitrary number of  agents (common in chemical reaction networks) by binary interactions. In previous work this simulation required to introduce ``reverse'' interactions that ``undo'' the effect of others, which led to a large slowdown \cite{BlondinEGHJ20}. We limit ourselves to interactions between two types of agents (but an arbitrary number), and provide a new simulation that avoids the use of ``reverse'' interactions.

Finally, we exploit that our computers for Presburger predicates are not only bounded, but satisfy an additional property, called \emph{rapidness}, to improve the stabilisation time bound 
to $\O(m^7 \cdot n^2)$ interactions.

\medskip

\noindent\textbf{Related work.} Our results are for the canonical population protocol model of~\cite{AADFP04,AADFP06}, where (a) agents have a constant number of states, independent of the size of the population; (b) there are no leader agents; (c) time complexity is measured in terms of the expected number of interactions until stabilisation; and (d) protocols decide the predicate with probability 1 for all inputs. As mentioned above, in this model every protocol for the majority predicate needs $\Omega(n^2)$ expected interactions~\cite{AlistarhAEGR17}, i.e. \(\Omega(n)\) parallel stabilisation time, and so our construction is optimal. However, there is a vast body of work concerning variants of the model in which one or more of (a)-(d) are relaxed in order to find faster protocols for specific tasks. We briefly discuss some of this work.

If condition (a) is relaxed, then protocols running in $\O(\Polylog(n))$ instead of $\Omega(n)$ parallel stabilisation time have been proposed for specific tasks like majority or leader election. In the first  such protocol, presented in \cite{AlistarhGV15}, the number of states still grew very rapidly in the number of agents. Much subsequent work led to protocols where the number of states grows much more slowly; for example, in 2018 two surveys were published devoted only to this question \cite{AlistarhG18,ElsasserR18}. An asymptotically optimal protocol for majority with $\Theta(\log n)$ states and expected $\Theta(n \log n)$ interactions was given in \cite{DotyEGSSU21}, and an optimal protocol for leader election with $\Theta(\log \log n)$ states per agent and expected $\Theta(n \log n)$ interactions was presented in \cite{BerenbrinkGK20}. However, the properties of the model in which the number of states can grow in the size of the population are very different from the canonical one. In particular, the decision power of the model may go beyond Presburger arithmetic, depending on the rate at which the number of states is allowed to grow with the size of the population, on whether the agents ``know'' an upper bound on the size of the population or not, etc. (For example, if agents know an upper bound on the size of the population and have enough memory to implement a counter that can count up to that number, then they can elect a leader, and then let the leader simulate a broadcast population protocol with only one broadcasting agent \cite{BlondinEJ19}. To simulate a broadcast, the leader counts the number of agents it interacts with until the bound is reached. These protocols can compute  all predicates lying in the complexity class NL, which properly contains Presburger arithmetic.) To the best of our knowledge, the expressive power of many variants is not even known, and so the question of generic constructions yielding a protocol from a specification of the predicate cannot even be formulated. Our results might be used to produce succinct protocols for Presburger predicates in some of these variants, but this question is beyond the scope of this paper. 

Angluin \etal\ consider in~\cite{AAE08} a model that relaxes (b) by allowing one leader agent, (c) by measuring time in terms of the number of interactions until \emph{convergence} (loosely speaking, an execution converges after $t$ interactions if all configurations reached during the execution after $t$ steps exhibit the same consensus, observe that at that point \(t\) it might still be theoretically possible to reach non-consensus configurations), and (d) by allowing a small probability of error. They exhibit a construction that, given any Presburger predicate, produces a protocol running in $\O(\Polylog(n))$ parallel convergence time. Further, they show that zero probability of error can be achieved by suitably combining a fast protocol with 
small probability of error and a slow but exact \emph{backup protocol} (a technique later used in other works, like \cite{BerenbrinkEFKKR21}). Our work provides the first \emph{succinct} backup protocol. In future work we plan to investigate if there also exist succinct protocols running in $\O(\Polylog(n))$ parallel convergence time, with small probability of error. 

Kosowski and Uzn\'anski  improve the construction of~\cite{AAE08} by showing that $\O(\Polylog(n))$ parallel convergence time can also be achieved without a leader, i.e.\ by relaxing only (c) and (d)~\cite{KosowskiU18}. Further, using the exact protocols of~\cite{AAE08}, they provide protocols that run in $\O(n^\epsilon)$ parallel convergence time for arbitrary $\epsilon$ (i.e.\ only (c) is relaxed). Again, an interesting question for future work is whether these constructions can be made succinct, and, again,  our results can be seen as a first step that exhibits a succinct backup protocol.

\smallskip\noindent\textbf{Organization of the paper.} 
We give preliminary definitions in Section \ref{sec:pre} and introduce population computers in Section \ref{sec:PCdef}. Section \ref{sec:motivation} describes why previous constructions were either not succinct or slow. Section \ref{sec:results} gives an overview of the rest of the paper and summarises our main results.  Section \ref{sec:construction} describes bounded population computers for every Presburger predicate. Section \ref{sec:conversion} shows that every bounded computer can be converted into 
a succinct population protocol. Section \ref{sec:speed} shows that the protocols obtained for the bounded computers of Section \ref{sec:construction} are not only succinct but also fast.

\section{Preliminaries} \label{sec:pre}

\noindent\textbf{Multisets.}
Let $E$ be a finite set. A multiset over $E$ is a mapping $E \rightarrow \N$, and $\N^E$ denotes the set of all multisets
over $E$. We sometimes write multisets using set-like notation, e.g.\ $\multiset{a, 2 \cdot b}$
denotes the multiset $v$ such that $v(a) = 1$, $v(b) = 2$ and $v(e) = 0$ for every $e \in E \setminus \Set{a,b}$.
The empty multiset $\multiset{ }$ is also denoted $\multiset{}$. 

For $E' \subseteq E$, $v(E') := \sum_{e\in E'} v(e)$ is the number of elements in $v$ that are in $E'$. 
The \emph{size} of $v \in \N^E$ is $\Abs{v} := v(E)$. The \emph{support} of $v \in \N^E$ is the set $\Support{v} := \{ e \in E \mid v(e) > 0 \}$. If $E \subseteq \Z$, then we let $\multisum{v}:= \sum_{e \in E} e \cdot v(e)$ denote the sum of all the elements of $v$. 
Given $u,v \in \N^E$, $u + v$ and $u-v$ denote the multisets given by $(u+v)(e):=u(e)+v(e)$ 
and $(u-v)(e):=u(e)-v(e)$ for every $e \in E$. The latter is only defined if $u \geq v$ (i.e.\ $u(e)\ge v(e)$ for all $e\in E$).

\smallskip\noindent\textbf{Multiset rewriting transitions.} A \emph{multiset rewriting transition}, or just a \emph{transition}, is a pair 
$(r, s) \in  \N^E\times\N^E$, also written $r \mapsto s$. A transition $t=(r,s)$ is \emph{enabled} at $v \in \N^E$ if $v \ge r$, and its \emph{occurrence} 
leads to $v':=v-r+s$, denoted $v \rightarrow_t v'$. We call $v \rightarrow_t v'$ a \emph{step}. 

Let $\delta\subseteq \N^E\times\N^E$ denote a finite set of transitions. The following definitions depend on $\delta$. In the paper it will always be clear from context which $\delta$ is meant, hence we leave this dependence implicit.
The multiset $v$ is \emph{terminal} if it does not enable any transition $t\in\delta$.
An \emph{execution} is a finite or infinite sequence $v_0, v_1, \ldots $ of multisets such that $v \rightarrow_{t_1}v_1\rightarrow_{t_2} \cdots$ for some sequence $t_1, t_2, \ldots\in\delta$ of transitions.  A multiset $v'$ is \emph{reachable} from $v$ if there is an execution  $v_0, v_1, \ldots, v_k$ with $v_0=v$ and $v_k = v'$; we also say that the execution \emph{leads from $v$ to $v'$}.  
An execution is a \emph{run} if it is infinite or it is finite and its last multiset is terminal.  A run $v_0, v_1, \ldots $ is \emph{fair} if it is finite, or it is infinite and for every multiset $v$, if $v$ is reachable from $v_i$ for infinitely many $i \geq 0$, then $v= v_j$ for some $j \geq 0$.

\smallskip\noindent\textbf{Presburger arithmetic and QFPA.}  Presburger arithmetic is the first-order theory of addition~\cite{Haase18}. A formula $\varphi(x_1, \ldots, x_v)$ of Presburger arithmetic with $x_1, \ldots, x_v$ as free variables 
induces a predicate $P_\varphi \colon \N^\Nvar \rightarrow \{0,1\}$ defined by: $P_\varphi(n_1, \ldots, n_v)=1$ if{}f $\varphi(n_1, \ldots, n_v)$ is true. A predicate $\N^\Nvar \rightarrow \{0,1\}$ is \emph{definable in Presburger arithmetic} or just a \emph{Presburger predicate} if it is induced by some formula of Presburger arithmetic. Presburger predicates are known to be the same as the \emph{semilinear} predicates~\cite{Haase18}. Using the quantifier elimination procedure of Presburger arithmetic, one can show that every Presburger predicate is equivalent to a boolean combination of threshold and remainder predicates, defined as follows: 
A predicate $\varphi \colon \N^\Nvar \rightarrow \{0,1\}$ is a \emph{threshold predicate} if $\varphi(x_1, \ldots, x_\Nvar) = \left( \sum_{i=1}^\Nvar a_i x_i \geq c \right) $, where $a_1, \ldots, a_\Nvar, c \in \Z$, and a \emph{remainder predicate} if $\varphi(x_1, \ldots, x_\Nvar) = \left( \sum_{i=1}^\Nvar a_i x_i \equiv_{\Modulus} c \right)$, where $a_1, \ldots, a_\Nvar \in \Z$, $\Modulus \geq 1$, $c \in \Set{0, \ldots, \Modulus{-}1}$, and 
$a \equiv_{\Modulus} b$ denotes that $a$ is congruent to $b$ modulo $\Modulus$~\cite{Haase18}.  

We call the set of boolean combinations of threshold and remainder predicates  \emph{quantifier-free Presburger arithmetic}, or QFPA.  We define the \emph{size} of a Presburger predicate as the number of bits of a shortest formula of QFPA representing it, with coefficients written in binary. As mentioned in the introduction, 
Angluin~\etal\ showed that population protocols can decide exactly the Presburger predicates, or, by the above, the predicates definable in QFPA~\cite{AAER07}.

\section{Population Computers} \label{sec:PCdef}

Population computers are a generalization of population protocols. They allow us to give very concise descriptions of protocols
for Presburger predicates.

\smallskip\noindent\textbf{Syntax.}  A \emph{population computer} is a tuple $\Prot=(Q,\delta,I,O,H)$, where:
\begin{itemize}
	\item $Q$ is a finite set of \emph{states}. Multisets over $Q$ are called \emph{configurations}.
	\item $\delta\subseteq\N^Q\times\N^Q$ is a finite set of multiset rewriting transitions $r \mapsto s$ over $Q$ such that $\Abs{r} = \Abs{s} \ge2$ and $\Abs{\Support{r}} \le 2$. Further, we require that $\delta$ is a partial function, i.e.\ if $(r \mapsto s_1), (r \mapsto s_2) \in \delta$ then $s_1=s_2$. The \emph{arity} of a transition $r \mapsto s$ is the size of the multiset $r$ (or $s$). A transition is \emph{binary} if it has arity two. A population computer is \emph{binary} if all its transitions are binary. 
	\item $I\subseteq Q$ is a set of \emph{input states}. An \emph{input} is a configuration $C$ such that $\Support{C} \subseteq I$.
	\item $O \colon 2^Q \rightarrow\{0,1, \bot\}$ is an \emph{output function}. The \emph{output} of a configuration $C$ is $O(\Support{C})$. An output function $O$ is a \emph{consensus output function} if there is a partition $Q = Q_0 \cup Q_1$ of $Q$ such that $O(Q') = 0$ if{}f $Q' \subseteq Q_0$, $O(Q') = 1$ if{}f $Q' \subseteq Q_1$, and $O(Q')=\bot$ otherwise, for all $Q'\subseteq Q$.
	\item $H\in\N^{Q\setminus I}$ is a multiset of \emph{helper agents} or just \emph{helpers}.  A \emph{helper configuration} is a configuration $C$ such that $\Support{C} \subseteq \Support{H}$ and $C\ge H$.
\end{itemize}
In particular, note that $\delta$ is restricted in that a transition can only involve two types of agents. For example,
	$\multiset{p,p,q} \mapsto \multiset{p,q,o}$ (which in the following is written simply as $p,p,q \mapsto p,q,o$) is allowed, but $p,q,o \mapsto p,p,q$ is not. 

\smallskip\noindent\textbf{Semantics.}  Intuitively, a population computer decides which output (0 or 1) corresponds to an input $C_I$ as follows. 
It adds to the agents of $C_I$ an arbitrary helper configuration $C_H$ of agents to produce the initial configuration $C_I + C_H$.  Then it starts to execute a run
and lets it stabilise to configurations of output $1$ or output $0$.  Formally, the \emph{initial configurations} of $\Prot$
for input $C_I$ are all configurations of the form $C_I + C_H$ for some helper configuration $C_H$. 
A run $C_0 \, C_1 \ldots$ \emph{stabilises to $b$} if there exists an $i \geq 0$ such that $O(\Support{C_i})=b$ and $C_i$ only reaches configurations $C'$ with $O(\Support{C'})=b$. 
An input $C_I$ \emph{has output $b$} if for every initial configuration $C_0=C_I+C_H$, every fair run starting at $C_0$ stabilises to $b$.  A population computer $\Prot$ \emph{decides} a predicate $\varphi \colon \N^I\rightarrow\{0,1\}$ if every input $C_I$ has output $\varphi(C_I)$. Observe that, crucially, the protocol has to work for \emph{any} helper configuration $C_H$.

\smallskip\noindent\textbf{Terminating and bounded computers.}  A population computer is \emph{bounded} if no run starting at any initial configuration $C$ is infinite, and \emph{terminating} if no fair run starting at $C$ is infinite, i.e.\ every fair run ends at a terminal configuration\footnote{This is the classical notion of termination under fairness in concurrent systems \cite{francez86}. It differs from recent notions of termination in the literature on population protocols, e.g.\ the one of \cite{DotyE19}.}.

\begin{example}
Consider a population computer with states $\{q, p\}$, input state $q$, and transitions $t=\multiset{2 \cdot q} \mapsto  \multiset{2 \cdot q}$ and $u=\multiset{2 \cdot q} \mapsto  \multiset{2 \cdot p}$. The computer is not bounded, because, for example, there is an infinite run from the initial configuration $C = \multiset{2 \cdot q}$, namely $C \, C \, C \, C \ldots$. However, it is terminating because every fair run eventually reaches a terminal configuration of the form  $\multiset{n \cdot p}$ or $\multiset{q, n \cdot p}$ for some $n \geq 0$.
\end{example}

\noindent\textbf{Graphical notation.} 
We visualise population computers as Petri nets (see e.g.\ the left part of Figure \ref{fig:population_computer_modulo} in page \pageref{fig:population_computer_modulo}). 
Places (circles) and transitions (squares) represent respectively states and transitions. The number of agents currently occupying a state is written within the place representing the state. 
For example, the computer on the left of Figure \ref{fig:population_computer_modulo} has states 0,1,2,4,8,16 and $\multiset{4 \cdot \text{16}} \mapsto \multiset{2 \cdot \text{0}, \text{1}, \text{8}}$ is one of its transitions. 
Currently, there are $12$ agents in state 0 and no agents elsewhere.

\smallskip\noindent\textbf{Size and adjusted size.} Let \(\Prot=(Q,\delta,I,O,H)\) be a population computer. We assume that \(O\) is described as a boolean circuit with \(\size(O)\) gates. 
For every transition \(t\) let \(\Abs{t}\) be the arity of $t$.
The \emph{size} of \(\Prot\) is \(\size(\Prot):=\Abs{Q}+\Abs{H}+\size(O)+\sum_{t \in \delta}\Abs{t}\). 
If $\Prot$ is binary, then (as for population protocols) we do not count the arities and define the \emph{adjusted size} \(\size_2(\Prot):=\Abs{Q}+\Abs{H}+\size(O)\). For a binary computer $\Prot$ we have  $\sum_{t \in \delta}\Abs{t} = 2 \Abs{\delta} \leq 2 \Abs{Q}^2$, and so in particular $\size(\Prot) \leq  2\size_2(\Prot)^2$. Observe that both the arity of a transition $r \mapsto s$ and the size of the helper multiset $H$ are defined as the number of elements of the multisets $r$ and $H$, respectively, and not as the number of bits of these numbers. In other words, we consider their size in unary. This makes our result about the existence of succinct population computers stronger.

\smallskip\noindent\textbf{Population protocols and speed of a protocol.} 
A population computer $\Prot=(Q,\delta,I,O,H)$ is a \emph{population protocol} if it is binary, has no helpers ($H = \emptymultiset
$), and $O$ is a consensus output. It is easy to see that this definition coincides with the one of  \cite{AADFP06}.

The speed of a binary population computer without helpers, and so in particular of a population protocol, is defined as follows. 
We assume a probabilistic execution model for a population protocol $\Prot$ in which at a configuration $C$ two agents are picked uniformly at random and execute a transition, if possible, moving to a configuration $C'$ (by assumption two agents enable at most one transition). This is called an \emph{interaction}. Repeated occurrences of interactions, starting from an input $C_0$, produce an \emph{execution} $C_0 \, C_1 \, C_2 \ldots$. An execution \emph{stabilises at time $t$} if every configuration $C$ reachable from $C_t$ satisfies $O(\Support{C})=O(\Support{C_t})$, and \emph{converges} after $t$ interactions if $C_{t'}$ satisfies $O(\Support{C_t'})=O(\Support{C_t})$ for every $t' \ge t$. 
Let $\Prot$ be a protocol that decides a given predicate $\varphi$. Given an input $C_I$, we say that $\Prot$ decides $\varphi(C_I)$ \emph{within $k$ interactions} if the expected value of the earliest stabilisation time of the executions starting at $C_I$ is at most $k$. (The earliest stabilisation time is  the random variable that assigns to each execution the smallest time at which it stabilises.) Let $T \colon \N \rightarrow \N$. We say that $\Prot$ decides $\varphi$ \emph{within $T$ interactions} if 
it decides $\varphi(C_I)$ within $T(n)$ interactions for every $n \geq 0$ and for every input $C_I$ of size $n$.
See e.g.\ \cite{AAE08} for more details. Notice that in this paper we study the stabilisation time, and not the convergence time \cite{KosowskiU18}.

\smallskip\noindent\textbf{Population computers vs.\ population protocols.} 
Population computers generalise population protocols in three ways:
\begin{itemize}
	\item They have non-binary transitions, but only those in which the interacting agents populate at most two states. 
	\item They use a multiset $H$ of auxiliary helper agents, but the addition of more helpers must not change the output of the computation. Intuitively, contrary to the case of leaders, agents do not know any upper bound on the number of helpers, only the multiset $H$. Since, by definition, the initial configurations contain \emph{at least} $H$ helpers but possibly more, the agents only know a lower bound. 
	\item They have a more flexible output condition. A population protocol accepts or rejects by moving all agents to accepting or rejecting states, respectively. In contrast, population computers look at the states that are present in the current configuration, and then choose an output based on that set. 
\end{itemize}

\smallskip\noindent\textbf{Fast and succinct population protocols.} 
As announced in the introduction, the goal of this paper is to show that every Presburger predicate has a \emph{fast} and \emph{succinct} population protocol. We formalise these notions. 

\begin{definition}
\label{def:fastsuccinct}
Let $\mathcal{F}$ be a function that assigns to every predicate $\varphi\in\QFPA$ a population protocol $\mathcal{F}(\varphi)$ deciding $\varphi$. 
\begin{itemize}
\item $\mathcal{F}$ produces \emph{succinct} protocols if there exists a constant $a \geq 0$ such that for every $\varphi\in\QFPA$ the protocol $\mathcal{F}(\varphi)$ has \(\O(\Abs{\varphi}^a)\) states. 
\item $\mathcal{F}$ produces \emph{fixed-parameter fast} protocols if there exists a function $f \colon \N \rightarrow \N$ and a constant $b \geq 0$ such that for every $\varphi\in\QFPA$ the protocol $\mathcal{F}(\varphi)$ decides $\varphi$ within $\O(f(\Abs{\varphi}) \cdot n^b)$ interactions.
\item $\mathcal{F}$ produces \emph{fast} protocols if there exist constants $a,b \geq 0$ such that for every $\varphi\in\QFPA$ the protocol $\mathcal{F}(\varphi)$ decides $\varphi$ within $\O(\Abs{\varphi}^a \cdot n^b)$ interactions.
\end{itemize}
\end{definition}

We call an effectively computable function $\mathcal{F}$ a \emph{construction} or a \emph{procedure}. The formalization of ``every Presburger predicate has a fast and succinct protocol''
is ``there exists a construction that produces fast and succinct protocols''. The next section explains why none of the constructions in the literature produces fast and succinct protocols. The rest of the paper describes
a new construction that produces fast and succinct protocols. 

\section{Previous Constructions: Angluin \etal\ and Blondin \etal} \label{sec:motivation}

We show by means of some examples that the construction by Angluin \etal\ \cite{AADFP06} does not produce succinct protocols, and the construction of Blondin \etal\ \cite{BlondinEGHJ20} does not produce 
fast protocols, not even fixed-parameter fast.

\begin{example}\label{ex1}
Consider the protocol of \cite{AADFP06} for the predicate $\varphi =(x - y \geq 2^\h)$. The states are the triples 
$(\ell, b, u)$ where $\ell \in \{A,P\}$, $b \in \{Y,N\}$ and $-2^\h \leq u \leq 2^\h$. Intuitively, $\ell$ indicates whether
the agent is active (A) or passive (P), $b$ indicates whether it currently believes that $\varphi$ holds (Y) or not (N), and
$u$ is the agent's wealth, which can be negative. Agents for input $x$ are initially in state $(A, N, 1)$, and agents for $y$ in $(A, N, -1)$. 
If two passive agents meet their encounter has no effect. If at least one agent is active, then the result of the encounter  is given by the transition
$(*, *, u), (*, *, u') \mapsto (A, b, q), (P, b, r)$
where $b = Y$ if $u+u' \geq 2^\h$ else $N$; $q= \max(-2^\h, \min(2^\h, u+u'))$; and $r = (u+u') - q$. 
The protocol stabilises after $\O(n^2 \log n)$ expected interactions \cite{AADFP06}, but it has $2^{\h+1}+1$ states, exponentially many in $\Abs{\varphi} \in \Theta(\h)$.
\end{example}

\begin{example}\label{ex2}
We give a protocol for $\varphi =(x - y \geq 2^\h)$ with a polynomial number of states, very similar to the protocol of \cite{BlondinEGHJ20}. The procol is defined in two steps. First, we remove states and transitions from the protocol of \Cref{ex1}, retaining only the states $(\ell, b, u)$ such that $u$ is a power of $2$, and some of the transitions involving these states:
\begin{gather*}
\begin{array}{rcll}
(*, *, 2^i),  (*, *, 2^i) &\mapsto &(A,N, 2^{i+1}),  (P, N, 0) &\mbox{for every $0 \leq i \leq \h-2$} \\
(*, *, 2^{\h-1}),  (*, *, 2^{\h-1}) &\mapsto & (A, Y, 2^\h), (P, Y, 0) \\
(*, *, -2^i),  (*, *, -2^i) &\mapsto & (A, N, -2^{i+1}), (P, N, 0) &\mbox{for every $0 \leq i \leq \h-1$}\\
(*, *, 2^i),  (*, *, -2^i) &\mapsto & (A, N, 0),  (P, N, 0) &\mbox{for every $0 \leq i \leq \h-1$} 
\end{array}
\end{gather*}
This protocol is not yet correct. For example, for $\h=1$ and the input $x=2, y=1$, the protocol can reach in one step the configuration in which the three agents (two $x$-agents and one $y$-agent) are in states $(A, Y, 2), (P,Y,0), (A, N, -1)$, after which it gets stuck. In \cite{BlondinEGHJ20} this is solved in a second step that adds the following ``reverse'' transitions:
\begin{gather*}
\begin{array}{rcll}
(A, N, 2^{i+1}),  (P, N, 0) &\mapsto &  (A, N, 2^i),  (P, N, 2^i)   &\mbox{for every $0 \leq i \leq \h-2$} \\
(A, Y, 2^\h), (P, Y, 0) &\mapsto & (A, N, 2^{\h-1}),  (P, N, 2^{\h-1})   \\
(A, N, -2^{i+1}), (P, N, 0) &\mapsto & (A, N, -2^i),  (A, N, -2^i)   &\mbox{for every $0 \leq i \leq \h-1$}\\
\end{array}
\end{gather*}
The protocol has only $\Theta(\h)$ states and transitions, but runs within $\Omega(n^{2^\h-2})$ interactions. Consider the inputs $x, y$ such that $x -y = 2^\h$, and let $n:=x + y$. Say that an agent is \emph{positive} at a configuration if it has positive wealth at it. The protocol can only stabilise if it reaches a configuration with exactly one positive agent with wealth $2^\h$. Consider a configuration with $i < 2^\h$ positive agents. The next configuration can have $i-1$, $i$, or $i+1$ positive agents. 
One can see that the probability of $i+1$ positive agents is $\Omega(1/n)$, the probability of $i-1$ positive agents is only $\O(1/n^2)$, and the expected number of interactions needed to go from $2^\h$ positive agents to only $1$ is $\Omega(n^{2^\h-1})$. 
Let use see why. First, let us analyse the probabilities of $i+1$ and $i-1$ positive agents:
\begin{itemize}
\item $i \rightarrow i+1$. This happens whenever a non-zero agent with wealth different from $1$ or $-1$ meets a zero agent. Since $i$ is the number of positive agents, the configuration $C$ has $n-i > n -2^\h$ zero agents. 
Further, since the total wealth is $2^\h$ and there are less than $2^\h$ non-zero agents, at least one agent has wealth bigger than $1$. So the probability is at least $p^+:= 2(n-2^\h)/n(n-1)$, and so $\Omega(1/n)$.
\item $i \rightarrow i-1$. This can only happen if two non-zero agents meet. Since there are less than $2^\h$ non-zero agents, $p^-:=2^\h (2^\h-1) / n(n-1)$ is an upper bound, and so the probability is $\O(1/n^2)$ for fixed $\h$.
\end{itemize}
So we obtain a random walk with states $\{2^\h, 2^{\h -1}, \dots,1\}$, initial state $2^\h$, target \(1\), and probabilities \(p_1:=p^+ \in \mathcal{O}(1)\) of moving towards $2^\h$, and probability \(p_2 := 1-p^+ \in \Theta(1)\) of moving towards \(1\). The 
expected time to state $1$ underapproximates the expected stabilisation time of the protocol, because in the walk one cannot stay in a state and must instead move towards \(1\).  Standard results on the Gambler's Ruin problem yield $E[T_{2^d}] \in \Omega(n^{2^\h-2})$ \cite{spitzer}. 

\end{example}

Recall that predicates of quantifier-free Presburger arithmetic are boolean combinations of threshold and remainder predicates. Therefore,
the size of a predicate depends on the number $d$ of bits of the largest coefficient, and the number $s$ of predicates
of the boolean combination. Examples~\ref{ex1} and~\ref{ex2} show that in the constructions of~\cite{AADFP06} and~\cite{BlondinEGHJ20} 
the number of states, respectively the expected number of iterations, grows exponentially in $d$. The next example shows that the same holds for 
the parameter $s$.

\begin{example}\label{ex3}
Given protocols $\Prot_1, \Prot_2$ with $n_1$ and $n_2$ states deciding predicates $\varphi_1$ and $\varphi_2$, Angluin \etal\ construct in \cite{AADFP06} a protocol $\Prot$ for $\varphi_1 \wedge \varphi_2$ with $n_1 \cdot n_2$ states. 
(The states of $\Prot$ are all possible pairs of states $(q, r)$, where $q$ and $r$ are states of $\Prot_1$ and $\Prot_2$, respectively.)
It follows that the number of states of a protocol for $\varphi := \varphi_1 \wedge \cdots \wedge \varphi_\Npred$ grows exponentially in $\Npred$, and so in $\Abs{\varphi}$. 

Blondin \etal\ give an alternative construction with polynomially many states~\cite[Section 5.3]{BlondinEGHJ20}. However, the protocol contains transitions that, as in the previous example, reverse the effect of other transitions, and make the protocol very slow. The problem is already observed in the toy protocol with states $q_1, q_2$ and transitions $q_1, q_1 \mapsto   q_2, q_2$ and $q_1, q_2 \mapsto   q_1, q_1$. (Similar transitions are used in the initialisation of \cite{BlondinEGHJ20}.) Starting with an even number $n\geq 2$ of agents in $q_1$, eventually all agents move to $q_2$ and stay there. We show that the expected number of interactions is $\Omega(2^{n/10})$. 

Let \((C_0,C_1,\dots)\) be the stochastic process induced by the toy protocol, where $C_i$ indicates the configuration after $i$ interactions. Since at every step agents are chosen independently and uniformly at random, the process is a Markov chain. We can identify the state space of the chain with the set \(\{0,1,2,\dots,n\}\) via the mapping \(C_t \mapsto C_t(q_1)\). At state $i$,  three transitions can happen, leading to states  \( i+1\), \( i\) and \(i-2\). The probabilities of moving to \(i+1\) and \(i-2\) are \(\frac{i(n-i)}{n(n-1)}\) and \(\frac{i(i-1)}{n(n-1)}\), respectively. The goal is to reach state \(0\) from state \(n\).

In order to obtain a lower bound on the number of steps, let us reduce the states to \(\{0,\dots,\lfloor n/5 \rfloor\}\), replacing the transition \(\lfloor n/5 \rfloor \mapsto \lfloor n/5 \rfloor+1\) by a self-loop at \(\lfloor n/5 \rfloor\), and starting at state \(\lfloor n/5 \rfloor\) instead of $n$. This only reduces the number of steps to the goal. In this new chain, the quotient of the probabilities of moving to \(i+1\) and \(i-2\) is \(\frac{i(i-1)}{i(n-i)}\leq \frac{i}{n-i}\leq \frac{1}{4}\) for all states $i$ such that $i+1$ and $i-2$ exist. 
It is easy to see that we can simplify the chain further, without increasing the number of steps to the goal, into a chain with probability \(4/5\)  and \(1/5\) of moving to from \(i\) to \(i+1\)  and to \(i-2\), respectively.
The expected number of steps to the goal in this chain is the same as for a random walk with states  \(\{0,\dots,\lfloor n/10 \rfloor\}\), biased by a factor of \(2\) in the ``wrong'' direction. (Indeed, the fact that in the chain we move from \(i\) to \(i-2\), while in the random walk we move from \(i\) to \(i-1\), is compensated by the probability in the chain being lower by a factor of \(4\)). This biased random walk needs \(\Omega(2^{n/10})\) steps in expectation until it reaches \(0\) from \( \lfloor n/10 \rfloor \)  \cite{spitzer}.
\end{example}

\section{Constructing Fast and Succinct Protocols: Overview} \label{sec:results}

Given a predicate $\varphi\in\QFPA$, in the rest of the paper we show how to construct a fast and succinct protocol deciding $\varphi$.
We give an overview of the procedure, which first constructs a population computer for a different predicate, called $\Double(\varphi)$, and then 
transforms it into a protocol for $\varphi$. We start by defining $\Double(\varphi)$.

\begin{definition}\label{def:double}
Let $\varphi\in\QFPA$ be a predicate over variables $x_1, \ldots, x_{\Nvar}$. The predicate $\Double(\varphi) \in \mathit{QFPA}$ over variables $x_1, \ldots, x_{\Nvar}, x_1' , \ldots, x_{\Nvar}'$ is
defined as follows: For every $i \in \{1, \ldots, \Nvar\}$, replace every occurrence of $x_i$ in $\varphi$ by $x_i + 2 x_i'$. 
\end{definition}

For example, if $\varphi = (x -y \geq 0)$ then $\Double(\varphi) = (x+2x'-y - 2y' \geq 0)$. Observe that 
$\size{\Double(\varphi)} \in \O(\size{\varphi})$. The procedure consists of the following steps:
\begin{enumerate}
\item Construct a succinct bounded population computer $\Prot$ deciding $\Double(\varphi)$.
\item Convert $\Prot$ into a fixed-parameter fast and succinct population protocol $\Prot'$ deciding $\varphi$ for inputs of size 
$\Omega(\Abs{\varphi})$.
\item Prove that $\Prot'$ is not only fixed-parameter fast, but even fast. 
\item Convert $\Prot'$ into a  fast and succinct protocol deciding $\varphi$ for all inputs.
\end{enumerate}

\begin{remark}
The restriction to inputs of size $\Omega(\Abs{\varphi})$ is mild. Indeed, in
the intended applications of population protocols, like molecular programming,  
the number of agents is typically much larger than $\Abs{\varphi}$.
In these applications the behaviour of the protocol for small inputs is irrelevant, and so Step 4 is of little interest. We include it for completeness.
\end{remark}

We describe each of the steps in some more detail, and the results we obtain.

\medskip\noindent \textbf{Step 1 (Section \ref{sec:construction}).} We exhibit a procedure that constructs succinct and bounded 
population computers for all Presburger predicates (and so, in particular, for any predicate of the form $\Double(\varphi)$). More precisely, the section proves the following theorem:

\begin{restatable}{theorem}{thmmainA}\label{thm:mainA}
For every predicate $\varphi\in\QFPA$ there exists a bounded population computer of size $\O(\Abs{\varphi})$ that decides $\varphi$.
\end{restatable}

\medskip\noindent \textbf{Step 2 (Section~\ref{sec:conversion}).} Loosely speaking, the section shows that every succinct bounded computer for $\Double(\varphi)$ can be transformed into a fixed-parameter fast and succinct protocol for $\varphi$. Formally, it proves the following theorem:

\begin{restatable}{theorem}{thmmainB}\label{thm:mainB1}
Every bounded population computer of size $m$ deciding $\Double(\varphi)$ can be converted into a terminating population protocol with $\O(m^2)$ states that decides $\varphi$ within $2^{\O(m^2 \log m)}\,n^3$ interactions for all inputs of size $\Omega(m)$.
\end{restatable}

Observe that this theorem relates boundedness, a \emph{qualitative} property of population computers that can be proved using classical techniques like ranking functions, to the \emph{quantitative} property of stabilising in $2^{\O(m^2 \log m)}\,n^3$ expected interactions. This greatly simplifies the task of designing fixed-parameter fast protocols. The theorem is proved by means of a sequence of conversions enforcing the conditions that make a population computer a population protocol: only binary transitions, no helpers, and consensus output. 

\medskip\noindent \textbf{Step 3 (Section~\ref{sec:speed}).}  Theorem  
\ref{thm:mainB1} does not yet prove the existence of succinct and fast protocols, because of the $2^{\O(m^2 \log m)}$ term. 
On the other hand, it holds for \emph{arbitrary} bounded population computers, not only the ones defined in Section \ref{sec:construction}. So we trade generality against speed. We show that for the protocols of Section \ref{sec:construction}
the conversion of Section~\ref{sec:conversion} yields fast protocols; more precisely, we reduce the  $2^{\O(m^2 \log m)}$ term to $m^7$. Moreover, we also reduce the dependence on $n$. It is known that population protocols deciding majority need $\Omega(n^2)$ interactions in expectation~\cite{AlistarhAEGR17}\footnote{In fact, $\Omega(n^2)$ interactions are required for “most” semilinear predicates~\cite{BellevilleDS17}} and so, since Theorem \ref{thm:mainB1} only gives a $\O(n^3)$ upper bound, there is still a gap. Our refined analysis closes the gap.  Formally, Section \ref{sec:speed} proves:

\begin{restatable}{theorem}{thmmainc}\label{thm:mainc}
For every predicate $\varphi\in\QFPA$ there exists a terminating population protocol of size $\O(\Abs{\varphi})$ that decides $\varphi$ in $\O(\Abs{\varphi}^7\,n^2)$ interactions for inputs of size $\Omega(\Abs{\varphi})$.
\end{restatable}

\noindent So Theorem \ref{thm:mainc}  shows that our construction is optimal in $n$. Regarding the number of states, an $\Omega(\Abs{\varphi}^{1/4})$ lower bound was shown in 
\cite{BlondinEJ18}, which leaves a polynomial gap. We conjecture that the lower bound  of \cite{BlondinEJ18} can be improved, but this question exceeds the scope of this paper and is left for future research. 

\medskip\noindent \textbf{Step 4.} It remains to obtain succinct protocols that are fast for all inputs not only for those of size $\O(m)$. This step is carried out by direct application of a technique of \cite{BlondinEGHJ20} that, given a predicate $\varphi$ and  a constant $\ell \in \O(\Abs{\varphi}^3)$, constructs a succinct protocol deciding $\varphi$ for inputs of size at most $\ell$ (see Section  6 of \cite{BlondinEGHJ20}).
This protocol can be combined with the one obtained in Step 3  to yield a succinct protocol that decides $\varphi$ for all inputs (see Section 3 of \cite{BlondinEGHJ20}), and has speed $\O(\Abs{\varphi}^7\,n^2)$ for all inputs of size \(\Omega(m)\), and so $\O(\Abs{\varphi}^7\,n^2)$ asymptotic speed. Applying Theorem \ref{thm:mainc} we directly obtain the following result:

\begin{theorem}\label{thm:final}
For every \(\varphi \in \QFPA\) there exists a succinct terminating population protocol of size $\O(\Abs{\varphi}^a)$, for some constant $a \geq 0$, that decides \(\varphi\) in at most $\O(\Abs{\varphi}^7 \,n^2)$ interactions.
\end{theorem}


\section{Succinct Bounded Population Computers for Presburger Predicates} \label{sec:construction}
This section is structured as follows. \Cref{subsec:method} introduces a generic method to construct succinct computers for predicates whose coefficients are powers of $2$. \Cref{subsec:construction_modulo} and \Cref{subsec:construction_threshold} apply the method to remainder and threshold predicates, respectively. \Cref{subsec:construction_combination} shows how to generalise the construction to remainder and threshold predicates with arbitrary integer coefficients, and how to construct computers for boolean combinations of remainder and threshold predicates.

\subsection{A generic method to construct succinct computers}
\label{subsec:method}
We introduce a method to construct computers for remainder predicates 
$\sum_{i=1}^\Nvar a_i x_i \equiv_{\Modulus} c$ and threshold predicates $\sum_{i=1}^\Nvar a_i x_i \geq c$. We call $\{a_1, \ldots, a_\Nvar\}$ the set of \emph{coefficients} of the predicate.

The states of the computer for a predicate $\varphi$ are a finite set of integers, including the coefficients of $\varphi$ and $0$. The initial states are the coefficients of $\varphi$, and all helpers (in a number to be determined) are initially in state $0$. With this choice, a configuration $C$ is a multiset of integers, and we define its \emph{value} as $\multisum{C}$. For example, a configuration that puts one agent in state $8$, three agents in state $2$, and two agents in state $0$ has value $8+ 3 \cdot 2 + 2 \cdot 0 = 14$.  Observe that helpers have value $0$, and so all initial configurations $C_I+C_H$ for a given input $C_I$ have the same value.

We introduce some terminology. 
A configuration $C$ \emph{satisfies} a remainder predicate $\sum_{i=1}^\Nvar a_i x_i \equiv_{\Modulus} c$  if $\multisum{C} \equiv_{\Modulus} c$, and a threshold predicate $\sum_{i=1}^\Nvar a_i x_i \geq  c$ if $\multisum{C} \geq  c$. For initial configurations \(C\), this definition coincides with \(\varphi(C)=1\), hence the terminology satisfy. However, this definition of satisfying a predicate is now also applicable if \(C\) includes states other than the coefficients of \(\varphi\). While this is the obvious way to perform this extension, it is important to highlight that with this definition, whether a configuration \(C\) satisfies a predicate \(\varphi\) is \emph{independent} of the coefficients of \(\varphi\). Instead satisfying a predicate only depends on the modulus \(\Modulus\) or threshold \(c\). 

Satisfying a predicate induces an equivalence relation: two configurations are \emph{equivalent with respect to $\varphi$} if both of them satisfy $\varphi$, or none of them does. (When $\varphi$ is clear from the context, we just say that the configurations are equivalent.) In particular, two configurations with the same value are equivalent with respect to any predicate. 

Recall that $\Support{C}$ is the configuration given by $\Support{C}(q) = 1$ if $C(q) \geq 1$ and 
$\Support{C}(q) = 0$ otherwise. A configuration $C$ is \emph{well-supported} w.r.t.\ $\varphi$ if it is equivalent to $\Support{C}$. Loosely speaking, whether well-supported configurations satisfy $\varphi$ or not depends only on their supports. In particular, if $C= \Support{C}$, i.e.\ if $C$ puts at most one agent in each state, then $C$ is well-supported. However, the converse does not hold:

\begin{example}
Consider the predicate  $\varphi = \sum_{i=1}^\Nvar a_i x_i \geq c$, and assume $a_1 \geq c$. The configuration $C=\multiset{a_1, a_1}$ is well-supported w.r.t.\ $\varphi$. Indeed, we have $\multisum{C}=2 a_1$ and  
$\multisum{\Support{C}}= a_1$, and so both of them satisfy $\varphi$. However, we have $\Support{C} = \{a_1\}$, and so $C \neq \Support{C}$.
\end{example}

Our generic method for constructing computers is based on the following simple fact:

\begin{proposition}
\label{prop:method}
Let $\varphi$ be a remainder or threshold predicate. Let $\Prot$ be a computer with integers as states, the coefficients of $\varphi$ as initial states, and all helpers initially in state $0$. If $\Prot$ satisfies the following four properties, then it decides $\varphi$:
\smallskip
\begin{enumerate}
\item $\Prot$ is bounded.  \label{spec:prop1}
\item Transitions preserve equivalence, i.e.\ if $C \mapsto C'$ then $C$ and $C'$ are equivalent w.r.t.\ $\varphi$. \label{spec:prop2}
\item Terminal configurations are well-supported. \label{spec:prop3}
\item The output function $O$ is given by $O(S)=1$ if{}f $S$ satisfies $\varphi$. \label{spec:prop4}
\end{enumerate}
\end{proposition}
\begin{proof}
Since $\Prot$ is bounded, every run starting at an initial configuration $C_0=C_I+C_H$ reaches a terminal configuration $ C_T$, and we have:
\begin{center}
\begin{tabular}{rlr}
       & $C_I$ satisfies $\varphi$ \\
if{}f & $C_0$ satisfies $\varphi$ & ($\multisum{C_0}=\multisum{C_I}$ because helpers have value $0$)  \\ 
if{}f & $C_T$ satisfies $\varphi$ & ($C_T$ is equivalent to $C_0$ w.r.t.\ $\varphi$ by (\ref{spec:prop2})) \\
if{}f & $\Support{C_T}$ satisfies $\varphi$  & (by (\ref{spec:prop3})) \\
if{}f & $O(\Support{C_T}) = 1$ & (by (\ref{spec:prop4})) \\
if{}f & $\Prot$ returns $1$.
\end{tabular}
\end{center}
\end{proof}

In the next two sections, we apply this method to remainder and threshold predicates whose coefficients are positive or negative powers of $2$. Given such a predicate, we define a computer satisfying the properties of \Cref{prop:method}.

\subsection{Population computers for remainder predicates} \label{subsec:construction_modulo}

Since every equivalence class modulo \(\Modulus\) has a representative \(r\) with \(0 \leq r \leq \Modulus-1\), every remainder predicate with integer coefficients is equivalent to a remainder predicate with coefficients in this range, hence we only consider this case. For example \( 7x - 2y \equiv_7 11\) can be rewritten to \(5y \equiv_7 4\). Further, we assume in this section that the coefficients are powers of $2$, a restriction lifted later in Section \ref{subsec:construction_combination}. So we let $\powp = \{ 2^i \mid i \geq 0 \}$, and for the rest of the section fix a remainder predicate

$$\varphi := \sum_{i=1}^\Nvar a_i x_i \equiv_{\Modulus} c   \quad  \mbox{ where $\{a_1, \ldots, a_\Nvar\} \subseteq \powp \cap \Set{1,...,\Modulus{-}1}$ }$$

\noindent Let  $\h :=\lceil \log_2 \Modulus \rceil$. We define the computer $\Prot_\varphi=(Q,\delta,I,O,H)$ as follows.

\smallskip\noindent\textbf{States and initial states. } 
The set of states of $\Prot_\varphi$ is $Q:= \powp_\h \cup \{ 0 \}$, where $\powp_\h  = \{2^i \mid 0 \leq i \leq d \}$.
The initial states are the coefficients of $\varphi$, i.e.\ $I:= \{a_1, \ldots,a_\Nvar\}$. 

\smallskip\noindent\textbf{Transitions.} The transitions of $\Prot_\varphi$ transform configurations into equivalent configurations  ``closer'' to being well-supported. Configurations $C$ that put at most one agent in each state of $\powp_d$ satisfy $\multisum{C} = \multisum{\Support{C}}$, hence are well-supported\footnote{Observe that $C$ and $\Support{C}$ may differ on the number of agents they put in state $0$, but such agents have value $0$.}. So for each state $2^i   \in \powp_d \setminus \{2^d\}$ we add to $\delta$ a transition that reduces the number of agents in $2^i$, if there are more than one:

\begin{itemize}
\item For $2^i \in \{2^0,\ldots, 2^{\h-1}\}$, we add to $\delta$ a transition that takes two agents from state $2^i$, and puts one agent each in the states $2^{i+1}$ and $0$:
\begin{gather*}
	\begin{aligned}
		2^i&, 2^i &&\mapsto& 2^{i+1}&,0  	&\qquad \text{for } 0 \leq i \leq  \h-1 
	\end{aligned} \TraName[mod:combine:main]{combine}
\end{gather*}
\end{itemize}

We still need a transition that reduces the number of agents in $2^d$. So we add to $\delta$ a transition
that replaces an agent in $2^\h$ by a multiset of agents $r$ satisfying $\multisum{r} = 2^\h - \Modulus$, preserving equivalence:

\begin{itemize}
\item Let $b_{\h-1}b_{\h-2} \ldots b_0$ be the binary encoding of $2^\h-\Modulus$, and let $\{i_1, \ldots, i_j\} \subseteq [0, \h-1]$ be the set of positions at which the binary encoding has a $1$. We add to $\delta$ a transition
\begin{gather*}
	\begin{aligned}
		2^\h&, \underbrace{0,\dots,0}_{j-1} &&\mapsto& 2^{i_1}, \dots , 2^{i_j}
	\end{aligned}\TraName[mod:modulo:main]{modulo}
\end{gather*}
\noindent For example, if $\Modulus = 19$, then $\h=5$, $2^\h-m = 13$,  $b_{\h-1}b_{\h-2} \ldots b_0 = 1101$, $\{i_1, \ldots, i_j\} = \{0, 2, 3\}$, and \TraRef{mod:modulo:main} is $2^5, 0, 0 \mapsto 2^3, 2^2, 2^0$.
\end{itemize}

As shown below, transitions  \TraRef{mod:combine:main} and \TraRef{mod:modulo:main} are enough for correctness. However, in order to make the protocol faster we also add to $\delta$ a last transition
that takes $\h$ agents from state $2^\h$ and replaces them by a multiset of agents with total value $\h \cdot 2^\h \bmod \Modulus$:

\begin{itemize}
\item Let $b_\h b_{\h-1} \ldots b_0$ be the binary encoding of $\h \cdot 2^\h \bmod \Modulus$, and let $\{i_1, \ldots, i_j\} \subseteq [0, d-1]$ be the set of positions at which the binary encoding has a $1$,
i.e.\ $b_{i_1}= \cdots = b_{i_j}=1$. We introduce the transition:
\begin{gather*}
	\begin{aligned}
		\underbrace{2^\h,\dots,2^\h}_{d} & &&\mapsto& 2^{i_1}&, \dots, 2^{i_j}, \underbrace{0, \ldots, 0}_{d-j}
	\end{aligned}\TraName[mod:fast-modulo:main]{fast modulo}
\end{gather*}
\end{itemize}
Note that $d$ agents in $2^\h$ can always represent their combined value modulo $\Modulus$.

\smallskip\noindent\textbf{Helpers.}  We set $H:= \multiset{3\h \cdot 0}$, i.e.\ state $0$ initially contains at least $3\h$ helpers.  (As shown in the proof of \Cref{thm:PCmod},  with $3d$ helpers all terminal configurations are well-supported.)

\smallskip\noindent\textbf{Output function.} For every set $S$ of states, $O(S) := 1$ if $S$ satisfies $\varphi$, else $O(S):=0$.

\begin{example}
Figure \ref{fig:population_computer_modulo} shows the population computer for $\varphi = (8x+2y+z \equiv_{11} 4)$.
\begin{figure}[h] 
	\centering 
\newcommand{\labelL}[1]{{#1}}
\begin{tikzpicture}[->, auto, node distance=0.5cm]
	\tikzset{every place/.append style={minimum size=0.4cm, niceblue,fill=nicebgblue}}
	\tikzset{every label/.append style={text=niceblue}}
	\tikzset{every transition/.style={minimum size=0.25cm}}
	\tikzset{every edge/.append style={font=\scriptsize,-stealth}}

	\tikzstyle{blankhint}=[dash pattern=on 2pt off 1pt]
	\tikzstyle{up}=[fill=gray!30]
	\tikzstyle{modulo}=[fill=red!30]
	\tikzstyle{fastmodulo}=[fill=green!30]
	\tikzstyle{distribution}=[]
	\tikzstyle{up}=[fill=gray!30]
	\tikzstyle{bdd}=[circle, draw=black, fill=gray!70]
	\tikzstyle{true}=[nicegreen]
	\tikzstyle{false}=[nicered]
	\tikzstyle{bddedge}=[auto=false]
	\tikzstyle{bddedgelabel}=[sloped,fill=notquitegray!18,rectangle, inner sep=1,outer sep=0]
	\tikzstyle{clustergrey}=[notquitegray!18]
	\tikzstyle{init}=[fill=nicebluelight,line width=0.05cm]
			
	\newcommand*{\distanceblankhint}{0.3cm}
	\newcommand*{\distanceuplabel}{-0.15cm}
	\newcommand*{\distancesubthrx}{1.3cm}
	\newcommand*{\distancesubmodx}{1.5cm}
	\newcommand*{\distancesuby}{2cm}
	\newcommand*{\distancetoR}{2cm}
	
	\newcommand*{\distancebddl}{-7.00cm}
	\newcommand*{\distancebddsplitl}{0.8cm}
	
	\node[place, label= below:$\labelL{1}$, init] (s1_l1) {}; 
	\node[transition, up] (s1_l1_up) [above left= of s1_l1] {}; \coordinate[above left= \distanceblankhint of s1_l1_up] (s1_l1_up_b);
	\node[place, label= below:$\labelL{2}$, init] (s1_l2)  [above right= of s1_l1_up] {}; 
	\node[transition, up] (s1_l2_up) [above left= of s1_l2] {};  
	\node[place, label= below:$\labelL{4}$] (s1_l4)  [above right= of s1_l2_up] {}; 
	\node[transition, up] (s1_l4_up) [above left= of s1_l4] {}; 
	\node[place, label= below:$\labelL{8}$, init] (s1_l8)  [above right= of s1_l4_up] {};  
	\node[transition, up] (s1_l8_up) [above left= of s1_l8] {};  
	\node[place, label= above:$\labelL{16}$] (s1_l16)  [above right= of s1_l8_up] {}; 
	
	\node[place, label= left:$0$, minimum size=0.85cm, left=\distancetoR of s1_l4] (R) {\textcolor{nicered}{12}};

	\node[transition, modulo] (s1_mod1) [right= 3.0cm of s1_l4_up] {}; 
	\node[transition, fastmodulo] (s1_mod2) [right= 3.0cm of s1_l2_up] {};  \coordinate[above= \distanceblankhint of s1_mod2] (s1_mod2_b);

	\path
	(s1_l1)  edge[] node[below left=\distanceuplabel] {2} (s1_l1_up)
	(s1_l1_up) edge[] node {} (s1_l2)
	(s1_l1_up) edge node {} (R)
	
	(s1_l2)  edge[] node[below left=\distanceuplabel] {2} (s1_l2_up)
	(s1_l2_up) edge[] node {} (s1_l4)
	(s1_l2_up) edge node {} (R)
	
	(s1_l4)  edge[] node[below left=\distanceuplabel] {2} (s1_l4_up)
	(s1_l4_up) edge[] node {} (s1_l8)
	(s1_l4_up) edge node {} (R)
	
	(s1_l8)  edge[] node[below left=\distanceuplabel] {2} (s1_l8_up)
	(s1_l8_up) edge[] node {} (s1_l16)
	(s1_l8_up) edge node {} (R);

	\path
	(s1_l16)  edge[] (s1_mod1)
	(R)  edge[out=80, in=100, looseness=1.8] (s1_mod1)
	(s1_mod1)  edge[] (s1_l4)
	(s1_mod1)  edge[] (s1_l1)
	
	(s1_l16)  edge[] node[pos=0.35,above right=\distanceuplabel]{4} (s1_mod2)
	(s1_mod2)  edge[] (s1_l1)
	(s1_mod2)  edge[] (s1_l8)
	(s1_mod2)  edge[out=-100, in=-80, looseness=1.8] node{2} (R)
	;
	
	\node[bdd] (bdd1_8) [left= \distancebddl of s1_l16] {};
	\node[bdd] (bdd1_4l) [left= \distancebddl + \distancebddsplitl of s1_l8] {};
	\node[bdd] (bdd1_4r) [left= \distancebddl - \distancebddsplitl of s1_l8] {};
	\node[bdd] (bdd1_2l) [left= \distancebddl + \distancebddsplitl of s1_l4] {};
	\node[bdd] (bdd1_2r) [left= \distancebddl - \distancebddsplitl of s1_l4] {};
	\node[bdd] (bdd1_1l) [left= \distancebddl + \distancebddsplitl of s1_l2] {};
	\node[bdd] (bdd1_1r) [left= \distancebddl - \distancebddsplitl of s1_l2] {};
	\node[bdd, true, label=above:$\color{nicegreen}t$] (bdd1_t) [left= \distancebddl of s1_l4] {};
	\node[bdd, false, label=below:$\color{nicered}f$] (bdd1_f) [left= \distancebddl of s1_l1] {};
	
	\path
	(bdd1_8) edge[bddedge] node[bddedgelabel] {$8 \in S$} (bdd1_4l)
	(bdd1_8) edge[bddedge] node[bddedgelabel] {$8 \not \in S$} (bdd1_4r)
	(bdd1_4l) edge[bddedge] node[bddedgelabel] {$4 \in S$} (bdd1_2l)
	(bdd1_4l) edge[false, bddedge, out=-125,in=165, pos=0.20] node[bddedgelabel] {$4 \not \in S$} (bdd1_f)
	(bdd1_4r) edge[bddedge] node[bddedgelabel] {$4 \in S$} (bdd1_2r)
	(bdd1_4r) edge[false, bddedge, out=-55,in=15, pos=0.20] node[bddedgelabel] {$4 \not \in S$} (bdd1_f)
	(bdd1_2l) edge[bddedge] node[bddedgelabel, pos=0.6] {$2 \in S$} (bdd1_1l)
	(bdd1_2l) edge[false, bddedge, out=-125,in=150, pos=0.20] node[bddedgelabel] {$2 \not \in S$} (bdd1_f)
	(bdd1_2r) edge[bddedge] node[bddedgelabel] {$2 \not \in S$} (bdd1_1r)
	(bdd1_2r) edge[false, bddedge, out=-55,in=30, pos=0.20] node[bddedgelabel] {$2 \in S$} (bdd1_f)
	(bdd1_1l) edge[true, bddedge] node[bddedgelabel] {$1 \in S$} (bdd1_t)
	(bdd1_1l) edge[false, bddedge] node[bddedgelabel] {$1 \not \in S$} (bdd1_f)
	(bdd1_1r) edge[true, bddedge] node[bddedgelabel] {$1 \not \in S$} (bdd1_t)
	(bdd1_1r) edge[false, bddedge] node[bddedgelabel] {$1 \in S$} (bdd1_f)
	;
	
	\path ($(bdd1_8) + (-0.7,0)$) edge[bddedge] (bdd1_8);
	
	\begin{pgfonlayer}{background}
		\filldraw [line width=11mm,join=round,clustergrey]
		(bdd1_8.south  -| bdd1_2l.west)  rectangle (bdd1_f.south  -| bdd1_2r.east)
		;
	\end{pgfonlayer}
\end{tikzpicture} 
	\caption{
		Graphical Petri net representation (see Section \ref{sec:PCdef}) of the population computer for the predicate  $\varphi = (8x+2y+z \equiv_{11} 4)$. 
		Recall that circles and squares represent states and transitions, respectively.
		Initial states are shown in darker colour: state $8$ for input $x$, state $2$ for input $y$, and state $1$ for input $z$. We have $\Modulus=11$, $\h =\lceil \log_2 \Modulus \rceil = 4$, and $c = 4$. State $0$ contains initially $3\h = 12$  helpers. Transitions \TraRef{mod:combine:main}, \TraRef{mod:modulo:main}, and \TraRef{mod:fast-modulo:main} are shown in grey, red, and green, respectively.
		The output function returns $1$ for the supports $S$ such that $\multisum{S} \equiv_{11} 4$, i.e.\ $O(S)=1$ for $S = \{4\}$ and $S=\{1,2,4,8\}$. A decision diagram for this function is shown on the right. For the support $S=\{2,4,8\}$, the decisions are: left (because $8 \in S$), right (because $4 \in S$), right (because $2 \in S$), and down (because $1 \not \in S$) leading to \textit{false} as $2+4+8 = 14 \not \equiv_{11} 4$.
		}
	\label{fig:population_computer_modulo}
\end{figure}
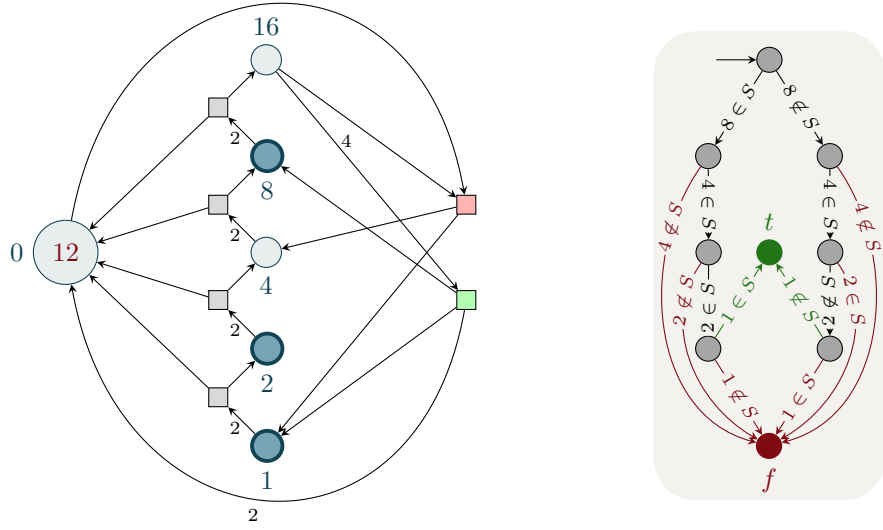
\end{example}

\begin{restatable}{lemma}{theoremPCmod}\label{thm:PCmod}
Let $\Modulus \in \N$, and let $\h := \lceil \log_2 \Modulus \rceil$. Let
$\varphi := \sum_{i=1}^\Nvar a_i x_i \equiv_{\Modulus} c$ be a remainder predicate such that 
$a_i \in \powp_{\h -1}$ for every $1 \leq i \leq \Nvar$ and $0 \leq c \leq  \Modulus-1$. The computer $\Prot_\varphi$ described above satisfies the conditions of  \Cref{prop:method}, and so decides $\varphi$. Further,  $\Prot_\varphi$ has size $\O(\h)$.
\end{restatable}

\begin{proof}
Let us prove that $\Prot_\varphi$ satisfies the conditions of \Cref{prop:method}.

\smallskip\noindent (\ref{spec:prop1}) $\Prot_\varphi$ is bounded. Let $C_0$ be an initial configuration with $n$ agents. We first claim that 
      every run starting at $C_0$ contains 
	at most $n$ occurrences of \TraRef{mod:modulo:main} or \TraRef{mod:fast-modulo:main} transitions.
      Recall that, given a configuration $C$, we have $\multisum{C}= \sum_{i=0}^d C(2^i) \cdot 2^i$. 
       So, since $\h :=\lceil \log_2 \Modulus \rceil$, we have $0 \leq \multisum{C} \leq \Modulus n$. Further, if $C \mapsto C'$, then $\multisum{C} \geq \multisum{C'}$;
       moreover, if $C'$ is obtained from $C$ by executing a \TraRef{mod:modulo:main} or a \TraRef{mod:fast-modulo:main} transition, then 
       $\multisum{C} \geq \multisum{C'} + \Modulus$,  i.e.\ the sum decreases by at  least $\Modulus$. This proves the claim.
      
       Since an occurrence of \TraRef{mod:combine:main}
	decreases the number of agents occupying the states $2^0, \ldots, 2^d$ by one, there are at most $n$ occurrences of \TraRef{mod:combine:main} transitions between any two consecutive occurrences of \TraRef{mod:modulo:main} or \TraRef{mod:fast-modulo:main} transitions. So, by the claim, every run starting at $C_0$ reaches a terminal configuration after $\O(n^2)$ steps, and we are done.
	
	\smallskip\noindent (\ref{spec:prop2}) Transitions preserve equivalence w.r.t.\ $\varphi$. Inspection of \TraRef{mod:combine:main}, \TraRef{mod:modulo:main}, and \TraRef{mod:fast-modulo:main} shows that  $C \mapsto C'$ implies $\multisum{C} \bmod \Modulus = \multisum{C'} \bmod \Modulus$. So
	$\multisum{C} \equiv_\Modulus c$ holds if{}f $\multisum{C'} \equiv_\Modulus c$ holds.
	
	\smallskip\noindent (\ref{spec:prop3}) Every terminal configuration $C_T$ of $\Prot_\varphi$ is well-supported.
	
	First of all, observe that every configuration \(C\) s.t. \(C(q)\leq 1\) for all \(q \neq 0\) is well-supported. This follows because it would imply \(\Support{C}(q)=C(q)\) for all \(q \neq 0\), and the state \(0\) trivially does not influence \(\multisum{C}\) and well-supportedness. Hence our strategy to prove (\ref{spec:prop3}) is to prove that every terminal configuration fulfills this stronger property.
	
	We have $C_T(2^i) \leq 1$ for $0 \leq i \leq \h-1$, because all \TraRef{mod:combine:main} transitions are disabled in $C_T$. It remains to prove \(C(2^\h)=0\). If it were the case that $C_T(2^\h) \geq \h$, then \TraRef{mod:fast-modulo:main} would be enabled, hence \(C_T(2^\h) \leq \h-1\). Since the number of agents is at least the number of helpers, i.e.\ at least \(3\h\), we have $C_T(0) + C_T(2^\h) \geq 2\h$, hence \(C_T(0) \geq \h\). This implies that if we had $1 \leq C_T(2^\h)$, then \TraRef{mod:modulo:main} would be enabled. So $C_T(2^\h) = 0$, and the claim is proved. 

      \smallskip\noindent (\ref{spec:prop4}) $O(S)=1$ if{}f $S$ satisfies $\varphi$. Holds by definition.

	\smallskip\noindent It remains to prove that $\Prot_\varphi$ has size $O(\h)$. The computer has $\O(\h)$ states and helpers. It has $\h$ \TraRef{mod:combine:main} transitions with size 2; further, $\Abs{\TraRef{mod:modulo:main}} \leq \h + 2$ and $\Abs{\TraRef{mod:fast-modulo:main}} \leq \h$. So the total size of the transitions is also $\O(\h)$.
	For the size of the output function, observe that for every set $S \subseteq Q$ we have $\multisum{S} \leq 2^{\h+1}$.
	So, since $2^\h \leq \Modulus \leq 2^{\h +1}$ by our choice of $\h$, either $\multisum{S} \bmod \Modulus = \multisum{S}$ or $\multisum{S} \bmod \Modulus = \multisum{S} - \Modulus$.
	Since $\multisum{S}$ and $c$ have $\h + 1$ bits, whether $\multisum{S} = c$ or $\multisum{S} - \Modulus= c $ can be decided by a boolean circuit with $O(\h)$ gates.
	Thus, $\size(\Prot_\varphi):=\Abs{Q}+\Abs{H}+\size(O)+\sum_{t\in \delta}\Abs{t} \in  \O(\h)$.
\end{proof}

\subsection{Population computers for threshold predicates} \label{subsec:construction_threshold}


We construct a population computer $\Prot_\varphi$ for a threshold predicate $\varphi:=\sum_{i=1}^\Nvar a_i x_i \geq c$.
Observe that, contrary to the case of remainder predicates, not every threshold predicate is equivalent to another one with positive coefficients. We also restrict ourselves to the case in which the coefficients are powers of $2$, i.e.\ elements of 
$\pow := \{2^i, -2^i \mid i \geq 0 \}$.

Let $\h_0 := \max \{ \lceil \log_2 c \rceil+1, \lceil\log_2 \Abs{a_1}\rceil, \ldots, \lceil\log_2 \Abs{a_\Nvar}\rceil \}$. We define a protocol $\Prot_\varphi$ for each $\h \geq \h_0$. (This  is used in Section \ref{subsec:construction_combination}, where we construct computers for boolean combinations of predicates.). The computer $\Prot_\varphi$ has $\pow_\h \cup \{0\}$ as set of states, where $\pow_\h:=\{2^i, -2^i \mid 1 \leq i \leq \h\}$.

The following lemma identifies a set of well-supported configurations w.r.t.\ $\varphi$, i.e.\ a set of configurations $C$ such that
$\multisum{C} \geq c \Leftrightarrow \multisum{\Support{C}} \geq c$. We design $\Prot_\varphi$ so that every terminal configuration belongs to this set. 

\begin{lemma}
\label{lem:wellsupportedthreshold}
Let $\varphi$ and $\h_0$ be defined as above and fix $\h \geq \h_0$. Every configuration $C$ over states $\pow_\h \cup \{ 0 \}$ satisfying the following three conditions is well-supported w.r.t.\ $\varphi$:

\begin{enumerate}
\item $C(2^i) \leq 1$ and $C(-2^i) \leq 1$ for every $1 \leq i \leq \h-1$; 
\item $C(2^i) = 0$ or $C(-2^i) = 0$ for every $1 \leq i \leq \h$; 
\item $C(2^\h)=0$ or $C(-2^{\h-1}) = 0$, and $C(-2^\h)=0$ or $C(2^{\h-1}) = 0$.
\end{enumerate}
\end{lemma}
Before proving the lemma, observe that the third condition is necessary. Let $c = 5$ and $\h = 3$. The configuration
$C=\multiset{2^3, 2^3, -2^2, -2^1}$ satisfies conditions 1 and 2, but is not well-supported; indeed, $\multisum{C}=10 \geq 5$, but
$\multisum{\Support{C}} =  2 < 5$. On the contrary, the configuration $\multiset{2^3, 2^3, -2^1}$ is well-supported.
\begin{proof}
Let $C$ be a configuration fulfilling the conditions. We prove that $C$ satisfies $\varphi$ if{}f  $\Support{C}$ satisfies $\varphi$, i.e.\  that $\multisum{C} \geq c$ holds if{}f $\multisum{\Support{C}} \geq c$ holds. For clarity, in the rest of the proof we abbreviate  $\Support{C}$ to $C_S$. We consider three cases:
\renewcommand{\Support}[1]{{#1}_S}

\begin{itemize}
\item $C(2^\h) > 0$. We prove $\multisum{C} \geq \multisum{\Support{C}} \geq c$, which shows that $C$ and  $\Support{C}$ satisfy $\varphi$. By definition we have $C(q) \geq \Support{C}(q)$ for every state $q$. Further, by conditions 1 and 2 and $C(2^\h) > 0$, we have $C(q) = C_S(q)$ for every state $q \neq 2^\h$. So $\multisum{C} \geq \multisum{\Support{C}}$. Now we prove $\multisum{\Support{C}} \geq c$:
\begin{align*}
 \multisum{\Support{C}} 
 & =  \sum_{i=1}^\h \Support{C}(2^i) \cdot 2^i - \sum_{i=1}^\h \Support{C}(-2^i) \cdot 2^i \\
 & \geq    2^\h - \sum_{i=1}^\h \Support{C}(-2^i) \cdot 2^i & \mbox{($C(2^\h) > 0$, and so $\Support{C}(2^\h) =1$)} \\
 & \geq    2^\h - \sum_{i=1}^{\h-2} \Support{C}(-2^i) \cdot 2^i & \mbox{(conditions 2 and 3)} \\
 & \geq 2^{\h-1} & \mbox{(condition 1)} \\
 & \geq c  & \mbox{(definition of $\h$)}
\end{align*}

\item $C(-2^\h) > 0$. Symmetric.
\item $C(2^\h) = 0 = C(-2^\h)$. By condition 1 we have $\multisum{C}= \multisum{\Support{C}}$, and we are done.
\end{itemize}

\renewcommand{\Support}[1]{\mathrm{supp}({#1})}
\end{proof}

We proceed to the formal description of the computer $\Prot_\varphi =(Q,\delta,I,O,H)$ for a predicate
$$\varphi:=\sum_{i=1}^\Nvar a_i x_i \geq c  \quad \mbox{where $a_i \in \pow$ for every $1 \leq i \leq \Nvar$ and $c \in \N$} $$
and  for a fixed $\h \geq \h_0$, where $\h_0 := \max \{ \lceil \log_2 c \rceil+1, \lceil\log_2 \Abs{a_1}\rceil, \ldots, \lceil\log_2 \Abs{a_\Nvar}\rceil \}$.

\smallskip\noindent\textbf{States and initial states. } 
The set of states of $\Prot_\varphi$ is $Q := \pow_\h \cup \{ 0 \}$. 
The initial states are the coefficients of $\varphi$, i.e.\ $I:= \{a_1, \ldots, a_\Nvar\}$.

\smallskip\noindent\textbf{Transitions.}  Since configurations that are not well-supported violate at least one of the conditions of \Cref{lem:wellsupportedthreshold}, we define transitions 
that ``repair'' these violations. For every $i \in [0, \h-1]$ we add to $\delta$  the following transitions,
which intuitively ``repair'' a violation of conditions 1, 2, and 3, respectively:
\begin{align*}
       2^i\,&, 2^i &&\mapsto& 0\,&,2^{i+1}   	&&\hspace{0.4cm}&	-2^i\,&, -2^i &&\mapsto& 0\,&,-2^{i+1}  \TraName[thr:combine:app]{combine} \\
	-2^i\,&, 2^i &&\mapsto& 0\,&,0 &&&	-2^\h\,&, 2^\h &&\mapsto& 0\,&,0& \TraName[thr:cancel:app]{cancel} \\
	2^\h\,&, -2^{\h-1} &&\mapsto& 0\,&,2^{\h-1} &&& 	-2^\h\,&, 2^{\h-1} &&\mapsto& 0\,&,-2^{\h-1}  \TraName[thr:cancel 2nd highest:app]{cancel 2nd highest}
\end{align*} 

\smallskip\noindent\textbf{Helpers.}  We set $H:= \multiset{\h \cdot 0}$, i.e.\ state $0$ initially contains at least $\h$ helper agents. 
(While the computer works correctly even with no helpers, the helpers in $H$ are used later in Section \ref{subsec:construction_combination} when
constructing computers for boolean combinations of remainder and threshold predicates.)

\smallskip\noindent\textbf{Output function.} For every set $S$ of states, $O(S) := 1$ if $S$ satisfies $\varphi$, else $O(S):=0$.

\begin{example}
Figure \ref{fig:population_computer_threshold} shows the population computer for $\varphi = \left(-2x+y \geq 5 \right)$ with $\h = 4$.
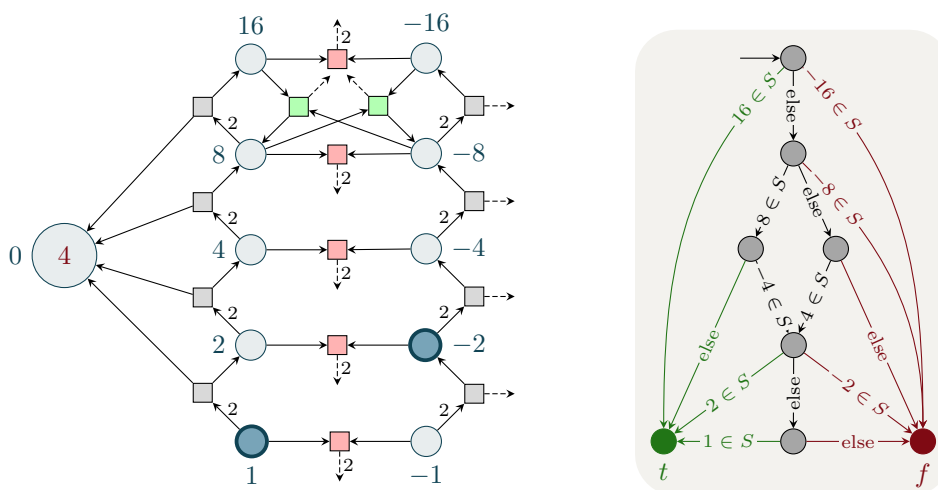
\begin{figure}[ht] 
	\centering 
\newcommand{\labelR}[1]{{#1}}
\begin{tikzpicture}[->, auto, node distance=0.5cm]
	\tikzset{every place/.append style={minimum size=0.4cm, niceblue,fill=nicebgblue}}
	\tikzset{every label/.append style={text=niceblue}}
	\tikzset{every transition/.style={minimum size=0.25cm}}
	\tikzset{every edge/.append style={font=\scriptsize,-stealth}}

	\tikzstyle{blankhint}=[dash pattern=on 2pt off 1pt]
	\tikzstyle{up}=[fill=gray!30]
	\tikzstyle{cancel}=[fill=red!30]	
	\tikzstyle{cancelsecond}=[fill=green!30]
	\tikzstyle{distribution}=[]
	\tikzstyle{bdd}=[circle, draw=black, fill=gray!70]
	\tikzstyle{true}=[nicegreen]
	\tikzstyle{false}=[nicered]
	\tikzstyle{bddedge}=[auto=false]
	\tikzstyle{bddedgelabel}=[sloped,fill=notquitegray!18,rectangle, inner sep=1,outer sep=0]
	\tikzstyle{clustergrey}=[notquitegray!18]
	\tikzstyle{initial}=[fill=nicebluelight,line width=0.05cm]
			
	\newcommand*{\distanceblankhint}{0.4cm}
	\newcommand*{\distanceuplabel}{-0.15cm}
	\newcommand*{\distancesubthrx}{2.0cm}
	\newcommand*{\distancesubmodx}{1.0cm}
	\newcommand*{\distancesuby}{2cm}
	\newcommand*{\distancebddr}{4.45cm}
	\newcommand*{\distancebddsplitr}{0.8cm}
	\newcommand*{\distancecancel}{0.8cm}

	\node[place, label= left:$0$, minimum size=0.85cm] (R) {\textcolor{nicered}{4}};

	\node[place, label= below:$\labelR{1}$,initial] (s2_l1) [below right= \distancesuby and \distancesubthrx of R] {};
	\node[transition, cancel] (s2_l1_c) [right= \distancecancel of s2_l1] {};
	\coordinate[below= \distanceblankhint of s2_l1_c] (s2_l1_c_b);
	\node[place, label= below:$\labelR{-1}$] (s2_lm1) [right= \distancecancel of s2_l1_c] {}; 
	\node[transition, up] (s2_l1_up) [above left= of s2_l1] {}; 
	\node[transition, up] (s2_lm1_up) [above right= of s2_lm1] {}; 
	\coordinate[left= \distanceblankhint of s2_l1_up] (s2_l1_up_b);
	\coordinate[right= \distanceblankhint of s2_lm1_up] (s2_lm1_up_b);
	\node[place, label= left:$\labelR{2}$] (s2_l2)  [above right= of s2_l1_up] {}; 
	\node[transition, cancel] (s2_l2_c) [right= \distancecancel of s2_l2] {};
	\coordinate[below= \distanceblankhint of s2_l2_c] (s2_l2_c_b);
	\node[place, label= right:$\labelR{-2}$,initial] (s2_lm2)  [above left= of s2_lm1_up] {}; 
	\node[transition, up] (s2_l2_up) [above left= of s2_l2] {}; 
	\node[transition, up] (s2_lm2_up) [above right= of s2_lm2] {}; 
	\coordinate[left= \distanceblankhint of s2_l2_up] (s2_l2_up_b);
	\coordinate[right= \distanceblankhint of s2_lm2_up] (s2_lm2_up_b);
	\node[place, label= left:$\labelR{4}$] (s2_l4)  [above right= of s2_l2_up] {}; 
	\node[transition, cancel] (s2_l4_c) [right= \distancecancel of s2_l4] {};
	\coordinate[below= \distanceblankhint of s2_l4_c] (s2_l4_c_b);
	\node[place, label= right:$\labelR{-4}$] (s2_lm4)  [above left= of s2_lm2_up] {}; 
	\node[transition, up] (s2_l4_up) [above left= of s2_l4] {}; 
	\node[transition, up] (s2_lm4_up) [above right= of s2_lm4] {}; 
	\coordinate[left= \distanceblankhint of s2_l4_up] (s2_l4_up_b);
	\coordinate[right= \distanceblankhint of s2_lm4_up] (s2_lm4_up_b);
	\node[place, label= left:$\labelR{8}$] (s2_l8)  [above right= of s2_l4_up] {}; 
	\node[transition, cancel] (s2_l8_c) [right= \distancecancel of s2_l8] {};
	\coordinate[below= \distanceblankhint of s2_l8_c] (s2_l8_c_b);
	\node[place, label= right:$\labelR{-8}$] (s2_lm8)  [above left= of s2_lm4_up] {};  
	\node[transition, up] (s2_l8_up) [above left= of s2_l8] {}; 
	\node[transition, up] (s2_lm8_up) [above right= of s2_lm8] {}; 
	\coordinate[left= \distanceblankhint of s2_l8_up] (s2_l8_up_b);
	\coordinate[right= \distanceblankhint of s2_lm8_up] (s2_lm8_up_b);
	\node[place, label= above:$\labelR{16}$] (s2_l16)  [above right= of s2_l8_up] {}; 
	\node[transition, cancel] (s2_l16_c) [right= \distancecancel of s2_l16] {};
	\coordinate[above= \distanceblankhint of s2_l16_c] (s2_l16_c_b);
	\node[transition, cancelsecond] (s2_lt_c1) [above right= of s2_l8] {}; 
	\coordinate[above right= \distanceblankhint of s2_lt_c1] (s2_lt_c1_b);
	\node[transition, cancelsecond] (s2_lt_c2) [above left= of s2_lm8] {}; 
	\coordinate[above left= \distanceblankhint of s2_lt_c2] (s2_lt_c2_b);
	\node[place, label= above:$\labelR{-16}$] (s2_lm16)  [above left= of s2_lm8_up] {};

	\path
	(s2_l1)  edge[] node[above right=\distanceuplabel] {2} (s2_l1_up)
	(s2_l1_up) edge[] node {} (s2_l2)
	(s2_l1_up) edge[] node {} (R)
	
	(s2_l2)  edge[] node[above right=\distanceuplabel] {2} (s2_l2_up)
	(s2_l2_up) edge[] node {} (s2_l4)
	(s2_l2_up) edge[] node {} (R)
	
	(s2_l4)  edge[] node[above right=\distanceuplabel] {2} (s2_l4_up)
	(s2_l4_up) edge[] node {} (s2_l8)
	(s2_l4_up) edge[] node {} (R)
	
	(s2_l8)  edge[] node[above right=\distanceuplabel] {2} (s2_l8_up)
	(s2_l8_up) edge[] node {} (s2_l16)
	(s2_l8_up) edge[] node {} (R)
	
	(s2_lm1)  edge[] node[above left=\distanceuplabel] {2} (s2_lm1_up)
	(s2_lm1_up) edge[] node {} (s2_lm2)
	(s2_lm1_up) edge[blankhint] node {} (s2_lm1_up_b)
	
	(s2_lm2)  edge[] node[above left=\distanceuplabel] {2} (s2_lm2_up)
	(s2_lm2_up) edge[] node {} (s2_lm4)
	(s2_lm2_up) edge[blankhint] node {} (s2_lm2_up_b)
	
	(s2_lm4)  edge[] node[above left=\distanceuplabel] {2} (s2_lm4_up)
	(s2_lm4_up) edge[] node {} (s2_lm8)
	(s2_lm4_up) edge[blankhint] node {} (s2_lm4_up_b)
	
	(s2_lm8)  edge[] node[above left=\distanceuplabel] {2} (s2_lm8_up)
	(s2_lm8_up) edge[] node {} (s2_lm16)
	(s2_lm8_up) edge[blankhint] node {} (s2_lm8_up_b)
	;
	
	\path
	(s2_l1)  edge[] (s2_l1_c)
	(s2_lm1) edge[] (s2_l1_c)
	(s2_l1_c) edge[blankhint] node[xshift=-0.08cm,pos=0.45] {2} (s2_l1_c_b)
	
	(s2_l2)  edge[] (s2_l2_c)
	(s2_lm2) edge[] (s2_l2_c)
	(s2_l2_c) edge[blankhint] node[xshift=-0.08cm,pos=0.45] {2} (s2_l2_c_b)
	
	(s2_l4)  edge[] (s2_l4_c)
	(s2_lm4) edge[] (s2_l4_c)
	(s2_l4_c) edge[blankhint] node[xshift=-0.08cm,pos=0.45] {2} (s2_l4_c_b)
	
	(s2_l8)  edge[] (s2_l8_c)
	(s2_lm8) edge[] (s2_l8_c)
	(s2_l8_c) edge[blankhint] node[xshift=-0.08cm,pos=0.45] {2} (s2_l8_c_b)
	
	(s2_l16)  edge[] (s2_l16_c)
	(s2_lm16) edge[] (s2_l16_c)
	(s2_l16_c) edge[blankhint] node[swap,xshift=-0.08cm,pos=0.4] {2} (s2_l16_c_b)
	;

	\path
	(s2_l16)  edge[] (s2_lt_c1)
	(s2_lm8) edge[] (s2_lt_c1)
	(s2_lt_c1)  edge[] (s2_l8)
	(s2_lt_c1)  edge[blankhint] (s2_lt_c1_b)
	
	(s2_lm16)  edge[] (s2_lt_c2)
	(s2_l8) edge[] (s2_lt_c2)
	(s2_lt_c2)  edge[] (s2_lm8)
	(s2_lt_c2)  edge[blankhint] (s2_lt_c2_b)
	;
	
	\node[bdd] (bdd2_16) [right= \distancebddr of s2_lm16] {};
	\node[bdd] (bdd2_8) [right= \distancebddr of s2_lm8] {};
	\node[bdd] (bdd2_4l) [right= \distancebddr + 0.7*\distancebddsplitr of s2_lm4] {};
	\node[bdd] (bdd2_4h) [right= \distancebddr - 0.7*\distancebddsplitr of s2_lm4] {};
	\node[bdd] (bdd2_2) [right= \distancebddr of s2_lm2] {};
	\node[bdd] (bdd2_1) [right= \distancebddr of s2_lm1] {};
	\node[bdd, true, label=below:$\color{nicegreen}t$] (bdd2_t) [left = 1.7*\distancebddsplitr of bdd2_1] {};
	\node[bdd, false, label=below:$\color{nicered}f$] (bdd2_f) [right = 1.7*\distancebddsplitr of bdd2_1] {};
	
	\path
	(bdd2_16) edge[bddedge] node[bddedgelabel] {else} (bdd2_8)
	(bdd2_16) edge[bddedge, true, out=-135, in=90, pos=0.1] node[bddedgelabel] {$16 \in S$} (bdd2_t)
	(bdd2_16) edge[false, bddedge, out=-45, in=90, pos=0.1] node[bddedgelabel] {${-}16 \in S$} (bdd2_f)
	(bdd2_8) edge[false, bddedge, out=-45, in=90, pos=0.15] node[bddedgelabel] {${-}8 \in S$} (bdd2_f)
	(bdd2_8) edge[bddedge] node[bddedgelabel] {$8 \in S$} (bdd2_4h)
	(bdd2_8) edge[bddedge] node[bddedgelabel] {else} (bdd2_4l)
	(bdd2_4l) edge[bddedge] node[bddedgelabel] {$4 \in S$} (bdd2_2)
	(bdd2_4l) edge[false, bddedge] node[bddedgelabel] {else} (bdd2_f)
	(bdd2_4h) edge[bddedge] node[bddedgelabel] {${-}4 \in S$} (bdd2_2)
	(bdd2_4h) edge[true, bddedge] node[bddedgelabel] {else} (bdd2_t)
	(bdd2_2) edge[false, bddedge] node[bddedgelabel] {${-}2 \in S$} (bdd2_f)
	(bdd2_2) edge[bddedge] node[bddedgelabel] {else} (bdd2_1)
	(bdd2_2) edge[true, bddedge] node[bddedgelabel] {$2 \in S$} (bdd2_t)
	(bdd2_1) edge[false, bddedge] node[bddedgelabel] {else} (bdd2_f)
	(bdd2_1) edge[true,bddedge] node[bddedgelabel] {$1 \in S$} (bdd2_t)
	;
	
	\path ($(bdd2_16) + (-0.7,0)$) edge[bddedge] (bdd2_16);

	\begin{pgfonlayer}{background}

		\filldraw [line width=11mm,join=round,clustergrey]
		(bdd2_16.south  -| bdd2_t.east)  rectangle (bdd2_f.south  -| bdd2_f.west)
		;
	\end{pgfonlayer}

\end{tikzpicture} 
	\caption{
		Graphical Petri net representation (see Section \ref{sec:PCdef}) of the population computer for the predicate  $\varphi = \left(-2x+y \geq 5 \right)$ with $\h = 4$. 
		Initial states are shown in darker colour: state $-2$ for input $x$, and state $1$ for input $y$. State $0$ contains initially $\h = 4$  helpers; all dashed edges leaving transitions implicitly lead to this state. Observe that the number of agents in $0$ can only increase. Transitions \TraRef{thr:combine:app}, \TraRef{thr:cancel:app}, and \TraRef{thr:cancel 2nd highest:app} are shown in grey, pink, and green, respectively. 
		The output function $O$ returns $1$ for the supports $S$ such that $\multisum{S} \geq 5$. A decision diagram accepting these supports is shown on the right. For $S=\{8,-4,2\}$ the decisions are: centre (because $16\not\in S$ and $-16\not\in S$), left (because $8 \in S$), right (because $-4 \in S$), and left (because $2 \in S$) leading to \textit{true} as $8-4+2 = 6 \geq 5$.}
	\label{fig:population_computer_threshold}
\end{figure}
\end{example}

\begin{restatable}{lemma}{throremPCthr}\label{thm:PCthr}
	Let $\varphi := \sum_{i=1}^\Nvar a_i x_i \geq  c$, where $a_i \in \{2^j, 2^{-j} \mid j \geq 0\}$ for every $1 \leq i \leq \Nvar$. For every $\h \geq  \max \{ \lceil \log_2 c \rceil+1, \lceil\log_2 \Abs{a_1}\rceil, \ldots, \lceil\log_2 \Abs{a_\Nvar}\rceil \}$, the computer $\Prot_\varphi$ described above satisfies the conditions of  \Cref{prop:method}, and so decides $\varphi$. Further,  $\Prot_\varphi$ has size $\O(\h)$.
\end{restatable}

\begin{proof}
We first prove that $\Prot_\varphi$ satisfies the conditions of \Cref{prop:method}.

\smallskip\noindent (\ref{spec:prop1}) $\Prot_\varphi$ is bounded. Every transition increases the number of agents in state $0$.
	Therefore, every run starting at an initial configuration with $n$ agents has length at most $n$.
	
\smallskip\noindent (\ref{spec:prop2})  Transitions preserve equivalence w.r.t.\ $\varphi$, i.e.\ if \(C \rightarrow C'\), then $\multisum{C} \ge c \iff \multisum{C'} \ge c$. In fact, even the stronger property \(\multisum{C}=\multisum{C'}\) holds by simple inspection of the transitions. For example for \TraRef{thr:combine:app} we check \(-2^i + (-2^i) =0 + (-2^{i+1})\).

\smallskip\noindent (\ref{spec:prop3}) Every terminal configuration $C_T$ of $\Prot_\varphi$ is well-supported. Terminal configurations satisfy all conditions of \Cref{lem:wellsupportedthreshold}, because every configuration violating at least one condition enables at least one transition. 

\smallskip\noindent (\ref{spec:prop4}) $O(S)=1$ if{}f $S$ satisfies $\varphi$. Holds by definition.

\smallskip\noindent It remains to prove that $\Prot_\varphi$ has size $\O(\h)$.   Observe that $\Prot_\varphi$ is binary and has $\O(\h)$ transitions.
Thus, $\size(\Prot_\varphi):=\Abs{Q}+\Abs{H} + \size(O) + \sum_{t \in \delta} \Abs{t} \in (2\h+3) + \h + \size(O) + \O(\h) \subseteq \O(\h) + \size(O)$.
So it remains to describe a boolean circuit of size $O(\h)$ that decides whether a given  terminal configuration $C_T$ with support $S$ satisfies $\varphi$, i.e.\ 
whether $\multisum{S} \geq c$. For this, abbreviate $p_i:= S(2^i)$ and $n_i = S(-2^i)$. We have $p_i, n_i \in \{0,1\}$, and so $\multisum{S}$ is  the difference of the binary numbers $p_\h p_{\h-1}\ldots p_0$ and $n_\h n_{\h-1}\ldots n_0$. Whether this difference is bigger than or equal to $c$ can be decided by a circuit with $\O(\log c) = \O(\h)$ gates. 
\end{proof}

\subsection{Population computers for all Presburger predicates} \label{subsec:construction_combination}

We present a construction that, given threshold or remainder predicates $\varphi_1, \ldots, \varphi_\Npred$ over a common set $X= \{x_1, \ldots, x_\Nvar\}$ of variables, yields a population computer $\Prot_\varphi$ deciding an arbitrary given boolean combination $\varphi=B(\varphi_1, \ldots, \varphi_\Npred)$ of $\varphi_1, \ldots, \varphi_\Npred$. The construction has a number of technical details, but its essence is simple: The computer ``distributes'' a given input to ``subcomputers'' $\Prot_1, \ldots, \Prot_\Npred$ deciding $\varphi_1, \ldots, \varphi_\Npred$, and lets them run concurrently. Each subcomputer, say $\Prot_j$, reaches a terminal configuration with support $S_j$ such that $O_j(S_j)=1$ if{}f the input satisfies $\varphi_j$. The output function of the computer is defined as the boolean combination of the output functions of the subcomputers, i.e.\ $O(S) := B(O_1(S_1), \ldots, O_\Npred(S_\Npred))$, modulo some technical details.

We remark that our notion of subcomputers is similar to the technique of \emph{population splitting} (see e.g.~\cite{AlistarhG18}). However, while the eventual goal is the same (to have multiple subpopulations that work on distinct tasks), our construction differs in that \emph{every} agent must be distributed. This must hence occur concurrently with the rest of the computation. In contrast, population splitting is usually employed as a separate phase in the beginning, and it suffices to distribute “most” agents into “roughly equal” parts.

\begin{example}
\label{ex:runningbool}
We use $\varphi_1 = \left(8x+5y \equiv_{11} 4\right)$, $\varphi_2 = \left(-2x+y \geq 5 \right)$ and $\varphi = \varphi_1 \lor \varphi_2$ as running example. 
Observe that the coefficient $5$ of $\varphi_1$ is not a power of $2$; in fact, the construction also shows how to deal with general remainder and threshold predicates.
The Petri net representation of the computer for $\varphi$ is shown in Figure \ref{fig:population_computer}. The input is placed in states $x$ and $y$. Intuitively, the two pink transitions distribute it
to subcomputers for $\varphi_1$ (left) and $\varphi_2$ (right). Helpers help to run the subcomputers, but also to distribute the input.
For example, the top pink transition takes only one agent from the input state $x$, but sends two agents to the subcomputers, hence it needs one helper.  Careful choice of the exact 
set of states of each subcomputer and the number of helpers  guarantee that the computer distributes all the input, i.e.\ that no terminal configuration puts agents in any input state.
\begin{figure}[h] 
	\centering 
\newcommand{\labelinfig}[2]{(#1)_{#2}}
\newcommand{\labelL}[1]{\labelinfig{#1}{1}}
\newcommand{\labelR}[1]{\labelinfig{#1}{2}}
\begin{tikzpicture}[->, auto, node distance=0.5cm]
	\tikzset{every place/.append style={minimum size=0.4cm, niceblue,fill=nicebgblue}}
	\tikzset{every label/.append style={text=niceblue}}
	\tikzset{every transition/.style={minimum size=0.25cm}}
	\tikzset{every edge/.append style={font=\scriptsize,-stealth}}

	\tikzstyle{blankhint}=[dash pattern=on 2pt off 1pt]
	\tikzstyle{cancle}=[fill=gray!30]
	\tikzstyle{distribution}=[fill=red!30]
	\tikzstyle{up}=[fill=gray!30]
	\tikzstyle{modulo}=[fill=gray!30]
	\tikzstyle{fastmodulo}=[fill=gray!30]
	\tikzstyle{cancel}=[fill=gray!30]	
	\tikzstyle{cancelsecond}=[fill=gray!30]
	\tikzstyle{bdd}=[circle, draw=black, fill=gray!70]
	\tikzstyle{true}=[nicegreen]
	\tikzstyle{false}=[nicered]
	\tikzstyle{bddedge}=[auto=false]
	\tikzstyle{bddedgelabel}=[sloped,fill=notquitegray!18,rectangle, inner sep=1,outer sep=0]
	\tikzstyle{clustergrey}=[notquitegray!18]
	\tikzstyle{initial}=[fill=nicebluelight,line width=0.05cm]
			
	\newcommand*{\distanceblankhint}{0.3cm}
	\newcommand*{\distanceuplabel}{-0.15cm}
	\newcommand*{\distancesubthrx}{2.3cm}
	\newcommand*{\distancesubmodx}{2.0cm}
	\newcommand*{\distancesuby}{2cm}
	\newcommand*{\distancetoR}{3cm}
	\newcommand*{\distancecancel}{0.7cm}

	\node[place, label= right:$0$, minimum size=0.85cm] (R) {\textcolor{nicered}{18}};
	
	\node[transition, distribution] (d_x) [above= of R] {};
	\node[place, label= above:$x$, initial] (x) [above= of d_x] {};
	\node[transition, distribution] (d_y) [below= of R] {};
	\node[place, label= below:$y$, initial] (y) [below= of d_y] {};

	\node[place, label= below:$\labelL{1}$] (s1_l1) [below left= \distancesuby and \distancesubthrx of R]{}; 
	\node[transition, up] (s1_l1_up) [above right= of s1_l1] {}; \coordinate[above left= \distanceblankhint of s1_l1_up] (s1_l1_up_b);
	\coordinate[right= \distanceblankhint of s1_l1_up] (s1_l1_b);
	\node[place, label= below:$\labelL{2}$] (s1_l2)  [above left= of s1_l1_up] {}; 
	\node[transition, up] (s1_l2_up) [above right= of s1_l2] {};  
      \coordinate[right= \distanceblankhint of s1_l2_up] (s1_l2_b);
	\node[place, label= below:$\labelL{4}$] (s1_l4)  [above left= of s1_l2_up] {}; 
	\node[transition, up] (s1_l4_up) [above right= of s1_l4] {}; 
	\coordinate[right= \distanceblankhint of s1_l4_up] (s1_l4_b);
	\node[place, label= below:$\labelL{8}$] (s1_l8)  [above left= of s1_l4_up] {};  
	\node[transition, up] (s1_l8_up) [above right= of s1_l8] {};  
	\coordinate[right= \distanceblankhint of s1_l8_up] (s1_l8_b);
	\node[place, label= above:$\labelL{16}$] (s1_l16)  [above left= of s1_l8_up] {}; 

	\node[transition, modulo] (s1_mod1) [left= 2.5cm of s1_l4_up] {}; \coordinate[above= 1.3*\distanceblankhint of s1_mod1] (s1_mod1_b);
	\node[transition, fastmodulo] (s1_mod2) [left= 2.5cm of s1_l2_up] {};  \coordinate[below= 1.3*\distanceblankhint of s1_mod2] (s1_mod2_b);

	\path
	(s1_l1)  edge[] node[below right=\distanceuplabel] {2} (s1_l1_up)
	(s1_l1_up) edge[] node {} (s1_l2)
	(s1_l1_up) edge[blankhint] node {} (s1_l1_b)
	
	(s1_l2)  edge[] node[below right=\distanceuplabel] {2} (s1_l2_up)
	(s1_l2_up) edge[] node {} (s1_l4)
	(s1_l2_up) edge[blankhint] node {} (s1_l2_b)
	
	(s1_l4)  edge[] node[below right=\distanceuplabel] {2} (s1_l4_up)
	(s1_l4_up) edge[] node {} (s1_l8)
	(s1_l4_up) edge[blankhint] node {} (s1_l4_b)
	
	(s1_l8)  edge[] node[below right=\distanceuplabel] {2} (s1_l8_up)
	(s1_l8_up) edge[] node {} (s1_l16)
	(s1_l8_up) edge[blankhint] node {} (s1_l8_b);

	\path
	(s1_l16)  edge[] (s1_mod1)
	(s1_mod1_b)  edge[blankhint] (s1_mod1)
	(s1_mod1)  edge[] (s1_l4)
	(s1_mod1)  edge[] (s1_l1)
	
	(s1_l16)  edge[] node[pos=0.35,below right=\distanceuplabel]{4} (s1_mod2)
	(s1_mod2)  edge[] (s1_l1)
	(s1_mod2)  edge[] (s1_l8)
	(s1_mod2)  edge[blankhint] node[swap]{2} (s1_mod2_b)
	;
	
	
	\node[place, label= below:$\labelR{1}$] (s2_l1) [below right= \distancesuby and \distancesubthrx of R] {};
	\node[transition, cancel] (s2_l1_c) [right= \distancecancel of s2_l1] {};
	\coordinate[below= \distanceblankhint of s2_l1_c] (s2_l1_c_b);
	\node[place, label= below:$\labelR{-1}$] (s2_lm1) [right= \distancecancel of s2_l1_c] {}; 
	\node[transition, up] (s2_l1_up) [above left= of s2_l1] {}; 
	\node[transition, up] (s2_lm1_up) [above right= of s2_lm1] {}; 
	\coordinate[left= \distanceblankhint of s2_l1_up] (s2_l1_up_b);
	\coordinate[right= \distanceblankhint of s2_lm1_up] (s2_lm1_up_b);
	\node[place, label= left:$\labelR{2}$] (s2_l2)  [above right= of s2_l1_up] {}; 
	\node[transition, cancel] (s2_l2_c) [right= \distancecancel of s2_l2] {};
	\coordinate[below= \distanceblankhint of s2_l2_c] (s2_l2_c_b);
	\node[place, label= right:$\labelR{-2}$] (s2_lm2)  [above left= of s2_lm1_up] {}; 
	\node[transition, up] (s2_l2_up) [above left= of s2_l2] {}; 
	\node[transition, up] (s2_lm2_up) [above right= of s2_lm2] {}; 
	\coordinate[left= \distanceblankhint of s2_l2_up] (s2_l2_up_b);
	\coordinate[right= \distanceblankhint of s2_lm2_up] (s2_lm2_up_b);
	\node[place, label= left:$\labelR{4}$] (s2_l4)  [above right= of s2_l2_up] {}; 
	\node[transition, cancel] (s2_l4_c) [right= \distancecancel of s2_l4] {};
	\coordinate[below= \distanceblankhint of s2_l4_c] (s2_l4_c_b);
	\node[place, label= right:$\labelR{-4}$] (s2_lm4)  [above left= of s2_lm2_up] {}; 
	\node[transition, up] (s2_l4_up) [above left= of s2_l4] {}; 
	\node[transition, up] (s2_lm4_up) [above right= of s2_lm4] {}; 
	\coordinate[left= \distanceblankhint of s2_l4_up] (s2_l4_up_b);
	\coordinate[right= \distanceblankhint of s2_lm4_up] (s2_lm4_up_b);
	\node[place, label= left:$\labelR{8}$] (s2_l8)  [above right= of s2_l4_up] {}; 
	\node[transition, cancel] (s2_l8_c) [right= \distancecancel of s2_l8] {};
	\coordinate[below= \distanceblankhint of s2_l8_c] (s2_l8_c_b);
	\node[place, label= right:$\labelR{-8}$] (s2_lm8)  [above left= of s2_lm4_up] {};  
	\node[transition, up] (s2_l8_up) [above left= of s2_l8] {}; 
	\node[transition, up] (s2_lm8_up) [above right= of s2_lm8] {}; 
	\coordinate[left= \distanceblankhint of s2_l8_up] (s2_l8_up_b);
	\coordinate[right= \distanceblankhint of s2_lm8_up] (s2_lm8_up_b);
	\node[place, label= above:$\labelR{16}$] (s2_l16)  [above right= of s2_l8_up] {}; 
	\node[transition, cancel] (s2_l16_c) [right= \distancecancel of s2_l16] {};
	\coordinate[above= \distanceblankhint of s2_l16_c] (s2_l16_c_b);
	\node[transition, cancelsecond] (s2_lt_c1) [above right= of s2_l8] {}; 
	\coordinate[above right= \distanceblankhint of s2_lt_c1] (s2_lt_c1_b);
	\node[transition, cancelsecond] (s2_lt_c2) [above left= of s2_lm8] {}; 
	\coordinate[above left= \distanceblankhint of s2_lt_c2] (s2_lt_c2_b);
	\node[place, label= above:$\labelR{-16}$] (s2_lm16)  [above left= of s2_lm8_up] {};

	\path
	(s2_l1)  edge[] node[above right=\distanceuplabel] {2} (s2_l1_up)
	(s2_l1_up) edge[] node {} (s2_l2)
	(s2_l1_up) edge[blankhint] node {} (s2_l1_up_b)
	
	(s2_l2)  edge[] node[above right=\distanceuplabel] {2} (s2_l2_up)
	(s2_l2_up) edge[] node {} (s2_l4)
	(s2_l2_up) edge[blankhint] node {} (s2_l2_up_b)
	
	(s2_l4)  edge[] node[above right=\distanceuplabel] {2} (s2_l4_up)
	(s2_l4_up) edge[] node {} (s2_l8)
	(s2_l4_up) edge[blankhint] node {} (s2_l4_up_b)
	
	(s2_l8)  edge[] node[above right=\distanceuplabel] {2} (s2_l8_up)
	(s2_l8_up) edge[] node {} (s2_l16)
	(s2_l8_up) edge[blankhint] node {} (s2_l8_up_b)
	
	(s2_lm1)  edge[] node[above left=\distanceuplabel] {2} (s2_lm1_up)
	(s2_lm1_up) edge[] node {} (s2_lm2)
	(s2_lm1_up) edge[blankhint] node {} (s2_lm1_up_b)
	
	(s2_lm2)  edge[] node[above left=\distanceuplabel] {2} (s2_lm2_up)
	(s2_lm2_up) edge[] node {} (s2_lm4)
	(s2_lm2_up) edge[blankhint] node {} (s2_lm2_up_b)
	
	(s2_lm4)  edge[] node[above left=\distanceuplabel] {2} (s2_lm4_up)
	(s2_lm4_up) edge[] node {} (s2_lm8)
	(s2_lm4_up) edge[blankhint] node {} (s2_lm4_up_b)
	
	(s2_lm8)  edge[] node[above left=\distanceuplabel] {2} (s2_lm8_up)
	(s2_lm8_up) edge[] node {} (s2_lm16)
	(s2_lm8_up) edge[blankhint] node {} (s2_lm8_up_b)
	;
	
	\path
	(s2_l1)  edge[] (s2_l1_c)
	(s2_lm1) edge[] (s2_l1_c)
	(s2_l1_c) edge[blankhint] node[xshift=-0.08cm,pos=0.45] {2} (s2_l1_c_b)
	
	(s2_l2)  edge[] (s2_l2_c)
	(s2_lm2) edge[] (s2_l2_c)
	(s2_l2_c) edge[blankhint] node[xshift=-0.08cm,pos=0.45] {2} (s2_l2_c_b)
	
	(s2_l4)  edge[] (s2_l4_c)
	(s2_lm4) edge[] (s2_l4_c)
	(s2_l4_c) edge[blankhint] node[xshift=-0.08cm,pos=0.45] {2} (s2_l4_c_b)
	
	(s2_l8)  edge[] (s2_l8_c)
	(s2_lm8) edge[] (s2_l8_c)
	(s2_l8_c) edge[blankhint] node[xshift=-0.08cm,pos=0.45] {2} (s2_l8_c_b)
	
	(s2_l16)  edge[] (s2_l16_c)
	(s2_lm16) edge[] (s2_l16_c)
	(s2_l16_c) edge[blankhint] node[swap,xshift=-0.08cm,pos=0.4] {2} (s2_l16_c_b)
	;

	\path
	(s2_l16)  edge[] (s2_lt_c1)
	(s2_lm8) edge[] (s2_lt_c1)
	(s2_lt_c1)  edge[] (s2_l8)
	(s2_lt_c1)  edge[blankhint] (s2_lt_c1_b)
	
	(s2_lm16)  edge[] (s2_lt_c2)
	(s2_l8) edge[] (s2_lt_c2)
	(s2_lt_c2)  edge[] (s2_lm8)
	(s2_lt_c2)  edge[blankhint] (s2_lt_c2_b)
	;

	\path
	(x)  edge[] (d_x)
	(R)  edge[] (d_x)
	(d_x)  edge[bend right=10] (s2_lm2)
	(d_x)  edge[] (s1_l8)
	
	(y)  edge[] (d_y)
	(R)  edge[] node[] {2} (d_y)
	(d_y)  edge[bend right=17] (s2_l1)
	(d_y)  edge[out=-135,in=0] (s1_l1)
	(d_y)  edge[out=135,in=0] (s1_l4)
	;

\end{tikzpicture} 
	\caption{
		Graphical Petri net representation (see Section \ref{sec:PCdef}) of the population computer for the predicate  $\varphi= \left(8x+5y \equiv_{11} 4\right) \lor \left(-2x+y \geq 5 \right)$. 
		The left and right parts correspond to the computers for $8x+5y \equiv_{11} 4$, and $-2x+y \geq 5$ (see Figures \ref{fig:population_computer_modulo} and \ref{fig:population_computer_threshold}).
		The common reservoir state is shown in the centre, and the input states above and below it. All dashed edges lead to or from it. The output function returns $1$ for a support $S$ if{}f $\multisum{S} \equiv_{11} = 4$, i.e.\ $O(S)=1$ for $S = \{4\}$ and $S=\{1,2,4,8\}$. A decision diagram can be obtained as the conjunction
		of the diagrams in Figures \ref{fig:population_computer_modulo} and \ref{fig:population_computer_threshold}.
		}
	\label{fig:population_computer}
\end{figure}
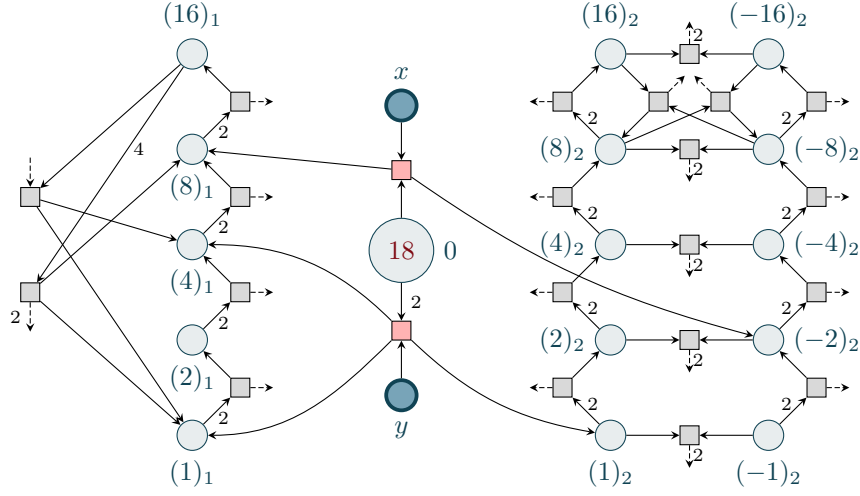
\end{example}

Let us now give a more detailed, but still informal description of the construction, which proceeds in six steps:

\smallskip \noindent\textbf{1. Rewrite the remainder and threshold predicates.}
	The constructions of Sections \ref{subsec:construction_modulo} and \ref{subsec:construction_threshold} only work for predicates where all coefficients are powers of 2. 
	We transform each predicate $\varphi_j$ into a new predicate $\varphi'_j$ where all coefficients are decomposed into their powers of 2.
	In our example, $\varphi'_1 := \varphi_1$ because all coefficients are already powers of 2. 
	However, $\varphi_2(x,y) = \left(8x+5y \equiv_{11} 4\right)$ is rewritten as $\varphi'_2(x,y_1,y_2) := \left(8x + 4y_1 + 1y_2 \equiv_{11} 4\right)$ because $5 = 4+1$.
	Note that $\varphi_2(x, y) = \varphi'_2(x,y,y)$ holds for every $x, y\in \N$.
	
\smallskip \noindent\textbf{2. Construct subcomputers.} 
	For every $1 \leq i \leq \Npred$, if $\varphi_j'$ is a remainder predicate, then let $\Prot_j$ be the computer defined in Section \ref{subsec:construction_modulo},
	 and if $\varphi_j'$ is a threshold predicate, then let $\Prot_j$ be the computer of Section~\ref{subsec:construction_threshold}, 
	with $\h =  \h_0 + \lceil \log_2 \Npred \rceil$. The computers $\Prot_1$ and $\Prot_2$ are shown in Figures 
	\ref{fig:population_computer_modulo} and \ref{fig:population_computer_threshold}, respectively.
	
\smallskip \noindent\textbf{3. Combine subcomputers.} 
	Take the disjoint union of $\Prot_j$, but merge their $0$ states. More precisely, rename all states $q \in Q_j$ to $\StateInProt{q}{j}$, with the exception of state $0$. 
	Construct a computer with the union of all the renamed states and transitions. We call the combined $0$ state \textit{reservoir state}, as it holds agents with value zero 
	needed for various tasks like input distribution. 
	
\smallskip \noindent\label{enum:constr_input_dist}\textbf{4. Distribute the input.} 
	For each variable $x_i \in X$, add a corresponding new input state $x_i$ and a \emph{distribution transition} that takes one agent from state $x_i$ and helpers from $0$,
	and sends them to the input states of the subcomputers $\Prot_1, \ldots, \Prot_s$. The destinations of the agents sent to $\Prot_j$ are determined by $a_i^j$, the coefficient 
	of $x_i$ in  $\varphi_j$. In \Cref{ex:runningbool}, the predicates $\varphi_1$ and $\varphi_2$ have two variables $x, y$, and so we add to the computer two new states $x$ and $y$. 
	Further, the coefficients of $x$ in $\varphi_1$ and $\varphi_2$ are $8$ and $-2$, respectively, and so we add the distribution transition  $x,0 \mapsto \StateInProt{8}{1}, \StateInProt{-2}{2}$. 
	In words, this transition takes one agent from $x$ and one helper agent, and sends them to state $8$ of $\Prot_1$ and state $-2$ of $\Prot_2$.
	More interestingly, the coefficients of $y$ in $\varphi_1$ and $\varphi_2$ are $5$ and $1$, and so we add the distribution transition $y,0,0 \mapsto \StateInProt{1}{1}, \StateInProt{4}{1}, \StateInProt{1}{2}$.
	This transition takes one agent from $x$ and \emph{two} helpers, and sends two of these agents to the states $4$ and $1$; in this way, $\Prot_1$ receives agents with a total value of $5$.  
	The third agent is sent to state $1$ of $\Prot_2$. Observe that the input is distributed to the subcomputers one agent at a time, and the distribution ends when the input states $x$ and $y$ become empty.
	
\smallskip \noindent\label{enum:constr_helpers}\textbf{5. Set the number of helpers.}  As we have seen,  distribution transitions need helpers. The initial number of helpers is chosen in order
to guarantee that every run of the computer distributes all the input, i.e.\ eventually reaches a configuration where the input states are empty. 
Let $\splitsize$ be the maximum arity of the distribution transitions for the input variables.  In our example, the distribution transitions for $x$ and $y$ have arity two and three, respectively, and so $\splitsize=3$. Intuitively, with $\splitsize-1$ helpers
the computer can distribute one agent from any of the input states $x_1, \ldots,x_\Nvar$. Initially, we put in the combined state $0$ all helpers from all subcomputers, plus $\splitsize-1$ additional helpers. 
In \Cref{ex:runningbool}, the subcomputers for $\varphi_1$ and $\varphi_2$ have $12$ and $4$ helpers, respectively, and $\splitsize=3$. So the final number of helpers is $12+4+2=18$.
\Cref{lem:fulldistribution} proves that this number of helpers guarantees the complete distribution of the input.
       
	
\smallskip \noindent \textbf{6. Combine the output functions.} Recall that $\Prot_\varphi$ combines the outputs of the subcomputers $\Prot_1,...,\Prot_\Npred$ according to $B(\varphi_1, \ldots, \varphi_\Npred)$.
In \Cref{ex:runningbool}, we set the output to 1 if and only if the output of $\Prot_1$ or $\Prot_2$ is 1.

\medskip

\subsubsection{Formal definition}
\label{subsec:formaldefboolean}

We define the population computer $\Prot_\varphi$ for a boolean combination of threshold or remainder predicates 
$\varphi_1, \ldots, \varphi_\Npred$. Formally, $\varphi :=B(\varphi_1, \ldots, \varphi_\Npred)$, where $B(\varphi_1, \ldots, \varphi_\Npred)$ is a boolean formula over  variables
$\varphi_1, \ldots, \varphi_\Npred$, e.g.\ $( \varphi_1 \wedge \varphi_2)\vee (\varphi_3 \wedge (\varphi_1 \vee \varphi_4))$.
We assume w.l.o.g. that each $\varphi_j$ is a predicate over the same set $X=\{x_1, \ldots, x_\Nvar\}$ of variables, and that it is either a remainder predicate
$\big(\sum_{i=1}^{\Nvar} a^j_i x_i \equiv_{\Modulus_j} c_j\big)$, where $0 \leq a^j_i < \Modulus_j$ and $0 \leq c_j < \Modulus_j$, or a threshold predicate
$\big(\sum_{i=1}^{\Nvar} a^j_i x_i \geq c_j\big)$.

\subparagraph{Rewriting the predicates.} We give the formal definition of the predicates $\varphi_j'$ for every $1 \leq j \leq s$.
We use an auxiliary function $\Bin{x}$ that maps an integer $x$ to the multiset over $\pow$ 
corresponding to the binary representation of $x$. For example, $\Bin{-13} = \multiset{-2^3, -2^2, -2^0}$ and $\Bin{10} = \multiset{2^3, 2^1}$. Formally, let $\sign{x} \colon \Z \rightarrow \{-1,0,1\}$ 
be the function that assigns $-1,0,1$ to the negative integers, $0$, and the positive integers, respectively, and define
\begin{align*}
	\Bin{x} &:= \multiset{\sign{x} \cdot 2^i \ \mid\ \text{$i$-th bit in the binary encoding of $\Abs{x}$ is $1$.}}
\end{align*}
Note that $\multisum{\Bin{x}} = x$ for all $x \in \Z$. We rewrite each predicate $\varphi_j$ into:
\begin{align*}
	\varphi'_j := 
	\begin{cases}
		\sum_{i=1}^\Nvar \sum_{e \in \Support{\Bin{a_i^j}}} e \cdot x_{i,e} \geq c_j &\text{ if $\varphi_j$ is a threshold predicate} \\
		\sum_{i=1}^\Nvar \sum_{e \in \Support{\Bin{a_i^j}}} e \cdot x_{i,e} \equiv_{\Modulus_j} c_j &\text{ if $\varphi_j$ is a remainder predicate}
	\end{cases}
\end{align*}

\subparagraph{Construction of subcomputers.} We define the population subcomputer $\Prot_j = (Q_j, \delta_j, I_j, O_j, H_j)$ for each $\varphi'_j$ as follows:
\begin{itemize}
\item  If $\varphi'_j$ is a remainder predicate, we use the construction in Section~\ref{subsec:construction_modulo}, setting  $Q_j := \powp_{\h_j} \cup \{0\}$  for $\h_j := \lceil\log_2 \Modulus_j \rceil$.
\item If $\varphi'_j$ is a threshold predicate, we use the construction in Section~\ref{subsec:construction_threshold}, setting  $Q_j := \pow_{\h_j} \cup \{0\}$ for
$\h_j := \max \{ \lceil \log_2 c_j \rceil+1, \lceil\log_2 (\Npred \cdot a^j_\text{max})\rceil \}+4$, where $a^j_\text{max} := \max \{ \Abs{a_1^j}, \ldots, \Abs{a_\Nvar^j} \}$. (The addition of $4$ is not necessary for correctness, but we will later use it to show that the protocol is fast.)
\end{itemize}

\subparagraph{Definition of $\Prot_\varphi$.} 

We proceed to formally define the computer $\Prot_\varphi = (Q,\delta,I,O,H)$.

\medskip\noindent\textit{States and initial states. } Define for each subcomputer $\Prot_j$ a mapping $\nu_j$ that renames the states of $Q_j$
as follows: $\nu_j(0)=0$ and $\nu_j(q) = \StateInProt{q}{j}$ for every $q \neq 0$. After renaming, the states of $\Prot_1, \ldots, \Prot_\Npred$ are 
pairwise disjoint,  with the exception of the common reservoir state $0$. The set of states of $\Prot_\varphi$ is $Q:=X \cup \bigcup_{j=1}^s \nu_j(Q_j) \cup \{0\}$.
The set of initial states is $I:=X$.

\smallskip\noindent\textit{Transitions.}  The set $\delta$ of transitions of $\Prot_\varphi$ contains:
\begin{itemize}
\item For each subcomputer $\Prot_j$,  all transitions of $\Prot_j$, suitably renamed: 
\begin{alignat*}{2}
\nu_j(r)&\mapsto\nu_j(s)&\qquad&\text{ for every }(r\mapsto s)\in\delta_j\TraName{subcomputer}
\end{alignat*}
\item Given a multiset $M\in\N^{Q_j}$, let $\nu_j(M)$ be the result of renaming the agents in $M$ according to $\nu_j$,
and let $b_i := \sum_{j=1}^\Npred |\Bin{a_i^j}|$. For each variable $x_i \in X$, the computer $\Prot_\varphi$ contains a transition  that distributes agents in state $x_i$ to the states of the subcomputers:
\begin{align*}
\begin{array}{rlr}
x_i,  \underbrace{0,\ldots,0}_{b_i -1}  & \mapsto  \sum_{j=1}^\Npred \nu_j(\Bin{a_i^j})  & \mbox{if $b_i > 1$}  \\
x_i, 0  & \mapsto  \sum_{j=1}^\Npred \nu_j(\Bin{a_i^j}), 0 &  \mbox{if $b_i = 1$}
\end{array}\TraName{distribute} 
\end{align*}
\end{itemize}

\smallskip\noindent\textit{Helpers.} Let $\splitsize := \max_{i=1}^\Nvar \sum_{j=1}^\Npred \Absbig{\Bin{a_i^j}}$. 
We set $H:=(\max(\splitsize,2) - 1)\cdot\multiset{0} + \sum_{j=1}^\Npred H_j$. So, loosely speaking, we put in state $0$ at least the total number of helpers of all subcomputers plus $\max(\splitsize,2) - 1$
additional helpers. This number of helpers guarantees that every run from a configuration that populates the initial states eventually enables some \TraRef{distribute} transition, and so that
terminal configurations do not populate the initial states (see \Cref{lem:fulldistribution}).

\smallskip\noindent\textit{Output function.} The output function is the boolean combination of the output functions of the subcomputers.
Formally, $O(S):= B(O_1(S \cap (\nu_1(Q_1))),...,O_\Npred(S \cap (\nu_\Npred(Q_\Npred))))$.

\newcommand{\proj}[2]{{#1}|_{#2}}
\subsubsection{Correctness and size}
\label{app:PC_general_correctness}

We prove that $\Prot_\varphi$ decides $\varphi = B(\varphi_1, \ldots, \varphi_s)$. We use the method that will be described in \Cref{prop:genmethod}.
This requires to generalise some notions of Section \ref{subsec:method} which were defined only for computers whose states are numbers (which is not the case for $\Prot_\varphi$, because the initial states are the variables in $X$ and these have no value), and only for remainder and threshold predicates, not for their boolean combinations.

Given a predicate $\varphi_j$ of the boolean combination, let us first define when a configuration $C$ of $\Prot_\varphi$ satisfies $\varphi_j$. Let $\proj{C}{j}$ denote the projection of $C$ onto $\nu_j(Q_j) \cup \{0\}$ (recall that $Q_j$ is the
set of states of the subcomputer for $\varphi_j$). Define $\multisum{C}_j := \multisum{\proj{C}{j}} + \sum_{x_i \in X} a_i^j C(x_i)$. Intuitively, this takes into
account that at $C$ the input may not have been completely distributed yet, and for the $j$-th subcomputer each agent in $x_i$ has value $a_i^j$.
We say that $C$ \emph{satisfies} $\varphi_j$ if $\varphi_j$ is a remainder predicate $\sum_{i=1}^\Nvar a_i^j x_i \equiv_{\Modulus_j} c_j$  and $\multisum{C}_j \equiv_{\Modulus_j} c_j$, or
if $\varphi_j$ is a threshold predicate $\sum_{i=1}^\Nvar a_i^j x_i \geq c_j$ and $\multisum{C}_j \geq c_j$. Similar to Section \ref{subsec:method}, this extends satisfying \(\varphi_j\) to states which are not variables. One important difference however is that in this case, \(\multisum{C}_j\) (and hence \(C\) satisfying \(\varphi_j\)) is \emph{not} independent of the coefficients of \(\varphi_j\), because of the \(\sum_{x_i \in X} a_i^j C(x_i)\) summand. 

We can now generalise the definitions of Section \ref{subsec:method} as follows: 

\begin{itemize}
\item $C$ \emph{satisfies} $\varphi$ if $B(b_1, \ldots, b_s)=1$, where $b_j=1$ if $C$ satisfies $\varphi_j$ and $b_j =0$ otherwise. 
\item Two configurations $C, C'$ are \emph{equivalent w.r.t.\ $\varphi$} if both $C$ and $C'$ satisfy $\varphi$, or none does.
\item A  configuration $C$  is \emph{well-supported} w.r.t.\ $\varphi$ if $C$ is equivalent to $\Support{C}$ w.r.t.\ $\varphi$.
\end{itemize}

We have the following result, proved exactly the same as \Cref{prop:method}:

\begin{proposition}
\label{prop:genmethod}
Let $\varphi = B(\varphi_1, \ldots, \varphi_s)$ be a boolean combination of remainder and modulo predicates $\varphi_1, \ldots, \varphi_s$. 
If $\Prot_\varphi$ satisfies the following properties, then it decides $\varphi$:
\smallskip
\begin{enumerate}
\item $\Prot$ is bounded.  \label{gspec:prop1}
\item Transitions preserve equivalence. \label{gspec:prop2}
\item Terminal configurations are well-supported. \label{gspec:prop3}
\item The output function $O$ is given by $O(S)=1$ if{}f $B(O_1(S \cap Q_1), \ldots, O_s(S \cap Q_s))=1$. \label{gspec:prop4}
\end{enumerate}
\end{proposition}

In the rest of the section we show that the protocol $\Prot_\varphi$ of Section \ref{subsec:formaldefboolean} satisfies the 
four properties of Proposition \ref{prop:genmethod}. The key of the proof is the following technical lemma, showing that every terminal 
configuration distributes the input completely.

\begin{restatable}{lemma}{lemmafulldistribution}\label{lem:fulldistribution}
Let $\Cterm$ be a terminal configuration of $\Prot_\varphi$ reachable from some initial configuration. Then $\Cterm(X)=0$.
\end{restatable}
\begin{proof}

%

Let $Q_j' = \{ (2^{\h_j})_j,  (-2^{\h_j})_j , \ldots, (2^{0})_j,  (-2^{0})_j\}$ be the states of $\Prot_j$ with non-zero value. 
Further, let $\Delta$ be the number of occurrences of \TraRef{distribute} transitions  in the run leading from $C_0$ to $\Cterm$. 
We proceed in three steps.

\smallskip\noindent (1) $\Cterm(Q_j') \leq H_j(0) + \frac{\Delta}{\Npred}$ holds for each subcomputer $\Prot_j$. 
Intuitively, this states that the number of agents in $Q_j$ is at most the number of helpers of $\Prot_j$ plus a \textit{``fair share''} (i.e.\ $\frac{1}{\Npred}$) of the processed agents.
For the proof, observe that the projection $\proj{\Cterm}{j}$ is a terminal configuration of $\varphi_j$.  If $\varphi_j$ is a remainder predicate, then $\Cterm(Q_j') \leq 3\h_j = H_j(0)$ because otherwise $\proj{\Cterm}{j}$ enables some transition of  $\Prot_j$ (see the proof of  \Cref{thm:PCmod}), and we are done. If $\varphi_j$ is a threshold predicate, then we proceed as follows. Observe that
$$\Cterm(Q_j') = \underbrace{\Cterm(\,(2^{\h_j})_j\, ) + \Cterm(\,(-2^{\h_j})_j\,)}_{\mbox{ $\alpha_j$ }}  + \underbrace{\sum_{i=0}^{\h_j-1} \Cterm(\,(2^{i})_j\,) + \sum_{i=0}^{\h_j-1} \Cterm(\,(-2^{i})_j\,)}_{\mbox{ $\beta_j$ }}$$
We show that $\alpha_j \leq \Delta / \Npred$ and $\beta_j \leq H_j(0)$.
\begin{itemize}
\item $\alpha_j \leq \Delta / \Npred$.  Let $a^j_\text{max}$ be the absolute value of the maximum coefficient of $\varphi_j$. Every occurrence of a \TraRef{distribute} transition increases the total absolute value of the agents in $Q_j'$ by at most $a^j_\text{max}$. Since this absolute value is initially equal to zero, we have $\multisum{\proj{\Cterm}{j}} \leq \Delta \cdot  a^j_\text{max}$. Further, $\Cterm$ populates at most one of the states $(2^{\h_j})_j$ and $(-2^{\h_j})_j$, because otherwise it enables  the \TraRef{thr:cancel:app} transition, contradicting that $\Cterm$ is terminal. Assume $\Cterm$ populates only $(2^{\h_j})_j$ (the other case is similar). Since each agent in this state has value $2^{\h_j} \geq a^j_\text{max}$ and the total absolute value of  $\proj{\Cterm}{j}$ is at most $\Delta \cdot  a^j_\text{max}$, at most $\Delta/\Npred$ agents of $\Cterm$ populate $(2^{\h_j})_j$.
\item $\beta_j \leq H_j(0)$. We have  $\beta_j \leq \h_j$ because, by \Cref{lem:wellsupportedthreshold}, terminal configurations of the computer for a threshold predicate put at most $\h_j$ agents in the states of $\pow_{\h_j-1}$, and $\h_j \leq H_j(0)$ by definition of the computer for a threshold predicate.
\end{itemize}

\smallskip\noindent (2)   $\Cterm(0) \geq \max(\splitsize,2){-}1$.  We start by collecting two facts:

\begin{enumerate}[a.]
\item By definition of $\Prot_\varphi$, the initial configuration $C_0$ puts at least  $\max(\splitsize,2){-}1 +  \sum_{j=1}^\Npred H_j(0)$ helpers in state $0$.
\item $\Cterm(X)= \Cterm(Q) - C_0(0) - \Delta$. This is a consequence of $\Cterm(X) = C_0(X) - \Delta$, which holds because each occurrence of a \TraRef{distribute} transition removes one agent from $X$,
and $C_0(X)  = C_0(Q) - C_0(0)$. 
\end{enumerate}

Now we proceed as follows:
\begin{align*}
\Cterm(0)  &= \Cterm(Q) - \big(\Cterm(X) + \sum_{j=1}^\Npred \Cterm(Q_j') \big)&   \mbox{$\{0\} = Q \setminus (X \cup  \bigcup_{j=1}^s Q_j' )$}\\
& = C_0(0) + \Delta - \sum_{j=1}^\Npred \Cterm(Q_j')  & \mbox{by (b)} \\
& \geq  \max(\splitsize,2){-}1 +  \sum_{j=1}^\Npred H_j(0) + \Delta - \sum_{j=1}^\Npred \Cterm(Q_j') & \mbox{by (a)}\\
& \geq  \max(\splitsize,2){-}1 +  \sum_{j=1}^\Npred H_j(0) + \Delta - \sum_{j=1}^\Npred \left(H_j(0) + \frac{\Delta}{s}\right) & \mbox{by (1)}\\
&\geq \max(\splitsize,2){-}1 
\end{align*}

\smallskip\noindent (3)  $\Cterm(X) = 0$. By contradiction. Assume $\Cterm(x_i) \geq 1$ for some $x_i \in X$. Then by (2) the \TraRef{distribute} transition for $x_i$ is enabled, contradicting that $\Cterm$ is a terminal configuration. 
\end{proof}

We are now able to prove our first main result.

\thmmainA*
\begin{proof}	
By definition, $\varphi$ is a boolean combination $\varphi = B(\varphi_1, \ldots, \varphi_s)$ of remainder and threshold predicates. 
Let $\Prot_\varphi$ be the population computer of Section \ref{subsec:formaldefboolean}.
We show that $\Prot_\varphi$ satisfies the conditions of \Cref{prop:genmethod} (and so decides $\varphi$) and has size $O(\Abs{\varphi})$. 
We begin with the conditions of \Cref{prop:genmethod}.

 \noindent (\ref{gspec:prop1}) $\Prot_\varphi$ is bounded.  The execution of a \TraRef{distribute} transition strictly reduces the number of agents in the input states, 
 and no \TraRef{subcomputer} transition puts agents in them. It follows that every run from an input configuration executes \TraRef{distribute} transitions only finitely often.
 Further, by \Cref{thm:PCmod} and \Cref{thm:PCthr}, each subcomputer $\Prot_j$ is bounded. So runs of $\Prot_\varphi$ also execute \TraRef{subcomputer}
 transitions finitely often, and we are done. 
 
 \smallskip\noindent (\ref{gspec:prop2}) Transitions preserve equivalence w.r.t.\ $\varphi$.  We have to show that if $C \mapsto C'$ then both $C$ and $C'$ satisfy $\varphi$, or none does.
 We prove a stronger property: for every $1 \leq j \leq s$, $C$ satisfies $\varphi_j$ if{}f $C'$ satisfies $\varphi_j$. Let $1 \leq j \leq s$. It suffices to prove 
 $\multisum{C}_j =  \multisum{C'}_j$.  If  $C \mapsto C'$ by a \TraRef{subcomputer} transition the result follows from the corresponding result for subcomputers.
 If  $C \mapsto C'$ by  a \TraRef{distribute} transition $x_i, 0,\ldots,0 \mapsto  \sum_{j=1}^\Npred \nu_j(\Bin{a_i^j})$, then we have
\begin{align*}
\multisum{C'}_j -  \multisum{C}_j & = (\multisum{\proj{C'}{j}} -  \multisum{\proj{C}{j}}) + \sum_{x_k \in X} a_k^j (C'(x_k) - C(x_k)) \\
 & = \sum_{j=1}^s{\Bin{a_i^j}} - a_i^j  = 0
 \end{align*}
 
 \noindent (\ref{gspec:prop3}) Terminal configurations are well-supported. This is the part requiring an application of \Cref{lem:fulldistribution}.
 Let $\Cterm$ be a terminal configuration. We prove that $\Cterm$ and  $\Support{\Cterm}$ are equivalent w.r.t.\  $\varphi_j$ for every $1 \leq j \leq s$,
 which implies that $\Cterm$ is equivalent to  $\Support{\Cterm}$ w.r.t.\ $\varphi$.  Pick $1 \leq j \leq s$, and assume  $\varphi_j$ is a threshold predicate 
 $\sum_{i=1}^\Nvar a_i^j x_i \geq c_j$; the remainder case \(\sum_{i=1}^\Nvar a_i^j x_i \equiv_{\Modulus_j} c_j\) is analogous. 
 
 Let $\proj{\Cterm}{j}$ be the projection of $C$ onto the subcomputer $\Prot_{\varphi_j}$.  
 By definition,  $\Cterm$ satisfies $\varphi_j$ if{}f  $\multisum{\Cterm}_j \geq  c_j$, and $\Support{\Cterm}$ satisfies $\varphi_j$ if{}f  $\multisum{\Support{\Cterm}}_j \geq  c_j$.
 So it suffices to show  $\multisum{\Cterm}_j \geq  c_j \Leftrightarrow \multisum{\Support{\Cterm}}_j \geq  c_j$.  
 We have $\multisum{\Cterm}_j = \multisum{\proj{\Cterm}{j}} + \sum_{x_i \in X} a_i^j \Cterm(x_i)$  
 by definition, and so, since $\Cterm(X)=0$ by \Cref{lem:fulldistribution}, we get $\multisum{\Cterm}_j = \multisum{\proj{\Cterm}{j}}$. Analogously, 
 $\multisum{\Support{\Cterm}}_j = \multisum{\proj{\Support{\Cterm}}{j}}= \multisum{\Support{\proj{\Cterm}{j}}}$.
 So it suffices to show $\multisum{\proj{\Cterm}{j}} \geq c_j \Leftrightarrow \multisum{\Support{\proj{\Cterm}{j}}} \geq c_j$, i.e.\ that $\proj{\Cterm}{j}$ is a well-supported configuration
 of $\Prot_j$ w.r.t.\ $\varphi_j$. But this holds by \cref{thm:PCthr}.
 
  \smallskip\noindent (\ref{gspec:prop4}) $O(S)=1$ if{}f $B(O_1(S \cap Q_1), \ldots, O_s(S \cap Q_s))=1$. By definition.

\medskip \noindent Now we show that $\Prot_\varphi$ has size $O(\Abs{\varphi})$.
By definition, the size of $\Prot_\varphi$ is $\size(\Prot_\varphi) :=  \Abs{Q} + \Abs{H} + \size(O) + \sum_{t\in \delta} \Abs{t}$. 
We show that each of $\Abs{Q}$, $\Abs{H}$, $\size(O)$ and $\sum_{t\in \delta} \Abs{t}$ is $\O(\Abs{\varphi})$.

For $Q$, recall that $Q:=X \cup \bigcup_{j=1}^\Npred \nu_j(Q_j) \cup \{0\}$. If $\varphi_j$ is a remainder predicate, then $\Abs{Q_j} = d_j := \lceil \log_2 \Modulus_j \rceil$.
If $\varphi_j$ is a threshold predicate, then $\Abs{Q_j} = 2 \h_j = 2 \h_{0j} + 2 \lceil \log_2 \Npred \rceil \in \O(\Abs{\varphi_j} + \log_2 \Npred)$.
So $\Abs{Q} \in \O(\Abs{\varphi} + \Npred \log_2 \Npred)$. Since $\varphi$ is a boolean formula over variables $\varphi_1, \ldots, \varphi_\Npred$, and each variable appears at least once in the formula,
$\varphi$ has size $\Omega(\Npred \log_2 \Npred)$, and so $\Abs{Q} \in \O(\Abs{\varphi})$.

For $H$, recall that $H:=(\max(\splitsize,2) - 1)\cdot\multiset{0} + \sum_{j=1}^\Npred H_j$, where $\splitsize := \max_{i=1}^\Nvar \sum_{j=1}^\Npred \Absbig{\Bin{a_i^j}}$
and $\Abs{H_j} \leq 3 \cdot d_j$ for every $1 \leq j \leq \Npred$. So $\Abs{H} \in O(\Abs{\varphi} + \Npred \log_2 \Npred) = \O(\Abs{\varphi})$.

For $O$, observe that given boolean circuits for functions $O_1, \ldots, O_\Npred$, with $\gamma_1, \ldots, \gamma_\Npred$ gates,  and a circuit $\gamma$ for 
a boolean formula $B(x_1, \dots, x_\Npred)$ with $\gamma$ gates, there is a circuit for $B(O_1, \dots, O_\Npred)$ with  $\sum_{j=1}^\Npred \Abs{\gamma_j} + \Abs{\gamma} + \O(1)$ gates.

For $\sum_{t\in \delta} \Abs{t}$, observe that $\sum_{t \in \delta_j} \Abs{t} = \O(d_j)$ for every subcomputer $\Prot_j$, and that the distribution transition for $x_i$ has arity 
$\sum_{j=1}^\Npred \Abs{\Bin{a_i^j}}$. So the total size is 
\begin{multline*}
\sum_{j=1}^\Npred \O(d_j) + \sum_{x_i \in X} \sum_{j=1}^\Npred\Abs{\Bin{a_i^j}} = \sum_{j=1}^\Npred \O(d_j) + \sum_{j=1}^\Npred\sum_{x_i \in X}\Abs{\Bin{a_i^j}}
\\ = \O(\Abs{\varphi} + \Npred \log_2 \Npred) + \sum_{j=1}^\Npred \Abs{\varphi_j} = \O(\Abs{\varphi} + \Npred \log_2 \Npred) + \O(\Abs{\varphi}) = \O(\Abs{\varphi}) 
\end{multline*}
\end{proof}

\section{From Bounded Population Computers to Fixed-Parameter Fast Population Protocols}  \label{sec:conversion}

We prove Theorem \ref{thm:mainB1}, i.e.\ we show that for every predicate $\varphi$, any bounded population computer of size $m$ deciding $\Double(\varphi)$ can be converted into a population protocol of size $\O(m^2)$ that decides $\varphi$ in $2^{\O(m^2 \log m)} \cdot n^3$ interactions for inputs of size $\Omega(m)$. 

We start in Section~\ref{ssec:fastoutputbroadcast} by proving that every bounded binary computer with no helpers terminates in \(2^{\O(m \log m)} \cdot n^3\) expected interactions. This relates to the bound in Theorem \ref{thm:mainB1} by replacing \(m\) by \(\O(m^2)\), the size of the computer we will use the statement on.

Hence it suffices to convert a bounded population computer for $\Double(\varphi)$ into a bounded population protocol for $\varphi$, with only a quadratic blow-up in size. We achieve this by applying a sequence of five conversion steps. For most of these conversions we have to prove that they preserve the decided predicate, and for this reason before presenting the steps we introduce a technique to prove equivalence of computers in Section~\ref{app:refinement}. The rest of Section \ref{sec:conversion} describes the five steps:
\begin{itemize}
\item Section~\ref{ssec:preconversion} converts a bounded computer $\Prot$ for $\Double(\varphi)$  into an equivalent bounded computer $\Prot_0$ of size $\O(\size(\Prot))$ satisfying two additional technical conditions. This is a very simple step that also serves as warm-up for the next ones.
\item Section~\ref{ssec:kwayconversion} converts $\Prot_0$ into an equivalent \emph{binary} bounded computer $\Prot_1$ of size $\O(\Abs{Q_0} \cdot \size(\Prot_0))$, where $Q_0$ is the set of states of $\Prot_0$.
\item Section~\ref{ssec:outputfunction1} converts $\Prot_1$ into an equivalent binary bounded computer $\Prot_2$ with a \emph{marked consensus output function} (a notion defined in the section) of adjusted size $\O(\size_2(\Prot_1))$. 
\item Section \ref{ssec:simulatehelpers} converts $\Prot_2$ into a  binary bounded computer $\Prot_3$ for $\varphi$ ― not $\Double(\varphi)$ ― with a marked consensus output function \emph{and no helpers} of adjusted size $\O(\size_2(\Prot_2))$. 
\item Section~\ref{ssec:converttoconsensusoutput}, converts $\Prot_3$ into a binary and terminating (not necessarily bounded) computer $\Prot_4$ for $\varphi$ with normal consensus output function and no helpers of adjusted size $\O(\size_2(\Prot_3))$. 
\item Section~\ref{ssec:puttogether} puts all steps together to show that $\Prot_4$ has the number of states and number of interactions given by Theorem \ref{thm:mainB1}.
\end{itemize}

Sections~\ref{ssec:preconversion} to \ref{ssec:converttoconsensusoutput} are self-contained and can be read in any order.

\subsection{A $2^{\O(m \log m)} \cdot n^3$ bound on the expected number of interactions}\label{ssec:fastoutputbroadcast}

We prove that a bounded binary computer with no helpers of size $m$ terminates within $2^{\O(m \log m)} \cdot n^3$ expected interactions. 

To prove this bound, we first introduce \emph{potential functions} in Definition \ref{defPotentialFunction}. A potential function assigns to every configuration a non-negative \emph{potential}, with the property that executing any transition strictly decreases the potential. In this paper we only consider linear potential functions.

Then, we show in Lemma \ref{lem:boundedpotential} that bounded population computers have (linear) potential functions, which allows us to show that every run of a bounded computer executes at most $2^{\O(m \log m)} \cdot n$ transitions. However, not every interaction between agents executes a transition.
For example, consider a computer with states $q, q', q''$, a single transition $\multiset{q,q'} \mapsto \multiset{q'',q''}$, and a configuration with one agent in each of $q$ and $q'$ and $n-2$ agents in state $q''$. If we choose two agents uniformly at random, the probability that one of them is in state $q$ and the other in state $q'$ is $2/n(n-1)$. In general, all we can say is that a transition is executed after $\O(n^2)$ interactions in expectation.  This leads to $2^{\O(m \log m)}  \cdot n^3$ interactions in expectation for the execution of the complete run. 

\begin{restatable}{definition}{defPotentialFunction}\label{defPotentialFunction}
A function $\Phi:\N^Q\rightarrow\N$ is \emph{linear} if there exist weights $w \colon Q\rightarrow\N$ s.t.\ $\Phi(C)=\sum_{q\in Q}w(q) \cdot C(q)$ for every $C\in\N^Q$. 
A \emph{potential function (for $\Prot$)} is a linear function $\Phi$ such that $\Phi(r)\ge\Phi(s)+\Abs{r}-1$ for all $(r\mapsto s)\in\delta$. 
\end{restatable}
\noindent Observe that $k$-way transitions reduce the potential by $k-1$, binary transitions by $1$. In this section we consider only binary computers, but in Section \ref{sec:speed} we will consider general ones.

If a population computer has a potential function with maximal weight \(W:=\max_{q\in Q} w(q)\), then every run executes at most $W \cdot n$ transitions, and so the computer is bounded. We prove that the converse holds for computers in which every state can be populated. That is, if a computer is bounded and every state can be populated, then the computer has a potential function. Observe that the condition that every state can be populated is very mild, since states that can never be populated can be deleted without changing the behaviour of the computer.


\newcommand{\Vzero}{\mathbf{0}}

\begin{lemma} \label{lem:boundedpotential}
Let $\Prot$ be a computer of size $m$ with set of states $Q$ such that for every $q \in Q$ some reachable configuration populates $q$. Then $\Prot$ is bounded if{}f there is a potential function $\Phi(C)=\sum_{q\in Q}w(q) \cdot C(q)$ for $\Prot$ such that $W:=\max_{q \in Q} w(q) \in 2^{\O(m \log m)}$.
\end{lemma}
\begin{proof}
Let the incidence matrix of $\Prot$ be the matrix $A\in\Z^{\delta\times Q}$ s.t.\ the $t$-th row is $A_t:=s-r$, for $t=(r\mapsto s)\in\delta$. In particular, given a vector $y\in\N^\delta$ which assigns each transition a count, $A^\top y$ is the change in the number of agents of each state after executing a sequence of transitions containing $y(t)$ times the transition $t$. In the following we write $\mathbf{1}$ for the all-ones vector of appropriate dimension. We prove the 
existence of a (linear) potential function for $\Prot$ by showing that the following statements are equivalent:

\begin{enumerate}[(a)]
\item\label{pb-tfaq:a} $\Prot$ is bounded.
\item\label{pb-tfaq:b} $A^\top y\ne\Vzero$ for all $y\in\N^\delta$ with $y\ne \Vzero$.
\item\label{pb-tfaq:c} $A^\top y\ne\Vzero$ for all $y\in\R_{\ge0}^\delta$ with $y\ne\Vzero$.
\item\label{pb-tfaq:d} $Ax\le-\mathbf{1}$ for some $x\in\R^Q$.
\item\label{pb-tfaq:e} There is a potential function $\Phi$ for $\Prot$.
\end{enumerate}

(Afterwards, we show the size bound on $\Phi$.)

Let us first give some brief intuition. Essentially, (a) states that $\Prot$ does not have a loop, i.e.\ no sequence of transitions leading from a configuration to itself. This is strengthened in (b), which says that no loop exists, even when the protocol is allowed to execute transitions at any time (i.e.\ the number of agents is allowed to go negative). It is further strengthened in (c), where the computer is also allowed to execute transitions ``fractionally''; for a transition $r \mapsto s$, the computer can now for example remove $0.5$ agents from each state in $r$ and add $0.5$ agents to each state of $s$. Statement (d) then says one can find real weights for each state s.t.\ the total weight of a configuration decreases with each transition. Finally, (e) strengthens this by requiring the weights to be natural numbers.

“$(\ref{pb-tfaq:a})\Rightarrow(\ref{pb-tfaq:b})$”: Assume that (\ref{pb-tfaq:b}) does not hold, so there is a nonempty multiset $y\in\N^\delta$ with $A^\top y=\Vzero$. Let $t_1,...,t_k\in\delta$ denote an enumeration of $y$. Due to the definition of $A$, $A^\top y=\Vzero$ means that executing the sequence $t_1t_2...t_k$ has no effect. Formally, for any $C,C'$ with $C\rightarrow_{t_1}...\rightarrow_{t_k}C'$ we get $C=C'$. It suffices to find such a configuration $C$ which is reachable; as then we can clearly construct an infinite run, contradicting (\ref{pb-tfaq:a}).
\newcommand{\Deltasize}{\lambda}
By assumption, for every state $q$ there exists an initial configuration $C_{Iq}$ and a  configuration $C_q$ reachable from $C_{Iq}$ such that
$C_q(q) > 0$. It follows that the configuration $C:=\sum_{q \in Q} C_q$ is reachable from the initial configuration $C_I := \sum_{q \in Q} C_{Iq}$,
and satisfies $C'(q) > 0$ for every $q \in Q$. Multiplying $C$ and $C_I$ by adequate constants, if necessary, we can 
assume w.l.o.g. that $C(q)\ge k\Deltasize$, where $\Deltasize$ is the maximum arity of any transition. A single transition moves at most $\Deltasize$ agents, and so the sequence $t_1,...,t_k$ can be executed at $C$.

\newcommand{\Scalefac}{\mu}
\smallskip “$(\ref{pb-tfaq:b})\Rightarrow(\ref{pb-tfaq:c})$”: We argue by contraposition and assume that $\{y\in\R_{\ge0}^\delta\setminus\{\Vzero\}:A^\top y=\Vzero\}$ is not empty. Then there is some $\varepsilon\in\Q$ with $\varepsilon>0$ s.t.\ $\{y\in\R_{\ge0}^\delta:A^\top y=\Vzero\wedge\mathbf{1}^\top y\ge\varepsilon\}$ is not empty either. This is a satisfiable system of linear inequalities and thus has a rational solution $y^*\in\Q^\delta$. As $y^*\ge\Vzero$ we can find a $\Scalefac>0$ with $\Scalefac y^*\in\N^\delta$, showing the negation of (\ref{pb-tfaq:b}).

\smallskip “$(\ref{pb-tfaq:c})\Rightarrow(\ref{pb-tfaq:d})$”: Due to $y\ge\Vzero$ the condition $y\ne \Vzero$ is equivalent to $\mathbf{1}^\top y>0$. In other words, the system $\{y\in\R_{\ge0}^\delta:A^\top y=\Vzero\wedge -\mathbf{1}^\top y<0\}$ has no solution. Applying one of the numerous versions of  Farkas' lemma (in this case \cite[Proposition~6.4.3iii]{GM07}), we obtain that the system $\{x\in\R^Q:Ax\le-\mathbf{1}\}$ does have a solution.

\smallskip “$(\ref{pb-tfaq:d})\Rightarrow(\ref{pb-tfaq:e})$”: Since $Ax\le-\mathbf{1}$ is a system of linear inequalities,  if it has a solution it also has a rational solution and so, after scaling with an adequate factor, also an integer solution $z^* \in\Z^Q$. Let $w \in \N^Q$ be the vector of natural numbers or \emph{weights} given by $w:= \Deltasize(z^*-z^*_\mathrm{min}\mathbf{1})$, where $z^*_\mathrm{min}:=\min_{q\in Q}z^*(q)$ and  $\Deltasize$ is the maximum arity of a transition. We define $\Phi$ as the linear function induced by $w$, that is, $\Phi(x) := w \cdot x$.  

We show that $\Phi$ is a potential function. Let $C, C'$ be configurations such that $C \rightarrow C'$.
Let  $t=(r\mapsto s)\in\delta$ be the transition whose execution leads from $C$ to $C'$. We prove $\Phi(C)-\Phi(C') > \Abs{r}$. By definition of  $C' = C - r +s$, and so $\Phi(C') - \Phi(C) = \Phi(s) - \Phi(r)$. 
Let $A_t$ be the $t$-th row of $A$. By the definition of the incidence matrix $A$ we have 
$$\Phi(s)-\Phi(r)=(s-r) \cdot w=A_t \cdot w= A_t  \cdot \left( \Deltasize(z^*-z^*_\mathrm{min}\mathbf{1}) \right)=  \Deltasize \left( A_t \cdot z^* - z^*_\mathrm{min}  A_t \cdot \mathbf{1} \right)$$
\noindent Since $z^*$ is a solution of $Ax\le-\mathbf{1}$, we have $A _t \cdot z^* \le-1$. Further, since $|r|=|s|$, we have $A_t \cdot \mathbf{1}=\Vzero$. So  $\Phi(C')-\Phi(C) \le - \Deltasize \le -|r|$, and we are done.

\smallskip “$(\ref{pb-tfaq:e})\Rightarrow(\ref{pb-tfaq:a})$”: For any initial configuration $C_0$ we have $\Phi(C_0) \leq W \cdot n$ as $\Phi$ is a linear function (Definition~\ref{defPotentialFunction}). Since, by definition, $\Phi(C)\ge0$ for all configurations $C$ and any transition strictly reduces $\Phi$, a run starting at $C_0$ can execute at most $\Phi(C_0)$ transitions.

\medskip It remains to prove $W \in 2^{\O(m \log m)}$. Recall that the vector of weights is
$w:= \Deltasize(z^*-z^*_\mathrm{min}\mathbf{1})$, where $z^* \in\Z^Q$ is a solution of the system $Ax\le-\mathbf{1}$ of linear inequalities.
By definition of $A$, each entry has absolute value at most $\Deltasize$. Using well-known results (see e.g.\ \cite[Lemma~1]{Papa81}), we have
$\Abs{z^*(q)} \in \O((\Deltasize \Abs{Q})^{2\Abs{Q}}) \subseteq \O(2^{4m \log_2 m})$, hence \(w(q) \in \O(2^{5 m \log_2 m})\subseteq 2^{\O(m \log m)}\), and we are done.
\end{proof}

\begin{proposition}
\label{prop:nthree}
Let \(\Prot\) be a bounded binary computer with no helpers of size $m$. Then \(\Prot\) terminates within \(2^{\O(m \log m)} \cdot n^3\) interactions. \label{corollary_bounded_n3}
\end{proposition}

\begin{proof}
Without loss of generality we assume that every state can be populated, since removing states which cannot be populated preserves boundedness, correctness and speed. Applying Lemma \ref{lem:boundedpotential} we obtain that \(\Prot\) has a linear potential function $\Phi(C)=\sum_{q \in Q} w(q) \cdot C(q)$. Let \(W:= \max_{q \in Q} w(q)\) be the maximal weight of $\Phi$. Since an initial configuration \(C_0\) with \(n\) agents fulfils \(\Phi(C_0) \leq W \cdot n\), and every transition reduces \(\Phi\) by at least 1, \(\Prot\) terminates after executing at most \(W \cdot n\) transitions. At every non-terminal configuration, at least one (binary) transition is enabled. The probability that two agents chosen uniformly at random enable this transition is $\Omega(1/n^2)$, and so a transition occurs within $\O(n^2)$ expected interactions. Hence \(\Prot\) terminates within $\O(W \cdot n^3)$ expected interactions. By Lemma \ref{lem:boundedpotential} we have
$W \in 2^{\O(m \log m)}$, and we are done.
\end{proof}

\subsection{Refinement: a technique for proving equivalence}\label{app:refinement}
We present a general framework for proving that two computers are equivalent, that is, decide the same predicate.

\newcommand{\Input}{\mathcal{I}}
\begin{definition}\label{def:refinement}
Let $\Prot=(Q,\delta,I,O,H)$ and $\Prot'=(Q',\delta',I',O',H')$ be population computers. $\Prot'$ \emph{refines} $\Prot$ if there is a mapping $\pi:\N^{Q'}\rightarrow\N^Q$ satisfying the following properties:
\begin{enumerate}
\item For all reachable configurations $C,D \in\N^{Q'}$, if $C \rightarrow D$ then $\pi(C)\rightarrow\pi(D)$.
\item $I=I'$; further, for every initial configuration $C$ of $\Prot'$ the configuration $\pi(C)$ is an initial configuration of $\Prot$ such that $\pi(C)(q)=C(q)$ for $q\in I$ (i.e.\ $C$ and $\pi(C)$ coincide on all initial states).
\item For every reachable terminal configuration $C\in\N^{Q'}$, the configuration $\pi(C)$ is terminal and $O(\Support{\pi(C)})=O'(\Support{C})$.
\end{enumerate}
\end{definition}

Usually \(\pi\) is chosen as a linear function of the form $\pi(C)(q) = \lambda_q C(q)$ for adequate coefficients $\lambda_q$. Intuitively, by choosing
$\lambda_q = 0$ the function $\pi$ \emph{discards} information about the number of agents in $q$, and so in this sense $\Prot'$ is a refinement of
$\Prot$: the configuration \(C'\) contains all the information of \(C\),  and more. 

We prove that if  $\Prot'$ is terminating and refines $\Prot$, then $\Prot$ and $\Prot'$ are equivalent. (Recall that a population computer is terminating if every fair run is finite, and bounded if every run, fair or not,  is finite.) 

\begin{lemma}[Refinement lemma]\label{lem:refinement}
Let $\Prot,\Prot'$ denote population computers. If $\Prot'$ is terminating and refines $\Prot$, then $\Prot'$ decides the same predicate as $\Prot$.
\end{lemma}
\begin{proof}
Let $C_0\in\N^{Q'}$ denote an arbitrary initial configuration of $\Prot'$. We decompose $C_0=:C_I+C_H$ into an input configuration $C_I\in\N^{I'}$ and a helper configuration $C_H\in\N^{\Support{H'}}$, $C_H\ge H$. As $\Prot'$ is terminating, there is a terminal configuration $C$ with $C_0\rightarrow C$. It now suffices to show $O'(\Support{C})=\varphi(C_I)$. For this, we use properties 1-3 of the definition of refinement.
By property 2, $\pi(C_0)$ is an initial configuration of $\Prot$ and $\pi(C_0)=C_I+D_H$, where $D_H\in\N^H$, $D_H\ge H$ is a helper configuration of $\Prot$. So we now only need to show that $\Prot$ outputs $O'(\Support{C})$ on $C_I$, i.e.\ that there is a terminal configuration $D\in\N^Q$ reachable from $\pi(C_0)$ with $O(\Support{D})=O'(\Support{C})$.
We set $D:=\pi(C)$. By property 1 we have $\pi(C_0)\rightarrow\pi(C)$, and by property 3 we get that $\pi(C)$ is terminal. Finally, property 3 also implies $O(\Support{\pi(C)})=O'(\Support{C})$.
\end{proof}

In the next sections we prove that some protocol $\Prot'$ is equivalent to $\Prot$ by exhibiting a suitable refinement function, and showing that $\Prot'$ is terminating or bounded (recall that bounded computers are terminating).

\subsection{A preprocessing step}\label{ssec:preconversion}
For translating a bounded population computer into an equivalent population protocol it is convenient to assume that the computer satisfies some technical conditions. We describe a conversion $\preprocessconv$ that, given any bounded population computer, outputs an equivalent bounded computer satisfying the conditions. The conversion is particularly simple, and so we also use it to illustrate how we will proceed in the coming sections. 
After giving a brief high-level overview, we describe the specification of the conversion, i.e.\ the assumptions on the input computer and the properties that the output computer must satisfy. Then we present $\preprocessconv$, and finally we prove that $\preprocessconv$ satisfies the specification. Among other properties, the specification of the conversion requires the input and output computers to be equivalent. For the equivalence  proof we use the notion of refinement introduced in Section~\ref{app:refinement}.


Let $\Prot=(Q,\delta,I,O,H)$ denote a bounded population computer deciding a predicate $\Double(\varphi)$. We need two additional conditions, namely that states in $I$ have no incoming transitions, and that every configuration in $\N^I$ is terminal. To achieve this, the idea is to add two types of information. 
\begin{enumerate}
\item Every agent gets an additional flag. As long as the flag is not set, the agent is not allowed to perform any transition of \(\Prot\).
\item Add a helper state \(h\) which gives a start signal to any agent \(q\) it meets (i.e. sets the flag), allowing \(q\) to start computing.
\end{enumerate} 
Without the helper, the computation cannot start, showing that \(C\in \N^I\) is always terminal. The new input states have no incoming transitions since the flag can never be unset. The refinement \(\pi\) in this case removes the extra information in form of the flag, and disregards helpers in \(h\) entirely.

\subsubsection{Specification}
\label{spec0}

\begin{specification}{\preprocess}\label{spec:preprocess}
{\normalsize \textbf{Input: }} & {\normalsize Bounded population computer \(\Prot=(Q,\delta,I,O,H)\).} \\[0.2cm]
{\normalsize \textbf{Output:}} &  
{\normalsize Equivalent bounded population computer \(\Prot'=(Q',\delta',I',O',H')\) of size \(\O(\size(\Prot))\) such that
\begin{enumerate}
\item states in $I'$ have no incoming transitions, 
\item all configurations in $\N^{I'}$ are terminal and 
\item $r(q)\le1$ for every $q\in I'$ and $(r\mapsto s)\in\delta'$.
\end{enumerate}
}
\end{specification}

\subsubsection{Conversion $\preprocessconv$}
\label{const0}

Given a population computer \(\Prot=(Q,\delta,I,O,H)\),  we define the computer $\Prot'=(Q',\delta',I',O',H')$ as follows:
\begin{itemize}
\item $Q':=Q\cup\{h\}\cup\{x_*:x\in I\}$,
\item $\delta':=\delta\cup\{t_{x}:=(x_*,h\mapsto x,h):x\in I\}$,
\item $I':=\{x_*:x\in I\}$,
\item $O'(S):=O(S\cap Q)$, for $S\subseteq Q'$, and
\item $H'(q):=H(q)$ for $q\in Q$ and $H(h):=1$.
\end{itemize}

\subsubsection{Correctness}

\begin{proposition}\label{thm:correctness_preprocess}
$\preprocessconv$ satisfies its specification (page \pageref{spec:preprocess}).
\end{proposition}
\begin{proof}
We proceed in several steps.

\medskip\noindent\textbf{Claim 1}. \(\Prot'\) refines \(\Prot\). \\
We  define the refinement \(\pi\) as follows: \(\pi(q)=q\) for every \(q\in Q\), \(\pi(h)=0\) and \(\pi(x_*)=x\) for every \(x\in I\).
We prove that $\pi$ satisfies the three properties of Definition\ref{def:refinement}.
For property 1, observe that the new transitions \(t_x=(r \mapsto s)\) we added fulfil \(\pi(r)=\pi(s)\), and for all old transitions the result is clear because of \(\pi(q)=q\) for all \(q\in Q\). Property 2 follows immediately from the definition.
For property 3, let \(C \in \N^{Q'}\) be a reachable terminal configuration. Since the helper cannot leave \(h\), we have \(C(h)>0\). If \(C(x_*)>0\) for any \(x\in I\), then \(t_x\) is enabled, contradiction to \(C\) being terminal. So \(C(x_*)=0\) for every \(x\in I\). Since \(\pi(h)=0\) and \(\pi(q)=q\) for all \(q\in Q\), we obtain \(\pi(C)=C|_Q\). Since \(C\) is terminal in \(\Prot'\), the smaller configuration \(C|_Q\) is also terminal in \(\Prot'\). Then \(C|_Q\) is terminal in \(\Prot\), since \(\delta \subseteq \delta'\). The outputs agree by definition of \(O'\).

\medskip\noindent\textbf{Claim 2}. $\Prot'$ is bounded. \\
Every occurrence of some \(t_x\) reduces the number of agents in \(I'\), therefore \(t_x\) occurs finitely often. Between any two of these occurrences, only finitely many other steps can occur, since \(\Prot\) is bounded.

\medskip\noindent\textbf{Claim 3}. $\Prot'$ decides $\varphi$. \\
By claims 1 and 2, and the Refinement lemma (Lemma \ref{lem:refinement}).

\medskip\noindent\textbf{Claim 4}.  $\Prot'$ has size \(\O(\size(P))\)  and satisfies properties 1.-3. of the specification. \\
By direct inspection of the  transition function.
\end{proof}

\subsection{Removing multiway transitions}\label{ssec:kwayconversion}


We transform a bounded population computer with $k$-way transitions $r \mapsto s$ such that $\Abs{\Support{r}}\leq 2$
into a binary bounded population computer. Let us first explain why the conversion introduced in \cite[Lemma 3]{BlondinEJ18}, which works for arbitrary transitions $r \mapsto s$, is too slow. In \cite{BlondinEJ18}, the 3-way  transition $t \colon q_1, q_2, q_3 \mapsto q_1', q_2', q_3'$ is simulated by the transitions 
\newcommand{\Space}{\hspace{5mm}}
$$
t_1 \colon q_1, q_2 \mapsto w, q_{12} \Space
t_2 \colon q_{12}, q_3 \mapsto  c_{12}, q_3' \Space
t_3 \colon c_{12}, w \mapsto q_1', q_2' \Space \overline{t}_1 \colon w, q_{12} \mapsto q_1, q_2
$$
\noindent Intuitively, the occurrence of $t_1$ indicates that two agents in $q_1$ and $q_2$ want to execute $t$, and are waiting for an agent in $q_3$. If the agent arrives, then all three execute $t_2 t_3$, which takes them to $q_1', q_2', q_3'$.  Otherwise, the two agents must be able to return to $q_1, q_2$ to possibly execute other transitions. This is achieved by the ``revert'' transition $\overline{t}_1$. The construction for a $k$-way transition has ``revert'' transitions $\overline{t}_1, \ldots,\overline{t}_{k-2}$. As in Example \ref{ex2} and Example \ref{ex3}, these transitions make the final protocol very slow.

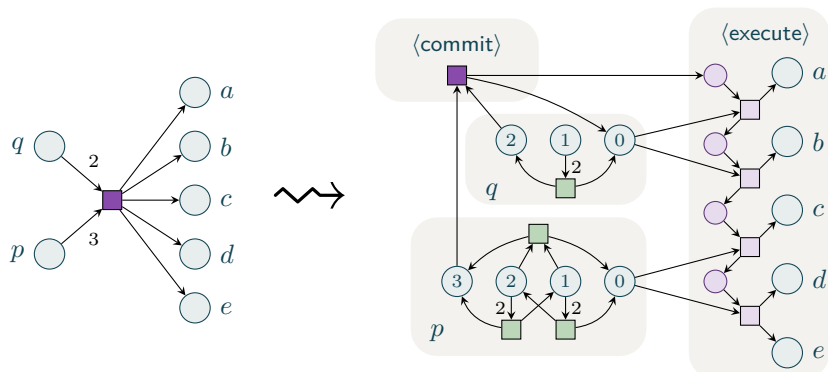
\begin{figure}[ht] 
	\centering
	\scalebox{1}{

	\begin{tikzpicture}[->, auto, node distance=0.30cm]
		\tikzset{every place/.append style={minimum size=0.4cm, niceblue,fill=nicebgblue}}
		\tikzset{every label/.append style={text=niceblue}}
		\tikzset{every transition/.style={minimum size=0.25cm}}
		\tikzset{every edge/.append style={font=\scriptsize,-stealth}}
		\tikzset{every token/.append style={scale=0.85}}

		\tikzstyle{blankhint}=[dash pattern=on 2pt off 1pt]
		
		\tikzstyle{other}=[fill=gray!50]
		\tikzstyle{commit}=[fill=nicepurple]
		\tikzstyle{commithelp}=[fill=nicepurple!20]
		\tikzstyle{commithelpstate}=[black!30!nicepurple, fill=nicepurple!20, minimum size=0.30cm]
		\tikzstyle{splittedstatelabel}=[niceblue]
		\tikzstyle{clustergrey}=[notquitegray!18]
		\tikzstyle{stack}=[fill=nicegreen!30]

		\newcommand*{\distancearrowleft}{0.4cm}
		\newcommand*{\distancearrowright}{1cm}
		\newcommand*{\distancearrowy}{0.1cm}
		\newcommand*{\distancecommitx}{1cm}
		\newcommand*{\distancecommity}{0.6cm}
		
		\node[place, label= right:$c$] (c) {};
		\node[place, label= right:$b$, above = of c] (b) {};
		\node[place, label= right:$a$, above= of b] (a) {};
		\node[place, label= right:$d$, below = of c] (d) {};
		\node[place, label= right:$e$, below= of d] (e) {};
		
		\node[transition, commit, left=0.75cm of c] (t) {};
		
		\node[place, label= left:$q$, left=1.5cm of b] (q) {};
		
		\node[place, label= left:$p$, left=1.5cm of d] (p) {};
		
		\path
		(q) edge[] node[above right=-0.5mm] {2} (t)
		(p) edge[] node[below right=-0.5mm] {3} (t)
		(t) edge[]  (a)
		(t) edge[]  (b)
		(t) edge[]  (c)
		(t) edge[]  (d)
		(t) edge[]  (e)

%
		;
		
		\node[scale=3] (arrow) [right=\distancearrowleft of c]  {$\rightsquigarrow$}; 
		
		\node (q3) [above right=\distancearrowy and \distancearrowright of arrow]  {}; 
		\node[place] (q2) [right= of q3]  {\scriptsize$2$}; 
		\node[place] (q1) [right= of q2]  {\scriptsize$1$}; 
		\node[place] (q0) [right= of q1]  {\scriptsize$0$}; 
		
		\node[splittedstatelabel] (q_label) [below left = 3mm and -1mm of q2] {$q$};
		
		\node[transition, below= of q1, stack] (q_up) {};
								
		\node[place, above right = 6mm and 10mm of q0, commithelpstate] (tmp1) {};
		\node[transition, below right = of tmp1, commithelp] (exe1) {};
		\node[place, below left = of exe1, commithelpstate] (tmp2) {};
		\node[transition, below right = of tmp2, commithelp] (exe2) {};
		\node[place, below left = of exe2, commithelpstate] (tmp3) {};
		\node[transition, below right = of tmp3, commithelp] (exe3) {};
		\node[place, below left = of exe3, commithelpstate] (tmp4) {};
		\node[transition, below right = of tmp4, commithelp] (exe4) {};
		
		\node[splittedstatelabel, above right = 0.6cm and -0.65cm of exe1] (exe_label) {\footnotesize \TraRef{execute:multiway}};		
		
		\node[place, label= right:$a$, above right= of exe1] (an) {};
		\node[place, label= right:$b$, above right= of exe2] (bn) {};
		\node[place, label= right:$c$, above right= of exe3] (cn) {};
		\node[place, label= right:$d$, above right= of exe4] (dn) {};
		\node[place, label= right:$e$, below right= of exe4] (en) {};

		\node[place] (p1) at (q1 |- tmp4) {\scriptsize$1$}; 
		\node[place] (p0) [right= of p1]  {\scriptsize$0$}; 
		\node[place] (p2) [left= of p1]  {\scriptsize$2$}; 
		\node[place] (p3) [left= of p2]  {\scriptsize$3$}; 
		
		\node[transition, commit] (commit) at (p3 |- tmp1) {};
		\node[splittedstatelabel, above = 0cm of commit] (commit_label) {\footnotesize \TraRef{commit}};		
		
		\path (p1) -- (p2) node[midway,transition, above=5mm, stack] (p_up1) {};
		
		\node[splittedstatelabel] (p_label) [below left = 3mm and -1mm of p3] {$p$};
		
		\node[transition, below= of p1, stack] (p_up2) {};
		\node[transition, below= of p2, stack] (p_up3) {};
		
%
		
		\path
		(q1) edge[] node[xshift=-0.05cm] {2} (q_up)
		(q_up) edge[bend left] (q2)
		(q_up) edge[bend right] (q0)
			
		(p1) edge[] (p_up1)
		(p2) edge[] (p_up1)
		(p_up1) edge[bend right=15] (p3)
		(p_up1) edge[bend left=15] (p0)
		
		(p1) edge[] node[xshift=-0.05cm] {2} (p_up2)
		(p_up2) edge[] (p2)
		(p_up2) edge[bend right] (p0)
		
		(p2) edge[] node[swap, xshift=0.05cm] {2} (p_up3)
		(p_up3) edge[bend left] (p3)
		(p_up3) edge[] (p1)
		
		(p3) edge[] (commit)
		(q2) edge[] (commit)
		(commit) edge[bend left=10] (q0)
		(commit) edge[] (tmp1)
		
		(tmp1) edge[] (exe1)
		(q0) edge[] (exe1)
		(exe1) edge[] (an)
		(exe1) edge[] (tmp2)
		
		(tmp2) edge[] (exe2)
		(q0) edge[] (exe2)
		(exe2) edge[] (bn)
		(exe2) edge[] (tmp3)
		
		(tmp3) edge[] (exe3)
		(p0) edge[] (exe3)
		(exe3) edge[] (cn)
		(exe3) edge[] (tmp4)
		
		(tmp4) edge[] (exe4)
		(p0) edge[] (exe4)
		(exe4) edge[] (dn)
		(exe4) edge[] (en)
		
%
		;

		\begin{pgfonlayer}{background}
			\filldraw [line width=7mm,join=round,clustergrey]
			(q2  -| q_label)  rectangle (q_up.north  -| q0)
			;
			\filldraw [line width=7mm,join=round,clustergrey]
			(p_up1.south  -| p_label)  rectangle (p_up2  -| p0)
			;
			\filldraw [line width=7mm,join=round,clustergrey]
			(commit_label  -| commit_label.west)  rectangle (commit  -| commit_label.east)
			;
			\filldraw [line width=7mm,join=round,clustergrey]
			(exe_label  -| tmp1)  rectangle (en  -| en.east)
			;
		\end{pgfonlayer}
	\end{tikzpicture}
	}
	\caption{Simulating the 5-way transition $\multiset{3 \cdot p, 2 \cdot q \mapsto a,b,c,d,e}$ by binary transitions.}
	\label{fig:k-way}
\end{figure}

We present a gadget without ``revert'' transitions that works for $k$-way transitions $r \mapsto s$ satisfying $\Abs{\Support{r}}\leq 2$. Figure \ref{fig:k-way} illustrates it, using Petri net notation, for the 5-way transition $t \colon \multiset{3 p, 2 q} \mapsto \multiset{a,b,c,d,e}$.
In the gadget, states $p$ and $q$ are split into $(p, 0),  \ldots, (p,3)$ and  $(q, 0), \ldots, (q, 2)$. 
Intuitively, an agent in $(q, i)$ acts as representative for a group of $i$ agents in state $q$. Agents in $(p, 3)$ and $(q, 2)$ commit to executing $t$ by executing the binary transition \TraRef{commit}. After committing, they move to the states $a, \ldots, e$ together with the other members of the group, who are ``waiting'' in the states $(p,0)$ and $(q, 0)$. Note that \TraRef{commit} is binary because of the restriction $\Abs{\Support{r}} \leq 2$ for multiway transitions.

To ensure correctness of the conversion, agents can commit to transitions if they represent more than the required amount. In this case, the initiating agents would commit to a transition and then elect representatives for the superfluous agents, before executing the transition. This requires additional intermediate states. 

The rest of this section is split into three parts. We first 
describe the formal specification of the conversion. Section \ref{app:multiway_construction} describes the conversion itself, which we call  \textit{Binarise}, formally.  Section~\ref{app:multiway_correctness} shows that \textit{Binarise} satisfies the specification. 

\begin{specification}{\binarise}\label{spec:binarise}
{\normalsize \textbf{Input: }} & {\normalsize Bounded population computer \(\Prot=(Q,\delta,I,O,H)\).} \\[0.2cm]
{\normalsize \textbf{Output:}} & {\normalsize Equivalent bounded binary population computer \(\Prot'=(Q',\delta',I',O',H')\) with adjusted size \(\O(\beta \cdot \size(\Prot))\), where \(\beta \leq \Abs{Q}\). Additionally:
\begin{enumerate}
\item If every state of $\Prot$ but one has at most $2$ outgoing transitions, then $\beta\le3$.
\item If no state in $I$ has incoming transitions, then neither do states in $I'$.
\item If all configurations in $\N^I$ are terminal and $r(q)\le1$ for $q\in I$ and $(r\mapsto s)\in\delta$, then all configurations in $\N^{I'}$ are terminal.
\end{enumerate}}
\end{specification}

\subsubsection{Conversion $\binarise$}\label{app:multiway_construction}

\newcommand{\Qstack}{Q_\mathrm{stack}}
\newcommand{\Qexec}{Q_\mathrm{exec}}
\newcommand{\Qreduce}{Q_\mathrm{reduce}}

Given a bounded population computer \(\Prot=(Q,\delta,I,O,H)\), we construct a binary population computer \(\Prot'=(Q',\delta',I',O',H')\). Let $m(q):=\max\{r(q):(r\mapsto s)\in \delta\}$ denote the maximum multiplicity of any outgoing transition of $q$. For each state $q$ we allow up to $m(q)$ agents to “stack” in $q$. 

Formally, we add states $\{(q,i):q\in Q,i=0,...,m(q)\}$ to $Q'$, and the following transitions, for $q\in Q$, $i,j\in\{1,...,m(q)-1\}$,  to $\delta'$:
\begin{gather*}
	\begin{alignedat}{2}
		(q,i),(q,j) &\mapsto (q, i+j),(q,0) &\qquad&\text{ if }i+j\le m(q) \\
		(q,i),(q,j) &\mapsto (q, m(q)),(q, i+j-m(q)) &&\text{ if }i+j\ge m(q).
	\end{alignedat}\TraName{stack}
\end{gather*}

\noindent Intuitively, an agent in state $(q,i)$ “owns” $i$ agents in state $q$, meaning that it certifies that $i-1$ additional agents are in $(q,0)$. Consider a transition $t=(r\mapsto s)\in \delta $ of $\Prot$ with $\Support{r}=\{q,p\}$. Executing $t$ in $\Prot$ requires $\Abs{r}$ agents. In $\Prot'$, the transition is simulated by a sequence of binary transitions. The simulation is started by any pair of agents that together own at least $r$ agents.
Assume these agents are in states $(q,i)$ and $(p, j)$ with $i+j \geq \Abs{r}$. The transition \TraRef{commit} initiating the simulation designates one of the agents, say $q$, as primary agent, and $p$ as secondary agent. The primary agent is responsible for executing the rest of the simulation. Transition \TraRef{commit}  moves the primary agent from $(q,i)$ to  $(q,i-r(q),t)$ and the secondary agent from $(p, j)$ to $(p,j-r(p))$; intuitively, the agents together ``designate'' $\Abs{r}$ agents to execute $t$.

%
Formally, we add states $\{(q,i,t):i=0,...,m(q)\}$ to $Q'$. For every $q, p$ in $Q$, if $q \neq p$ we add transitions
\begin{gather*}
	\begin{alignedat}{2}
		(q,i),(p,j) &\mapsto (q,i-r(q),t),(p,j-r(p)) &\qquad&\text{ for }i\ge r(q),j\ge r(p)
	\end{alignedat}\TraName{commit}
\end{gather*}
to $\delta$. If $p=q$ then for every $i,j$ with $i+j\ge r(q)$ we add transitions
\begin{gather*}
	\begin{alignedat}{2}
		(q,i),(q,j) &\mapsto (q,i+j-r(q),t),(q,0) &\qquad&\text{ if }i+j-r(q)\le m(q)\\
		(q,i),(q,j) &\mapsto (q,i+j-r(q)-m(q),t),(q,m(q)) &\qquad&\text{ else }
	\end{alignedat}\TraName[commit2]{commit}
\end{gather*}

After the execution of \TraRef{commit}, the primary agent transfers ownership of its remaining agents (if any) to another agent by means of a transition  \TraRef{transfer}. Formally, we add to $\delta'$ transitions 
\begin{gather*}
	\begin{alignedat}{2}
		(q,i,t),(q,0) &\mapsto (q,0,t),(q,i) &\qquad&\text{ for }i=1,...,m(q)
	\end{alignedat}\TraName{transfer}
\end{gather*}

\newcommand{\kwayL}{l}
\noindent The primary agent now proceeds with the simulation of the execution of $t$. Formally, we add states $\{(t,i):i=1,...,\Abs{r}\}$ to $Q'$. Let $s_1,...,s_\kwayL$ denote an enumeration of the multiset $s$ of $t$, with $\kwayL:=\Abs{s}$. Intuitively, an agent in $(t,i)$ moves one agent into $(s_i,1)$, and then goes to $(t,i+1)$. Instead of moving an agent into $(t,1)$ via a transition, we identify $(q,0,t)$ with $(t,1)$ directly. Additionally, we identify $(t,\kwayL)$ with $(s_\kwayL,1)$, so that we do not have to create a special transition for the last agent. Accordingly, we formally define the last set of transitions added to $\delta'$ as follows, for $i=1,...,\kwayL-1$.
\begin{gather*}
	\begin{alignedat}{2}
		(t,i),(p,0) &\mapsto (t,i+1),(s_i,1) &\qquad&\text{ if }i\le r(p)\\
		(t,i),(q,0) &\mapsto (t,i+1),(s_i,1) &\qquad&\text{ if }i>r(p)
	\end{alignedat}\TraName[execute:multiway]{execute}
\end{gather*}

\noindent (Observe that, as specified above, $\Prot'$ is not deterministic. For some of the transitions \TraRef{stack} and \TraRef{commit} it may be the case that e.g.\ \(\TraRef{stack}=(r \mapsto s_1)\) and \(\TraRef{commit}=(r \mapsto s_2)\) for the same \(r\) and $s_1\ne s_2$. However, if that happens we delete all but one of these transitions to ensure that the protocol is deterministic. When choosing which transition to keep, we prefer \TraRef{commit} to \TraRef{stack}, but otherwise pick an arbitrary one.)

We retain the original input states and helpers, by identifying each $q\in Q$ with $(q,1)$. For the output function we define $O'(S):=O(\{q:(q,i)\in S\})$ for any $S$.
Note, however,  that a circuit for $O$ grows by at most a factor of $3$, as $(q,i)\in\Support{C}\Rightarrow(q,0)\in\Support{C}$ for $i\ge2$ and any reachable configuration $C$ and state $q$, so it suffices to check for $(q,0)$ and $(q,1)$.

\subsubsection{Correctness}\label{app:multiway_correctness}

\begin{proposition}\label{thm:kwaycorrect}
$\binarise$ satisfies its specification (page~\pageref{spec:binarise}).
\end{proposition}
\begin{proof}
We first show that $\Prot'$ refines $\Prot$. 

\medskip\noindent\textbf{Claim 1}. $\Prot'$ refines $\Prot$. \\
To begin, let us introduce the mapping between configurations of $\Prot'$ and $\Prot$ describing the refinement. We define $\pi:Q'\rightarrow\N^Q$ by setting 
\begin{align*}
\pi((q,i))&:=q\cdot i\\
\pi((q,i,t))&:=q\cdot i+s&&\text{for all $q\in Q$, $t=(r\mapsto s)\in \delta $ and $i$}\\
\pi((t,i))&:=s_i+...+s_\kwayL
\end{align*}
This uses the same enumeration of $s$ as above. Clearly, $\pi$ is well-defined, as $\pi((q,0,t))=s=\pi((t,1))$ and $\pi((t,\kwayL))=s_\kwayL=\pi((s_\kwayL,1))$. Finally, we extend $\pi$ to a linear mapping $\pi:\N^{Q'}\rightarrow\N^Q$ in the obvious fashion.
Now we prove that $\pi$ fulfils the properties required by Definition~\ref{def:refinement}:
\begin{enumerate}
\item Note that $\pi$ is invariant under execution of \TraRef{stack}, \TraRef{transfer} and \TraRef{execute:multiway}. Additionally, for a $t\in \delta $ and a corresponding \TraRef{commit} transition $t'$, we find that $C\rightarrow_{t'} C'$ implies $\pi(C)\rightarrow_t\pi(C')$ for all $C,C'\in\N^{Q'}$. 
\item We identified $q\in Q$ with $(q,1)$ and set $\pi((q,1)):=q$, so $I=I'$ and $\pi(C)=C$ for all $C\in\N^Q$ follows.
\item Let  $C\in\N^{Q'}$ be a reachable terminal configuration. We have to show that $\pi(C)$ is terminal and satisfies $O(\Support{\pi(C)})=O'(\Support{C})$. The latter condition follows immediately from the definition of $O$. To show that $\pi(C)$ is terminal, We first observe that transition \TraRef{transfer} is always enabled if an agent is in state $(q,i,t)$, with $t=(r,s)\in \delta $ and $i\ge1$, as that state “owns” $i+r(q)$ agents in $q$. Hence there must be $i+r(q)-1\ge i\ge1$ agents in $(q,0)$. By the same line of argument, \TraRef{execute:multiway} is always enabled if an agent is in $(t,i)$, for $t\in \delta $ and $1\le i<|s|$. Now, assume $\pi(C)$ is not terminal, so there is some transition $t=(r\mapsto s)\in \delta $ which is enabled at $\pi(C)$. As we have just argued, all agents of $C$ are in states $(q,i)$, for $q\in Q$ and $i\in\{0,...,m(q)\}$. Transition \TraRef{commit} is not enabled at $C$, wherefore one of the states $q\in Q$ used by \(t\) fulfils $i<r(q)$ for all $i$ with $C((q,i))>0$. But $t$ is enabled at $\pi(C)$, so $\pi(C)(q)\ge r(q)$ and, by definition of $\pi$, there are $i,j>0$ s.t.\ $C$ contains both an agent in $(q,i)$ and one in $(q,j)$. Due to our choice of $m$, we have $r(q)\le m(q)$, and therefore $0<i,j<m(q)$, wherefore transition \TraRef{stack} is enabled, contradicting that $C$ is terminal. So $\pi(C)$ is terminal. 
\end{enumerate}

\noindent\textbf{Claim 2}. $\Prot'$ is bounded. \\
Assume an infinite run $C_0,C_1,...$ of $\Prot'$ exists. If transition \TraRef{commit} is executed infinitely often in that run, at steps $i_0,i_1,...\in\N$, then $\pi(C_{i_0}),\pi(C_{i_1}),...$ would be an infinite run of $\Prot$, contradicting that $\Prot$ is bounded. Hence there is an infinite suffix of $C_0,C_1,...$ in which \TraRef{commit} is never fired.

In this suffix the number of agents in a state $(q,i,t)\in Q'$ with $i>0$ cannot increase, but decreases whenever \TraRef{transfer} is executed. Hence this also happens only finitely often and the number of agents in a state $(t,i)$ cannot increase beyond a point. As \TraRef{execute:multiway} increases $i$, it too must occur only finitely often. The only transition left is \TraRef{stack}, which always increases the number of agents in either $(q,0)$ or $(q,m(q))$, for some $q\in Q$.

\medskip\noindent\textbf{Claim 3}. $\Prot'$ decides $\varphi$. \\
By claims 1 and 2, and the Refinement lemma (Lemma \ref{lem:refinement}).

\medskip\noindent\textbf{Claim 4}. $\Prot'$ has adjusted size $\O(\beta \cdot \size(\Prot))$ for some $\beta \leq \Abs{Q}$. \\
Let $\beta_q$ denote the number of transitions $t\in \delta$ for which $q\in Q$ is the primary agent, and set $\beta:=\max\{\beta_q:q\in Q\}$. It suffices to show \(\Abs{Q'} \leq (\beta+2) \size(\Prot)\). We begin by bounding the total value of $m(q)$. Clearly, $m(q)\le\sum_{(r\mapsto s)\in\delta}r(q)$ for $q\in Q$, and thus $\sum_{q\in Q}m(q)\le \sum_{t \in \delta} \Abs{t}$. For every $q\in Q$ we create $m(q)+1$ states and for every $t\in\delta$ we create $\Abs{t}$ states. Additionally, we create states for every transition $t$ using state $q$ as primary agent, so at most $\beta m(q)$. In total, we create at most $(\beta+2)\sum_{t \in \delta} \Abs{t}+\Abs{Q}$ states.

\medskip\noindent\textbf{Claim 5}.  $\Prot'$ satisfies conditions 1.-3. of the specification. \\
For condition 1., let $q\in Q$ denote the state with the most outgoing transitions. For our conversion, we can simply not choose $q$ as primary agent, for all transitions that also use a different agent. There is at most one other transition (using only agents in $q$), so every state is chosen as primary agent of at most $3$ transitions and $\beta\le3$. Condition 2. is obvious from the conversion. For condition 3., note that the condition implies $m(q)\le1$ for all $q\in I$. Therefore the only transition using an agent in $I'$ is \TraRef{commit}, which is not enabled at $C$ if $\pi(C)$ is terminal.
\end{proof}

\subsection{Converting output functions to marked-consensus output functions}\label{ssec:outputfunction1}

\newcommand{\Qterm}{Q_\mathrm{term}}
\newcommand{\Up}{\triangle}
\newcommand{\Down}{\triangledown}
\newcommand{\Qhelper}{Q_\mathrm{helper}}

\begingroup\def\TraNs{tra:output}

We convert a computer with an arbitrary output function into another one with a \emph{marked-consensus}  output function. An output function is a  \emph{marked-consensus} output function if there are disjoint sets of states $Q_0,Q_1\subseteq Q$ such that $O(S):=b$ if $S\cap Q_b\ne\emptyset$ and $S\cap Q_{1-b}=\emptyset$, for $b\in\{0,1\}$, and $O(S):=\bot$ otherwise. Intuitively, for every $S \subseteq Q$ we have $O(S)=1$ if all agents agree to avoid $Q_0$ (consensus), and at least one agent populates $Q_1$ (marked consensus). 

Our starting point is some bounded and binary computer $\Prot=(Q,\delta,I,O,H)$, e.g.\ as constructed in Section \ref{ssec:kwayconversion}. Let $(G,E)$ be a boolean circuit
with only NAND-gates computing the output function $O$. We simulate $\Prot$ by a computer $\Prot'$ with a marked consensus output and $\O(\Abs{Q}+\Abs{G})$ states.
This result allows us to bound the number of states of $\Prot'$ by applying well-known results on the complexity of boolean functions.

Intuitively, $\Prot'$ consists of two processes running asynchronously in parallel. The first one is (essentially, see below) the computer $\Prot$ itself. The second one is a gadget that simulates the execution of $G$ on the support of the current configuration of $\Prot$. Whenever $\Prot$ executes a transition, it raises a flag indicating that the gadget must be reset (for this, we duplicate each state $q \in Q$ into two states $(q, +)$ and $(q, -)$, indicating whether the flag is raised or lowered). Crucially, $\Prot$ is bounded, and so it eventually performs a transition  \emph{for the last time}. This resets the gadget for the last time, after which the gadget simulates $(G,E)$ on the support of the terminal configuration reached by $\Prot$.

\begin{specification}{\focalise}\label{spec:focalise}
{\normalsize \textbf{Input: }} & {\normalsize Bounded binary population computer \(\Prot=(Q,\delta,I,O,H)\).} \\[0.2cm]
{\normalsize \textbf{Output:}} & {\normalsize Equivalent bounded binary computer \(\Prot'=(Q',\delta',I',O',H')\) with adjusted size \(\O(\size_2(\Prot))\) using a marked consensus output. Additionally:
\begin{enumerate}
\item If no state in $I$ has incoming transitions, then neither do states in $I'$.
\item If all configurations in $\N^I$ are terminal, then so are all configurations in $\N^{I'}$.
\end{enumerate}}
\end{specification}

\subsubsection{Conversion \(\focalise\)}

Formally, the intuition above corresponds to a partition of the state space into four parts, \(Q=\Qorig \cup \Qsupp \cup \Qgate \cup \Qreset\), with the crucial property that no transition allows agents to change partition. The only exception are leader elections producing/removing reset agents. In the following we first describe these different parts of the partition, followed by the transitions. The set $\Qorig:=Q\times\{-,+\}$ was already described above.

\begin{figure}[!tb]
	\input{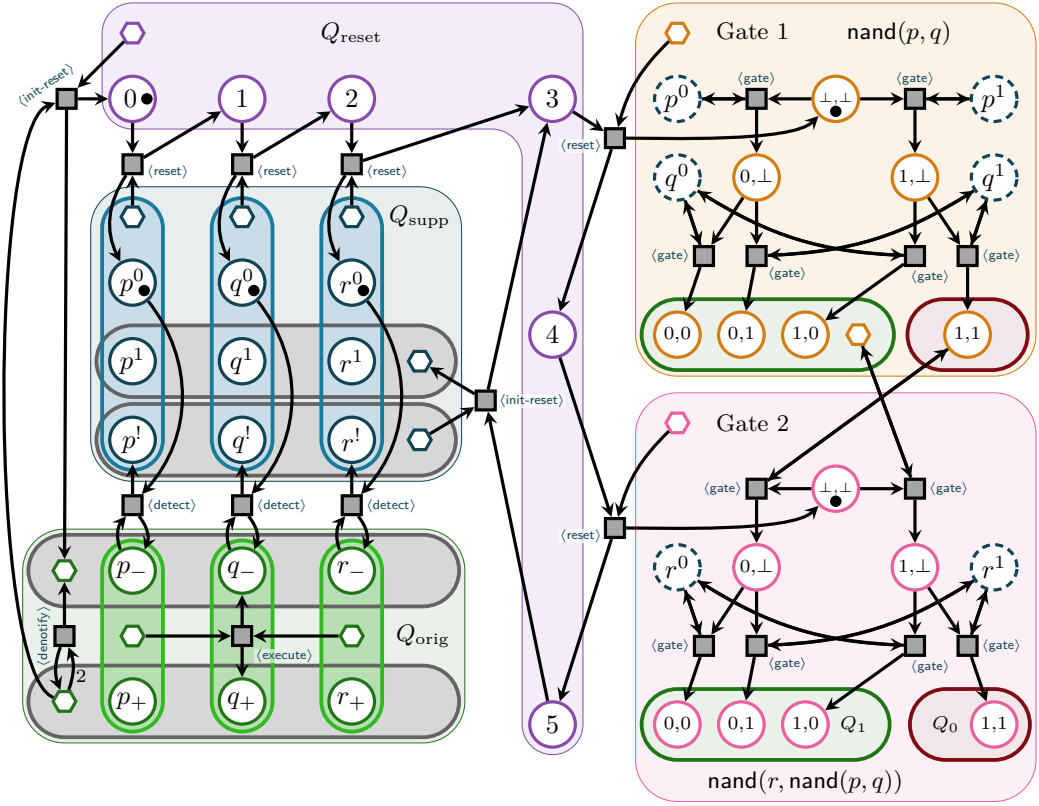}
	\caption[Q.Q]{Figure visualising the conversion to marked consensus output for a population computer with states $\{p,q,r\}$, a single transition $p,q \mapsto r,r$ and output function $\lnot(r \land \lnot (p \land q))$. 

    State names are abbreviated: $x^y$ and $x_y$ are used instead of $(x,y)$, and only the last two components of states in $\Qgate$ are shown. Transitions are drawn using Petri net notation. The same state may appear multiple times. Occurrences beyond the first are drawn with a dashed border.

    Hexagons indicate wildcards. Each hexagon is associated with a group of states (indicated by shaded areas). Used as input to a transition, it denotes that any agent of that group can be used (i.e.\ they create multiple copies of the transition). Transition may use hexagons as both input and output, in which case the output hexagon refers to the state corresponding to the state used for the input hexagon. For example, transition \TraRef{denotify} takes two agents $x_+,y_+$, with $x,y\in\{p,q,r\}$ and produces an agent in $x_-$ and one in $y_+$.

    We omit \TraRef{leader} transitions and the third case of \TraRef{init-reset} from the drawing. Please note, however, that the resulting computer is still correct under the assumption that no superfluous helpers are provided. (This is equivalent to using leaders instead of helpers.)
	}
	\label{fig:output_removal}
\end{figure}

The output gadget is designed to be operated by one \emph{state-helper} for each $q \in Q$, with set of states $\Qsupp(q)$, and a \emph{gate-helper} for each gate $g \in G$, with set of states 
$\Qgate(g)$, defined as follows:
\begin{itemize}
\item $\Qsupp(q):= \{q\} \times \{0,1,!\}$. These states indicate that $q$ belongs/does not belong to the support of the current configuration (states $(q, 0)$ and $(q, 1)$), or that the output has changed from $0$ to $1$ (state $(q,!)$). 
\item $\Qgate(g):= \{g\} \times \{0,1,\bot\}^3$ for each gate $g\in G$, storing the current values of the two inputs of the gate and its output. Uninitialised values are stored as $\bot$.  
\end{itemize}
The sets \(\Qsupp\) and \(\Qgate\) are now the disjoint union of the above over all states/gates, in total we therefore obtain $\Qsupp:= Q \times \{0,1,!\}$ and $\Qgate:= G \times \{0,1,\bot\}^3$. 

Recall that a population computer must also remain correct for a larger number of helpers. This is ensured by letting all helpers populating one of these sets, say $\Qsupp(q)$, perform a leader election; whenever two helpers in states of $\Qsupp(q)$ meet, one of them becomes a non-leader, and a flag requesting a complete reset of the gadget is raised. All resets are carried out by a \emph{reset-helper} with set of states $\Qreset:=\{0,...,\Abs{Q}+\Abs{G}\}$, initially in state $0$. Whenever a reset is triggered, the reset-helper contacts all other $\Abs{Q}+\Abs{G}$ helpers in round-robin fashion, asking them to reset the computation.

We now specify the required transitions. First, we need to refine the original protocol, requesting to recompute the support with each transition. Different occurrences of $\pm$ need not match.
\begin{alignat*}{2}
(q,\pm),(p,\pm)&\mapsto(q',+),(p',-)&\qquad&\text{ for }(q,p\mapsto q',p')\in T\TraName{execute}\\
\intertext{It suffices to reset once, so for the purpose of speed we clear superfluous flags.}
(q,+),(p,+)&\mapsto(q,+),(p,-)&\qquad&\text{ for }q,p\in Q\TraName{denotify}\\
\intertext{To keep the protocol deterministic, we remove all \TraRef{denotify} transitions which could also initiate another transition (in particular \TraRef{execute}). The support is computed by setting the stored bit to `$!$' once the corresponding state has been observed.}
(q,0),(q,-)&\mapsto(q,!),(q,-)&\qquad&\text{ for }q\in Q\TraName{detect}
\end{alignat*}
To define the transitions for $\Qgate$, we need to introduce some notation. First, we write $\Nand$ for the NAND function, i.e.\ $\Nand(i,j):=\neg(i\wedge j)$ for $i,j\in\{0,1\}$ and $\Nand(i,j):=\bot$ otherwise. We also use $E_1,E_2$ to denote the first and second component of $E$, and write $S_g^b$ for the set of states indicating that gate $g\in Q\cup G$ has truth value $b\in\{0,1\}$, i.e.\ $S_q^b:=\{(q,b)\}$ for $q\in Q$ and $S_g^b:=\{g,b\}\times\{0,1\}^2$ for $g\in G$. We add the following transitions, for any gate $g\in G$ and $b,i\in\{0,1\}$.
\begin{gather*}
\begin{alignedat}{2}
(g,\bot,\bot,\bot),q&\mapsto(g,\bot,b,\bot),q&\qquad&\text{ for }q\in S_{E_1(g)}^b\\
(g,\bot,i,\bot),q&\mapsto(g,\Nand(i,b),i,b),q&\qquad&\text{ for }q\in S_{E_2(g)}^b
\end{alignedat}\TraName{gate}
\end{gather*}
These transitions perform the computation of the gate, by initialising the first and then the second input. Once the second input is initialised, the output of the gate is set accordingly.

There are two kinds of resets; depending on whether the agents in $\Qsupp$ are affected. (The gates are always reset.) Both resets are executed by an agent in $\Qreset$, who goes through the other agents one by one. Let $q_1,...,q_{\Abs{Q}}$ denote an enumeration of $Q$. For \(G\) any enumeration is not enough, input gates have to be reset first. Therefore let \(g_1, \dots, g_{\Abs{G}}\) be a topological sorting of \(G\). 
\begin{gather*}
\begin{alignedat}{4}
i-1,\;&(q_i,b)&&\mapsto \;&i,\;&(q_i,0)&&\qquad\text{ for }(q_i,b)\in\Qsupp\\
\Abs{Q}+i-1,\;&(g_i,b_1,b_2,b_3)&&\mapsto \;&\Abs{Q}+i,\;&(g_i,\bot,\bot,\bot)&&\qquad\text{ for }(g_i,b_1,b_2,b_3)\in\Qgate\\
\end{alignedat}\TraName{reset}
\end{gather*}
There are three ways to initiate a reset:
\begin{gather*}
\begin{alignedat}{3}
(q,+),i&\mapsto(q,-),0&&\qquad\text{ for }(q,+)\in\Qorig,i\in\Qreset\\
(q,!),i&\mapsto(q,1),\min\{i,\Abs{Q}\}&&\qquad\text{ for }(q,!)\in\Qsupp,i\in\Qreset\\
i,j&\mapsto0,(q_h,-)&&\qquad\text{ for }i,j\in\Qreset
\end{alignedat}\TraName{init-reset}
\end{gather*}
First, an agent in $\Qorig$ may indicate that the support has changed, and everything will be reset. Second, an agent in $\Qsupp$ will request that all gates be reset whenever it changes its output. Third, if two agents are in $\Qreset$, the computation so far must be discarded, and we continue with only one of them. The other moves into an arbitrary state $q_h\in Q$ with $H(q_h)>0$, so it is given back to $\Prot$ to use for its computations. Picking a state with $H(q_h)>0$, i.e.\ a helper state, ensures that this does not affect the correctness of $\Prot$.

Finally, all states in $\Qsupp$ and $\Qgate$ also participate in a leader election, to ensure that there is only one agent in $\Qsupp$ for each $q\in Q$, and only one agent for each gate.
\begin{gather*}
\begin{alignedat}{3}
q,q'&\mapsto q,0&&\qquad\text{ for }q,q'\in\Qsupp\text{ s.t.\ }q_1=q'_1\\
g,g'&\mapsto g,0&&\qquad\text{ for }g,g'\in\Qgate\text{ s.t.\ }g_1=g'_1
\end{alignedat}\TraName{leader}
\end{gather*}
These transitions indirectly cause a reset, by producing an agent in $\Qreset$. 

It remains to define the inputs, helpers and outputs. For this, we identify a state $q\in Q$ with $(q,-)\in\Qorig$. We define \(I':=I\), as well as $H'(q):=H(q)$ for $q\in Q$ and $H(q):=1$ for $q\in Q\times\{0\}\cup G\times\{(\bot,\bot,\bot)\}\cup\{0\}$. To define the marked consensus output, we pick the special states $Q_0=\{(g_{\Abs{G}},1)\}\times\{0,1\}^2$ and $Q_1:=\{(g_{\Abs{G}},0)\}\times\{0,1\}^2$.



\subsubsection{Correctness} \label{app:correctness}

\begin{proposition}\label{thm:correctness_focalise}
$\focalise$ satisfies its specification.
\end{proposition}
\begin{proof}
We first show that $\Prot'$ refines $\Prot$, which by Lemma~\ref{lem:refinement} implies that they decide the same predicate. We introduce a mapping $\pi:\N^{Q'}\rightarrow\N^Q$ to describe the configuration that $\Prot'$ is representing. For all $C$ we define
\[\pi(C):=\sum_{q\in Q}q\cdot C((q,\pm))+q_h\cdot (C(\Qsupp\cup\Qgate\cup\Qreset)-\Abs{Q}-\Abs{G}-1)\]
Recall that $q_h\in Q$ is the state to which superfluous agents are moved, as defined in \TraRef{init-reset}. Eventually, we have exactly one agent for each state in $Q$, to detect the support, exactly one agent for each gate, and exactly one reset agent, so $\Abs{Q}+\Abs{G}+1$ in total. Everything beyond that is superfluous and will be returned to $q_h$ at some point. We prove the following claim:

\medskip\noindent\textbf{Claim 1}. Let $C$ denote a reachable configuration of $\Prot'$. Then $C(\Qreset)\ge1$, $C(\{q\}\times\{0,1,!\})\ge1$ for $q\in Q$, and $C(\{g\}\times\{0,1,\bot\}^3)\ge1$ for $g\in G$. If $C$ is terminal, the above hold with equality. \\
Let $S_q:=\{q\}\times\{0,1,!\}$ for $q\in Q$ and $G_g:=\{g\}\times\{0,1,\bot\}^3$ for $g\in G$. First, note that the sets of states $\Qreset$, $S_q$ and $G_g$ each contain at least one agent in an initial configuration (due to the choice of $H'$). Additionally, they cannot be emptied, as every transition removing agents from one of these sets also puts at least one agent back. (Using Petri net terminology, they are traps.)

If two agents are in $\Qreset$, the third part of \TraRef{init-reset} is active and $C$ is not terminal. Similarly, if two agents are in $S_q$, for some $q$, or two agents are in $G_g$, for some $g$, then \TraRef{leader} can be executed.

\medskip\noindent\textbf{Claim 2}. $\Prot'$ refines $\Prot$. \\
We show that $\pi$ fulfils the properties required by Definition~\ref{def:refinement}. The first two are simple.
\begin{enumerate}
\item Observe that $\pi(C)$ is changed only via transition \TraRef{execute}, and that this happens according to a transition $t\in T$.
\item $I=I'$ holds by construction and the remainder follows from $H(q_h)>0$ and the definition of $\pi$.
\end{enumerate}
Property 3 will take up the remainder of this proof. Let $C$ denote a reachable, terminal configuration of $\Prot'$. Using Claim 1 we get $\pi(C)=\sum_{q\in Q}q\cdot C((q,\pm))$. Therefore, if a transition $t\in\delta$ is enabled at $\pi(C)$, the corresponding \TraRef{execute} transition is enabled at $C$. As $C$ is terminal, so is $\pi(C)$.

Finally we have to argue $O'(\Support{C})=O(\Support{\pi(C)})$. Due to Claim 1 we know that in $C$ we have exactly one agent in either $(q,0)$, $(q,1)$, or $(q,!)$. It cannot be in $(q,!)$, as then transition \TraRef{init-reset} would be enabled.

If it were in $(q,0)$ but $\pi(C)(q)>0$, then transition \TraRef{detect} would be enabled, so that cannot be the case either. Conversely, if it were in $(q,1)$ but $\pi(C)(q)=0$, then we also arrive at a contradiction: after the last agent left $(q,\pm)$ via \TraRef{execute}, it must have triggered a reset, which moved the agent to $(q,0)$ (or, if there were multiple agents in $q\times\{0,1,!\}$, \TraRef{leader} would have triggered another reset later). But after that reset $\pi(C)(q)=0$, so it is impossible to leave $(q,0)$.

Therefore we find that the agents in $\Qsupp$ precisely indicate the support of $\pi(C)$. Whenever an agent in $\Qsupp$ changes its opinion (either due to a \TraRef{reset} or \TraRef{detect}), all gates will be reset. So there is some point at which the opinions of agents in $\Qsupp$ have stabilised (in particular, all inequalities of Claim 1 are tight, else there would be another reset) and the unique agent in $\Qreset$ is in state $\Abs{Q}$, i.e.\ it is in the process of resetting all gates. As the gates are reset in order of some topological sorting (so a gate is reset after its inputs are), a gate will only assume a value after its inputs have stabilised and therefore compute the correct value according to the circuit. As the circuit computes $O(\Support{\pi(C)})$, the statement follows.

\medskip\noindent\textbf{Claim 3}. $\Prot'$ is bounded. \\
Due to Claim 2 we know that $\pi(C)$ can change only finitely often, as $\Prot$ is bounded, and thus transition \TraRef{execute} can be executed only finitely often. After that, the number of agents in a state $Q\times\{+\}\subseteq\Qorig$ cannot increase, but can always decrease using the first \TraRef{init-reset} transition. So eventually no agents remain in those states.

Parallel to that, both $C(\Qsupp)$ and $C(\Qgate)$ cannot increase. Whenever $C(\Qsupp)>\Abs{Q}$ or $C(\Qgate)>\Abs{G}$, \TraRef{leader} is enabled and decreases one of them, until $C(\Qsupp)=\Abs{Q}$ and $C(\Qgate)=\Abs{G}$. Afterwards, \TraRef{leader} cannot fire again (note Claim 1), and $C(\Qreset)$ cannot increase. If $C(\Qreset)>1$, the third case of \TraRef{init-reset} will reduce this number, until $C(\Qreset)=1$.

To summarise, eventually no agents remain in $Q\times\{+\}$ and all inequalities of Claim 1 become tight. At that point, the first and third case of \TraRef{init-reset} are disabled, and it is not possible for the agent in $\Qreset$ to lower its value to below $\Abs{Q}$. Via \TraRef{reset}, it will thus eventually arrive at $\Abs{Q}$, and the first part of \TraRef{reset} cannot be executed again.

Once that happens, agents cannot enter states $Q\times\{0\}\subseteq\Qsupp$, such that \TraRef{detect} can never occur any more. This then causes \TraRef{init-reset} to eventually be fully disabled, and then \TraRef{reset} as well. Finally, transition \TraRef{gate} can then fire only finitely often, and the protocol terminates.

\medskip\noindent\textbf{Claim 4}. $\Prot'$ decides $\varphi$. \\
Follows from claims 2 and 3, and the Refinement lemma (Lemma \ref{lem:refinement}).

\medskip\noindent\textbf{Claim 5}. If no state in $I$ has incoming transitions, then neither do states in $I'$. \\
Note that $I'=I\times\{-\}$ by definition. If no state in $I$ has incoming transitions, then it is not possible to put an agent into $I\times\{+\}$, as that happens only via \TraRef{execute}. Therefore transitions \TraRef{denotify} or the first part of \TraRef{init-reset} cannot move an agent from $(q,+)$ to $(q,-)$, for $q\in I$. Finally, note that $q_h\notin I'$, as $\Support{H}\cap I=\emptyset$ by the definition of population computers. (So, to be precise, the statement only holds once we modify our construction to delete unused states and transitions.)

\medskip\noindent\textbf{Claim 6}. If all configurations in $\N^I$ are terminal, then so are all configurations in $\N^{I'}$. \\
Note that the only transition which can execute from a configuration in $\N^{I'}$ is \TraRef{execute}, but that requires a transition in $\Prot$ which can execute in a configuration in $\N^I$.
\end{proof}

\subsection{Removing helpers}\label{ssec:simulatehelpers}


We convert a bounded binary computer $\Prot$ deciding the predicate $\Double(\varphi)$ over variables $x_1, \ldots, x_k, x_1', \ldots, x_k'$  into a computer $\Prot'$ with no helpers deciding $\varphi$ over variables $x_1, \ldots, x_k$. In \cite{BlondinEGHJ20}, a protocol with helpers and set of states $Q$ is converted into a protocol without helpers with states $Q \times Q$. We provide a better conversion, called $\autarkify$, that avoids the quadratic blowup. 

Let us give some intuition first. All agents of an initial configuration of $\Prot'$ are in input states.
$\Prot'$ simulates $\Prot$ by \textit{liberating} some of these agents and transforming them into helpers, without changing the output of the computation. 
For this, two agents in an input state $x_i$ are allowed to interact, producing one agent in $x_i'$ and one ``liberated'' agent, which can be used as a helper. 
This does not change the output of the computation, because $\Double(\varphi)(\ldots, x_i, \ldots, x_i', \ldots) = \Double(\varphi)(\ldots, x_i-2, \ldots, x_i'+1, \ldots)$ holds by definition of  $\Double(\varphi)$.

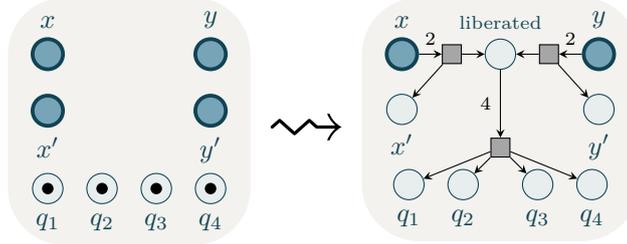
\begin{figure}[ht] 
	\centering
	\scalebox{1}{

	\begin{tikzpicture}[->, auto, node distance=0.3cm]
		\tikzset{every place/.append style={minimum size=0.4cm, niceblue,fill=nicebgblue}}
		\tikzset{every label/.append style={text=niceblue}}
		\tikzset{every transition/.style={minimum size=0.25cm, fill=gray!70}}
		\tikzset{every edge/.append style={font=\scriptsize,-stealth}}
		
		\tikzstyle{hidden}=[draw=none, fill=none]
		\tikzstyle{clustergrey}=[notquitegray!18]
		\tikzstyle{initial}=[fill=nicebluelight,line width=0.05cm]
		
		\newcommand*{\distancearrowx}{0.2cm}
		\newcommand*{\distancearrowy}{0.3cm}
		\newcommand*{\distanceoutputy}{1.35cm}
		\newcommand*{\distancedistributey}{0.9cm}
		
%
%

		
		\node[place, initial, label= above:$x$] (ooX) {};
		\node[place, initial, label= below:$x'$, below=of ooX] (oo2X) {};
		
		\node[place, below= \distanceoutputy of ooX, label= below:$q_1$, tokens=1] (oo1) {};
		\node[place, right= of oo1, label= below:$q_2$, tokens=1] (oo2) {};
		\node[place, right= of oo2, label= below:$q_3$, tokens=1] (oo3) {};
		\node[place, right= of oo3, label= below:$q_4$, tokens=1] (oo4) {};
		
		\node[place, initial, label= above:$y$, above= \distanceoutputy of oo4] (ooY) {};
		\node[place, initial, label= below:$y'$, below=of ooY] (oo2Y) {};

		\node[scale=3] (arrow2) [below right=\distancearrowy and \distancearrowx of ooY]  {$\rightsquigarrow$};

		\node[place, initial, label= above:$x$, above right=\distancearrowy and \distancearrowx of arrow2] (X) {};
		\node[place, label= below:$x'$, below=of X] (2X) {};
		\node[transition, right=of X] (scaleX) {};
		\node[place, label= above:{\scriptsize liberated}, right=of scaleX] (14) {};
		\node[transition, right=of 14] (scaleY) {};
		\node[place, initial, label= above:$y$, right=of scaleY] (Y) {};
		\node[place, label= below:$y'$, below=of Y] (2Y) {};
		
		\node[transition, below= \distancedistributey of 14] (t14) {};
		
		\node[place, below left = of t14, label= below:$q_2$] (2) {};
		\node[place, left=of 2, label= below:$q_1$] (1) {};
		\node[place, below right =of t14, label= below:$q_3$] (3) {};
		\node[place, right=of 3, label= below:$q_4$] (4) {};
		
		\path
		(X) edge[] node {2} (scaleX)
		(scaleX) edge[] (2X)
		(scaleX) edge[] (14)
		
		(Y) edge[] node[swap] {2} (scaleY)
		(scaleY) edge[] (2Y)
		(scaleY) edge[] (14)
		
		(14) edge[] node[swap] {4} (t14)
		(t14) edge[] (1)
		(t14) edge[] (2)
		(t14) edge[] (3)
		(t14) edge[] (4)
		;
		
		\begin{pgfonlayer}{background}
			\filldraw [line width=15mm,join=round,clustergrey]
			(ooX  -| ooX.east)  rectangle (oo4  -| ooY.west)
			;
			\filldraw [line width=15mm,join=round,clustergrey]
			(X  -| X.east)  rectangle (4  -| Y.west)
			;
		\end{pgfonlayer}
	\end{tikzpicture}
	}
	\caption{Illustration in graphical Petri net notation (see Section \ref{sec:PCdef}) of the conversion that removes helpers. Initial states are highlighted.}
	\label{fig:helper_removal}
\end{figure}

Figure \ref{fig:helper_removal} illustrates this idea. 
Assume $\Prot$ has input states $x, y, x', y'$ and helpers $H=\multiset{q_1, q_2, q_3, q_4}$, as shown on the left-hand side. Assume further that $\Prot$ computes a predicate $\Double(\varphi)(x, y, x', y')$.
The computer $\Prot'$ is shown on the right of the figure. The additional transitions liberate agents, and send them to the helper states $H$.
Observe that the initial states of $\Prot'$ are only $x$ and $y$. Let us see why $\Prot'$ decides $\varphi(x, y)$. As the initial configuration of $\Prot'$ for an input $x, y$
puts no agents in $x', y'$, the computer $\Prot'$ produces the same output on input $x, y$ as $\Prot$ on input $x, y, 0, 0$. Since $\Prot$ decides $\Double(\varphi)$
and $\Double(\varphi)(x, y,0, 0) = \varphi(x, y)$ by the definition of $\Double(\varphi)$, we are done. We make some remarks:
\begin{itemize}
	\item $\Prot'$ may liberate more agents than necessary to simulate the multiset $H$ of helpers of $\Prot$. 
	This is not an issue, because by definition additional helpers do not change the output of the computation. 
	\item If the input is too small, $\Prot'$ cannot liberate enough agents to simulate $H$. 
	Therefore, the new computer only works for inputs of size \(\Omega(\Abs{H})=\Omega(\Abs{\varphi})\). 
	\item Even if the input is large enough, $\Prot'$ might move agents out of input states before liberating enough helpers.
	This is where the assumption that all configurations in \(\N^{I}\) are terminal is used: Before the first full batch of liberated agents is dispatched, \(\Prot\) cannot execute any transition.
\end{itemize}

Recall the definition of \(\Double(\varphi)\) for some \(\varphi \in \QFPA\) (Definition~\ref{def:double}): for every variable \(x_i\) in \(\varphi\) create copies \(x_i\) and \(x_i'\); afterwards replace every occurrence of \(x_i\) by \(x_i + 2x_i'\). The specification of the construction is as follows:

\begin{specification}{\autarkify}\label{spec:autarkify}
{\normalsize \textbf{Input: }} & {\normalsize Bounded binary population computer \(\Prot=(Q,\delta,I,O,H)\) with marked consensus output deciding \(\Double(\varphi)\) for some \(\varphi \in \QFPA\); further, states in $I$ have no incoming transitions and every configuration in $\N^I$ is terminal.} \\[0.2cm]
{\normalsize \textbf{Output:}} & {\normalsize Bounded binary population computer \(\Prot'=(Q',\delta',I',O',\multiset{})\) of adjusted size \(\O(\size_2(\Prot))\) without helpers and with a marked consensus output deciding \(\varphi\) for inputs of size at least \(\Abs{I}+2\Abs{H}\).}
\end{specification}

\subsubsection{Conversion $\autarkify$}


If $\Prot'$ did not have to be binary, we could just add to $\Prot$ a state $\State{h}$ and transitions
\begin{alignat*}{2}
	x,x&\mapsto x',\State{h}&\qquad&\text{ for }x\in I\TraName{double}\\
	\Abs{H}\cdot \State{h}&\mapsto H \TraName[helper:idea]{helper}
\end{alignat*}
Instead, we inline the conversion from Section~\ref{ssec:kwayconversion} (simplified slightly), and define 
$\Prot'$ as follows. We add to $\Prot$ states $\Qhelper:=\{\Up_i,\Down_i:i=0,...,\Abs{H}\}$, and identify $\Up_1$ with $\State{h}$ and $\Down_1$ with $h_1$; here, $h_1,...,h_m$ is an enumeration of $H$. For $i,j\in\{1,...,\Abs{H}-1\}$ we also add transitions
\begin{gather*}
	\begin{aligned}
		\Up_i,\Up_j&\mapsto\Up_{i+j},\Up_0&&\text{ if }i+j<\Abs{H}\\
		\Up_i,\Up_j&\mapsto\Down_{\Abs{H}},\Up_{i+j-\Abs{H}}&&\text{ if }i+j\ge\Abs{H}\\
		\Down_{i+1},\Up_0&\mapsto\Down_{i},h_{i+1}
	\end{aligned}\TraName{helper}
\end{gather*}
\noindent Finally, we set $O'(S):=O(S\cap Q)$ for all $S\subseteq Q'$ (note that $Q\subseteq Q'$).

\subsubsection{Correctness} \label{app:helper_correctness}

\begin{proposition}\label{lem:helper-correctandbounded}
$\autarkify$ satisfies its specification (page~\pageref{spec:autarkify}).
\end{proposition}
\begin{proof}
As for the other conversions, we define a linear map $\pi:\N^{Q'}\rightarrow\N^Q$ to translate configurations of $\Prot'$ to ones of $\Prot$. Here, we simply choose $\pi(C)(q):=C(q)$ for $q\in Q$. We do not, however, show that $\Prot'$ refines $\Prot$, as that does not hold: transitions \TraRef{double} and \TraRef{helper} change $\pi(C)$ in a way that is not compatible with an execution of $\Prot$.  Instead, we start by showing that it suffices to consider only certain transition sequences, where the \TraRef{double} and \TraRef{helper} transitions occur only in the beginning. After that point, our proof proceeds just as for the refinement results.

Let $\sigma_1,\sigma_2,\sigma_3\in(\delta')^*$ denote (finite) sequences of transitions, where $\sigma_1$ contains only \TraRef{double} transitions, $\sigma_2$ only \TraRef{helper} transitions, and $\sigma_3\in\delta^*$. 
We call such a  sequence $\sigma:=\sigma_1\sigma_2\sigma_3$ \emph{good}. We first make the following claim:

\medskip\noindent\textbf{Claim}. If $C\in\N^{Q'}$ is reachable from an initial configuration $C_0\in\N^I$, then there is a good sequence $\sigma\in(\delta')^*$ with $C_0\rightarrow_\sigma C$. \\
Let us prove the claim. Since states $x\in I$ have no incoming transitions (precondition of the specification), the number of agents in $x$ is monotonically decreasing during a run. Hence a \TraRef{double} transition can always be moved to any earlier position in a transition sequence. Similarly, the states used as input by a \TraRef{helper} transition only have incoming \TraRef{double} or \TraRef{helper} transitions, so if a \TraRef{helper} transition is preceded by a $\delta$ transition, their order may be swapped. This proves the claim. 

\medskip Let $C\in\N^{Q'}$ denote a terminal configuration reachable from an input configuration $C_0$ with at least $2\Abs{H}+\Abs{I}$ agents. By the claim there are configurations $C_1,C_2$ with $C_0\rightarrow C_1\rightarrow C_2\rightarrow C$, s.t.\ going from $C_0$ to $C_1$ executes only \TraRef{double} transitions, going from $C_1$ to $C_2$ only \TraRef{helper} transitions, and from $C_2$ to $C$ only transitions in $\delta$. We now consider two cases.

If \TraRef{double} is not enabled at $C_1$, then $C_1(\State{h})\ge\Abs{H}$, as there can be at most $\Abs{I}$ agents left in $C_1(I)$. It is not possible to remove agents from $\Qhelper$ without executing \TraRef{helper}, so at some point at least $\Abs{H}$ agents in $C_1(\State{h})$ will be distributed to $h_1,...,h_{\Abs{H}}$ and we get $C_2\ge H$

Else, \TraRef{double} is enabled at $C_1$ (but not at $C$), thus some transition removing agents from $C(I)$ must have occurred between $C_2$ and $C$ (as \TraRef{helper} transitions cannot do so). Since every configuration of \(\N^{I}\) is terminal in $\Prot$ (precondition of the specification), we get $C_2(\Support{H})>0$, which, due to the construction of transition \TraRef{helper}, implies $C_2\ge H$. (In particular, helpers are distributed in batches of $H$.)

So in both cases we have $C_2\ge H$ and therefore find that $\pi(C_2)$ is an initial configuration (of $\Prot$). Between $C_0$ and $C_2$, only transition \TraRef{double} affects states $I\cup I'$, and it preserves the value of $\Double(\varphi)$. (Also note $\varphi(C_0)=\Double(\varphi)(C_0)$, as only agents in $I$ are present.)

As $C$ can be reached from $C_2$ by executing only transitions in $\delta$, we also get that $\pi(C)$ is a reachable configuration of $\Prot$. Moreover, $C$ is terminal w.r.t.\ $\delta'\supseteq\delta$, so $\pi(C)$ is a terminal configuration of $\Prot$. We have defined $O'$ s.t.\ $O'(C)=O(\pi(C))$ and thus get the correct output.

To argue that $\Prot'$ is bounded, we note that $\delta'\setminus\delta$ is acyclic. So if $\Prot'$ has arbitrarily long runs, then, by the claim, $\Prot$ has as well. But this contradicts that $\Prot$ is bounded.
\end{proof}

\endgroup

\subsection{Converting to consensus output}\label{ssec:converttoconsensusoutput}
The final step to produce a population protocol is to translate computers with marked-consensus output function into computers with standard consensus output function.

\begin{specification}{\distribute}\label{spec:distribute}
{\normalsize \textbf{Input: }} & {\normalsize Constant $k$, function $f$, bounded binary population computer \(\Prot=(Q,\delta,I,O,\multiset{})\) without helpers with marked consensus output function, and deciding a predicate \(\varphi\) for all inputs of size at least \(k\) in \(\O(f(n, \Abs{\varphi}))\) interactions. } \\[0.2cm]
{\normalsize \textbf{Output:}} & {\normalsize Terminating population protocol \(\Prot'\) with \(4\Abs{Q}\) states deciding \(\varphi\) for all inputs of size at least \(k\) in \(\O(n^2+f(n, \Abs{\varphi}))\) interactions. }
\end{specification}

\subsubsection{Conversion $\distribute$}

Let a marked agent be an agent in a state \(q \in Q_0 \cup Q_1\), where \(Q_0 \subset Q\) and \(Q_1 \subset Q\) are the sets of states as in the definition of marked consensus output. The obvious approach for the procedure \(\distribute\) would be to add an extra bit to every state, which will be the opinion of the agent, and is set whenever the agent meets a marked agent. As soon as the actual computation is done, all agents will be convinced of the correct opinion. However, convincing the $i$-th agent takes roughly $n^2/(n-i)$ steps (the inverse of the probability that the single marked agent meets one of the remaining $n-i$ agents with the wrong opinion); in total, this sums to $n^2\log n$ steps. We give a slightly different procedure that achieves $\O(n^2)$. Whenever an agent meets a marked agent, besides assuming the correct opinion, it will receive a token, which it can use once, to convince another agent.

Formally, let $Q_0,Q_1\subseteq Q$ denote the states defining the marked consensus output $O$, and set $Q_\bot:=Q\setminus (Q_0\cup Q_1)$. The set of states of $\Prot'$ is  $Q':=Q\times\{0,1\}^2$, where the second component denotes the current opinion of the agent, and the third whether it has a token. 

Let $q,p\in Q$. We want to execute $\Prot$ simultaneously to the following transitions. To write this down, we choose $q',p'$ as the result of a transition $(q,p\mapsto q',p')\in\delta$, if such a transition exists, else we set $q':=q,p':=p$.

For convenience, we write $*$ if the component does not matter. A $*$ in the result of the transition indicates that this component is left unchanged. Based on our definitions the agents of a transition have no order, so $(q,p\mapsto q',p')$ and $(p,q\mapsto q',p')$ are the same transition. Let $i\in\{0,1\}$.

If an agent meets a marked agent, then the former assumes the latter's opinion and receives a token.
\begin{alignat*}{2}
(q,*,*),(p,*,*)&\mapsto (q',i,1),(p',i,1)&\qquad&\text{ if }\{q',p'\}\cap Q_i\ne\emptyset\TraName{certify}\\
\intertext{If an agent with token meets a non-token agent of opposite opinion, the latter is convinced and the token consumed. Similarly, if two tokens held by agents with opposing opinions meet, the tokens are simply dropped.}
(q,i,1),(p,1-i,0)&\mapsto (q',i,0),(p',i,0)&&\TraName{convince}\\
(q,i,1),(p,1-i,1)&\mapsto (q',i,0),(p',1-i,0)&&\TraName{drop}\\
\intertext{Otherwise, nothing happens.}
(q,*,*),(p,*,*)&\mapsto (q',*,*),(p',*,*)&&\text{ if }(q,p)\ne (q',p')\TraName{noop}
\end{alignat*}
As defined above, our transitions are not deterministic. If multiple transitions are possible, we will pick a \TraRef{certify} transition, or (if there are none) a \TraRef{convince} or \TraRef{drop} transition.

Finally, we choose $O'$ as the consensus output given by the partition $O'_i:=Q\times\{i\}\times\{0,1\}$, for $i\in\{0,1\}$ and $I':=I$, where we identify $q\in I$ with $(q,0,0)$.

\subsubsection{Correctness} \label{app:consensus_correctness}

\begin{proposition}\label{thm:correctness_distribute}
$\distribute$ satisfies its specification (page~\pageref{spec:distribute}).
\end{proposition}
\begin{proof}
As for the other conversions, let $\pi:\N^{Q'}\rightarrow\N^Q$ denote a mapping from configurations of $\Prot'$ to ones of $\Prot$. We choose $\pi(C)(q):=C(q\times\{0,1\}^2)$, so $\pi$ simply projects onto the first component.

\medskip\noindent\textbf{Claim 1}. $\Prot'$ refines $\Prot$ (Definition~\ref{def:refinement}). \\
Properties 1 and 2 of Definition~\ref{def:refinement} follow immediately from the definition of $\autarkify$. For the third property, let $C$ denote a reachable, terminal configuration of $\Prot'$. If a transition of $\Prot$ were enabled at $\pi(C)$, the corresponding \TraRef{noop} transition would be enabled at $C$, so $\pi(C)$ must be terminal as well.

As $\Prot$ has a marked consensus output, there is an agent in a state $q\in Q_i$, where $i\in\{0,1\}$ is the output of $\Prot$ and $\pi(C)(q)>0$. By definition of $\pi$ this implies that there is a $q'\in\{q\}\times\{0,1\}^2$ with $C(q')>0$.

We now claim $\Support{C}\subseteq Q_i'$, so assume the contrary and pick a $p'\in Q'\setminus Q_{1-i}'$ with $C(p')>0$. But agents $q',p'$ can now execute transition \TraRef{certify}, which contradicts $C$ being terminal. Thus the claim is shown and our choice of $O'$ yields $O'(\Support{C})=i$.

\medskip\noindent\textbf{Claim 2}. $\Prot'$ is terminating. \\
Let $C$ denote a reachable configuration of $\Prot'$. As $\Prot'$ refines $\Prot$, $\pi(C)$ is reachable in $\Prot$; as $\Prot$ is bounded, $\pi(C)$ can reach some terminal configuration $D$. We can execute a corresponding sequence of transitions in $\Prot'$ and find a configuration $C'$ with $C\rightarrow C'$ and $\pi(C')$ terminal. At this point, $\pi(C)(Q_i)>0$ and $\pi(C)(Q_{1-i})=0$ for an $i\in\{0,1\}$, which corresponds to the output of $\Prot$. By executing at most $n-1$ \TraRef{certify} transitions, we can reach a configuration where all agents have opinion $i$ and a token; the resulting configuration is terminal.
To summarise, we have argued that any reachable configuration can reach a terminal configuration. Hence any infinite run can reach terminal configurations infinitely often and any fair run will eventually terminate.

\medskip\noindent\textbf{Claim 3}. $\Prot'$ decides $\varphi$ for inputs of size at least $k$. \\
To show that $\Prot'$ decides $\varphi$ for inputs of size at least $k$ we would like to apply the Refinement lemma (Lemma~\ref{lem:refinement}), together with Claims 1 and 2.
However, this does not work, as we would need to assume that $\Prot$ decides $\varphi$ for \emph{all} inputs. Fortunately, the proof can trivially be adapted to consider only inputs of size at least $k$.

\medskip\noindent\textbf{Claim 4}. $\Prot'$ runs in \(\O(n^2+f(n, \Abs{\varphi}))\) interactions for inputs of size $\Omega(k)$. \\
An inspection of $\distribute$ shows that a configuration \(C_1\) with \(\pi(C_1)\) terminal is reached after \(\O(f(n, \Abs{\varphi}))\) steps. It remains to argue that \(C_1\) reaches a terminal configuration within \(\O(n^2)\) steps.

Let $C_1$ denote a configuration s.t.\ $\pi(C_1)$ is terminal. In $C_1$, we have at least one marked agent for the correct answer $b:=O(\Support{C_1})$, and no marked agents for the wrong answer, and this will not change during the remainder of the computation.

Let $f_{ij}:=C(Q\times\{i\}\times\{j\})$, for $i,j\in\{0,1\}$ denote the number of agents with opinion $i$ and holding $j$ tokens. Then $F:=f_{b1}-f_{(1-b)0}-2f_{(1-b)1}$ counts the agents with correct opinion and a token, subtracting both the agents with the wrong opinion and the tokens held by agents with the wrong opinion. It is easy to see that this number cannot decrease, and will increase whenever a marked agent meets an agent that either has the wrong opinion, or does not have a token.

If $F\le\tfrac{2}{5}n$, we have $f_{b1}\le\tfrac{4}{5}n$, so in expectation we have to wait $5n$ steps for $F$ to increase. As $F\ge-2n$ by definition, we need $5n(\tfrac{2}{5}n+2n)\in\O(n^2)$ steps until we have $F\ge\tfrac{2}{5}n$.

As noted, $F$ cannot decrease, so after that point we always have $f_{b1}\ge F\ge\tfrac{2}{5}n$. Whenever an agent with opinion $1-b$ meets an agent in $f_{b1}$, the value $f_{(1-b)0}+2f_{(1-b)1}$ decreases. As long as agents with the wrong opinion exist, this has to happen after at most $\tfrac{5}{2}n$ steps in expectation. Noting $f_{(1-b)0}+2f_{(1-b)1}\le 2n$ we find that after $\O(n^2)$ steps no agents with opinion $1-b$ remain.
\end{proof}

\subsection{Proof of Theorem \ref{thm:mainB1}}\label{ssec:puttogether} 

Applying the conversions of the previous sections in sequence, we obtain the final result of the section:

\thmmainB*
\begin{proof}
Let $\varphi$ be a Presburger predicate, and let $\Prot$ be a bounded population computer of size $m$ deciding $\Double(\varphi)$.
We have:
\begin{itemize}
\item Applying $\preprocessconv$ to $\Prot$ yields a computer $\Prot_0$ of size $\O(\size(\Prot))$ for $\Double(\varphi)$. Moreover, $\Prot_0$ satisfies the preconditions of $\binarise$ and has size $\O(\size(\Prot))$. Additionally, states in $I$ have no incoming transitions, all configurations in $\N^{I}$ are terminal and $r(q)\le1$ for every $q\in I$ and $(r\mapsto s)\in\delta$, where $I$ are the initial states and $\delta$ is the transition function of $\Prot_0$ (\cref{thm:correctness_preprocess}).
\item Applying $\binarise$ to $\Prot_0$ yields a binary computer $\Prot_1$ of adjusted size $\O(\Abs{Q} \cdot \size(\Prot))\subseteq\O(\size(\Prot)^2)$ for $\Double(\varphi)$ that satisfies the preconditions of $\focalise$, with $Q$ being the states of $\Prot_1$. Additionally, states in $I$ have no incoming transitions and all configurations in $\N^{I}$ are terminal, where $I$ are the initial states of $\Prot_1$ (\cref{thm:kwaycorrect}).
\item Applying $\focalise$ to $\Prot_1$ yields a binary computer $\Prot_2$ of adjusted size $\O(\size(\Prot_2))=\O(\size(\Prot)^2)$ for $\Double(\varphi)$ that satisfies the preconditions of $\autarkify$ (\cref{thm:correctness_focalise}).
\item Applying $\autarkify$ to $\Prot_2$ yields a binary computer with marked consensus output $\Prot_3$ of adjusted size $\O(\size(\Prot)^2)$ that satisfies the preconditions of $\distribute$ (\cref{lem:helper-correctandbounded}).
\item Applying $\distribute$ to $\Prot_3$ yields a terminating population protocol $\Prot_4$ for $\varphi$ with $\O(\size(\Prot)^2)$ states (\cref{thm:correctness_distribute}).
\end{itemize}
By Proposition~\ref{prop:nthree}, $\Prot_3$ decides $\varphi$ in $2^{\O(\size(\Prot_3) \log(\size(\Prot_3)))} \cdot n^3=2^{\O(m^2 \log m)} \cdot n^3$ interactions, and so $\Prot_4$ does as well.
\end{proof}

\section{Fast and Succinct Population Protocols} \label{sec:speed}

Applying Theorem \ref{thm:mainB1} to any bounded population computer yields a fast population protocol stabilising within $2^{\O(m^2 \log m)} \cdot n^3$ expected interactions. This protocol is fixed-parameter fast, but not fast. We improve on this result for the family of bounded population computers constructed in Section \ref{sec:construction}. 
We show that applying the sequence of conversions $\binarise$-$\focalise$-$\autarkify$-$\distribute$ defined in Section~\ref{sec:conversion} to these computers yields fast protocols that stabilise in $\O(m^7n^2)$ expected interactions.\footnote{In the proof of Theorem~\ref{thm:mainB1} we also used the conversion $\preprocessconv$, but now it is no longer necessary.}
For this we continue to use potential functions, as introduced in Section~\ref{ssec:fastoutputbroadcast}, but improve our analysis as follows:
\begin{itemize}
\item We introduce \emph{rapidly decreasing} potential functions (Section~\ref{sec:rapidly decreasing}). Recall that the execution of any transition decreases the potential,  but not every interaction executes a transition. Indeed, interactions may be \emph{silent}, and change neither the states of the agents involved, nor the potential. 
Intuitively, rapidly decreasing potential functions certify that, at every non-terminal configuration, executing a transition is not only \emph{possible}, but also \emph{likely}. We introduce \emph{rapid} population computers as the computers with rapidly decreasing potential functions that also satisfy some technical conditions. 
\item We prove that the computers of Section~\ref{sec:construction} are rapid (Section~\ref{ssec:ours-are-rapid}).
\item Finally, we show that applying our conversions to rapid population computers results in population protocols that stabilise within $\O(\alpha m^4 n^2)$ interactions, where \(\alpha\) is a constant of a rapid computer (Sections~\ref{ssec:rapid-conv-1} to \ref{ssec:rapid-conv-4}). Loosely speaking, each of these sections shows that rapidness is preserved by one of the conversions.  
\end{itemize}


\subsection{Rapidly decreasing potential functions}
\label{sec:rapidly decreasing}

In order to define rapidly decreasing potential functions, we need a notion of “probability to execute a transition” that generalises to multiway transitions and is preserved by our conversions.
At a configuration $C$ of a protocol, the probability of executing a binary transition $t=(p,q\mapsto p',q')$ is $C(q)C(p)/n(n-1)$. Intuitively, leaving out the normalisation factor $1/n(n-1)$, the transition has “speed” $C(q)C(p)$, 
proportional to the \emph{product} of the number of agents in $p$ and $q$. But for a multiway transition like $q,q,p\mapsto r_1,r_2,r_3$ the situation changes. If $C(q)=2$, it does not matter how many agents are in 
$p$ – the transition is always going to take $\Omega(n^2)$ interactions (the time until the two agents in $q$ meet). We therefore define the speed of a transition as $\min\{C(q),C(p)\}^2$ instead of $C(q)C(p)$.

It is important to note that this is only an approximation for the sake of analysis. Up to constant factors, it always underestimates the “true” speed of the protocol (i.e.\ the speed in the standard execution model for population protocols, after the conversions have been applied), but the estimate is strong enough to show $\O(n^2)$ stabilisation time.

\newcommand{\Value}{\operatorname{value}}
\newcommand{\AbsValue}[1]{\operatorname{absvalue}(#1)}
\newcommand{\Tmin}[2]{\operatorname{tmin}_{#1}(#2)}
\newcommand{\Speed}{\operatorname{speed}}

For the remainder of this section, let $\Prot=(Q,\delta,I,O,H)$ denote a population computer.
We define formally the speed of a configuration and rapidly decreasing potential functions.

\begin{restatable}{definition}{defSpeed}
Given a configuration $C\in\N^Q$ and some transition $t=(r\mapsto s)\in\delta$, we let $\Tmin{t}{C}:=\min\{C(q):q\in\Support{r}\}$. For a set of transitions $T\subseteq \delta$, we define $\Speed_{T}(C):=\sum_{t\in T}\Tmin{t}{C}^2$, and write $\Speed(C):=\Speed_\delta(C)$ for convenience.
\end{restatable}

\begin{restatable}{definition}{defRapidlyDecreasing}\label{def:rapidlydecreasing}
Let $\Phi$ denote a potential function for $\Prot$ and let $\alpha\ge1$. We say that $\Phi$ is \emph{$\alpha$-rapidly decreasing} at a configuration $C$ if $\Speed(C)\ge(\Phi(C)-\Phi(\Cterm))^2/\alpha$ for all terminal configurations $\Cterm$ with $C\rightarrow\Cterm$.
\end{restatable}

Essentially, a potential function is rapidly decreasing at a configuration if the probability of reducing the potential is quadratic relative to the amount of potential which still has to be removed. In the formula, \(\Phi(\Cterm)\) describes the potential that will be left when the protocol terminates, the potential which still has to be removed is hence the difference to this term. 

Initially, the potential is at most linear in the number of agents $n$. (Recall that we only consider linear potential functions in this paper.) So, if the potential function is rapidly decreasing in all configurations, we get the rough estimate $\sum_{i=1}^{\infty} \alpha n^2/i^2\in \O(\alpha n^2)$ for the total amount of time until a terminal configuration has been reached.

However, no potential function is rapidly decreasing for all configurations of our protocols. Fortunately, we are able to overcome this problem. We show that, for computers satisfying some mild syntactic conditions, we only need the potential function to be rapidly decreasing  for well-initialised configurations:

\begin{restatable}{definition}{defWellInitialised}
$C\in\N^Q$ is \emph{well-initialised} if $C$ is reachable and $C(I)+\Abs{H}\le \frac{2}{3}n$.
\end{restatable}

(Observe that an initial configuration $C$ can only be well-initialised if $C(\Support{H})\in\Omega(C(I))$, i.e.\ the protocol has received a number of helpers linear in the sum of the input.)

We are now ready to present the structure of the rest of the paper. First, we introduce \emph{rapid} population computers as those satisfying some syntactic conditions, as well as having a rapidly decreasing potential function for well-initialised configurations:

\begin{definition}\label{def:rapid}
$\Prot$ is \emph{$\alpha$-rapid} if
\begin{enumerate}
\item it has a potential function $\Phi$ which is $\alpha$-rapidly decreasing in all well-initialised configurations,
\item every state of $\Prot$ but one has at most $2$ outgoing transitions,
\item all configurations in $\N^I$ are terminal, and
\item for all transitions \(t=(r\mapsto s)\), \(q\in I\) we have $r(q)\le1$ and \(s(q)=0\).
\end{enumerate}
\end{definition}

In the rest of the paper we prove the following two theorems:

\begin{restatable}{theorem}{thmmainCa}\label{thm:mainCa}
The population computers constructed in Section~\ref{sec:construction} are $\O(\Abs{\varphi}^3)$-rapid.
\end{restatable}

\begin{restatable}{theorem}{thmmainCb}\label{thm:mainCb}
Every $\alpha$-rapid population computer of size $m$ deciding $\Double(\varphi)$ can be converted into a terminating population protocol with $\O(m)$ states that decides $\varphi$ in $\O(\alpha\, m^4n^2)$ 
expected interactions for inputs of size $\Omega(m)$.
\end{restatable}

\noindent which together immediately lead to our last main result:

\thmmainc*

\begin{proof}
By Theorem \ref{thm:mainCa} there exists a construction producing population computers of size \(\O(\Abs{\varphi})\) which are \(\O(\Abs{\varphi}^3)\)-rapid. Concatenating this with the conversion procedure of Theorem \ref{thm:mainCb} gives a construction for population protocols of size \(\O(\Abs{\varphi})\) which decide \(\varphi\) in \(\O(\Abs{\varphi}^3 \Abs{\varphi}^4 n^2)=\O(\Abs{\varphi}^7 n^2)\) expected interactions for inputs of size \(\Omega(\Abs{\varphi})\).
\end{proof}


\begin{figure}[t]
\centering
\def\svgwidth{135mm}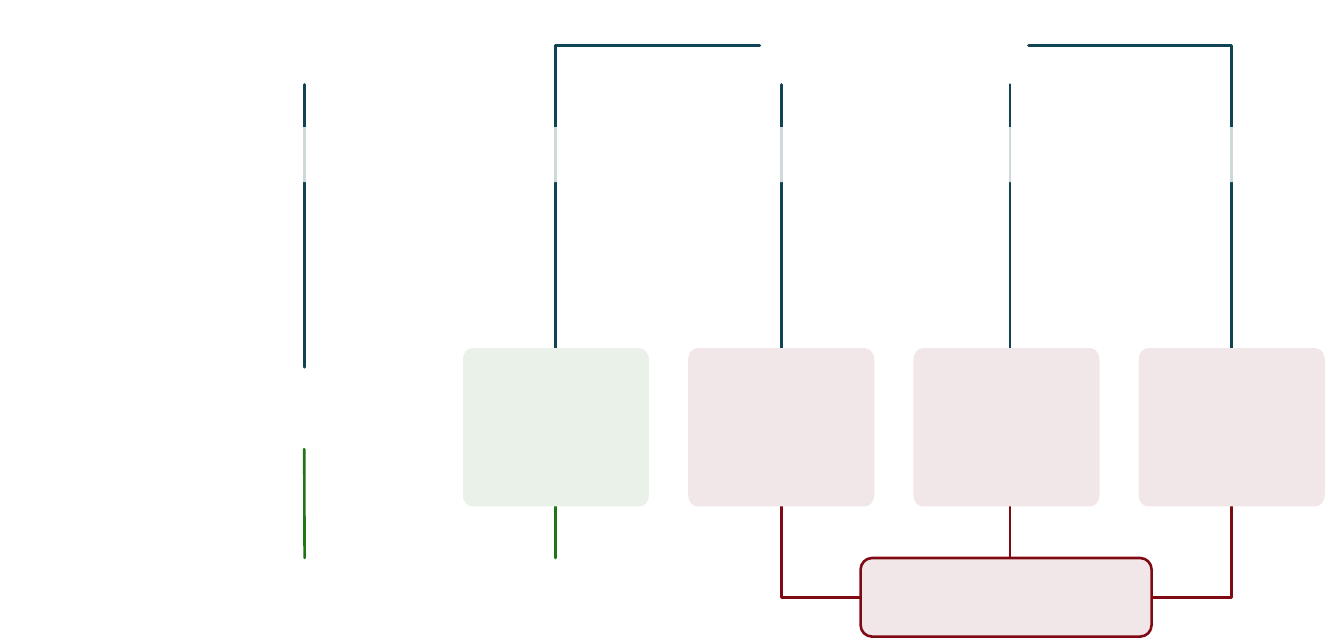
\caption{Overview of Section~\ref{sec:speed}.}\label{fig:section8}
\end{figure}

Figure~\ref{fig:section8} gives a graphical overview of Section~\ref{sec:speed}.
Theorem \ref{thm:mainCa} is proved in Section~\ref{ssec:ours-are-rapid}. The proof of Theorem \ref{thm:mainCb} is more involved. 
Recall that a population computer $\Prot=(Q,\delta,I,O,H)$ is a population protocol if (a) it is binary, (b) has no helpers ($H = \emptymultiset
$), and (c) $O$ is a consensus output.  In Sections \ref{ssec:kwayconversion}, \ref{ssec:outputfunction1}, \ref{ssec:simulatehelpers}, and \ref{ssec:converttoconsensusoutput} we introduced the $\binarise$, $\focalise$, 
$\autarkify$ and $\distribute$ conversions, which together transform a population computer deciding $\Double(\varphi)$ 
into one satisfying conditions (a), (b), and (c), and deciding $\varphi$. In Sections \ref{ssec:rapid-conv-1}-\ref{ssec:rapid-conv-4} we show that these conversions
have small impact on $\alpha$-rapidness, which proves Theorem \ref{thm:mainCb}. Section \ref{ssec:rapid-conv-1} proves the result for $\binarise$. Section \ref{ssec:speeddefinitions} generalises the notion of potential function. Sections \ref{ssec:rapid-conv-2}-\ref{ssec:rapid-conv-4} apply the generalisation to prove the result for $\focalise$, 
$\autarkify$ and $\distribute$, respectively.

\subsection{The population computers of Section~\ref{sec:construction} are rapid}\label{ssec:ours-are-rapid}

\newcommand{\Qhh}{Q_\h}
\newcommand{\Qinit}{I}
\newcommand{\Qother}{Q_\mathrm{self}}
\newcommand{\StateB}{0}

We prove Theorem~\ref{thm:mainCa} by showing that the four conditions of the definition of an $\alpha$-rapid protocol (Definition~\ref{def:rapid}) hold. The first condition, existence of a rapidly decreasing potential function for $\Prot$, is the key one.
While we already know that \emph{some} potential function exists ― $\Prot$ is bounded, so Lemma~\ref{lem:boundedpotential} applies ― we need to construct one that is rapidly decreasing. 

\subparagraph{The potential function $\Phi$.} Let $\Prot$ be the computer for a given predicate constructed as in Section~\ref{sec:construction}.
We give a potential function for $\Prot$. Recall that $\Prot$ consists of a number of subcomputers, say $s$, deciding threshold and remainder predicates. The sets of states of these subcomputers are disjoint, apart from a shared helper state $\StateB$. A state of $\Prot$ is either an input state, a state of one of the subcomputers, or the helper state $\StateB$. For each initial state there is a distribution transition that takes one agent from the state and $s-1$ helpers from state $\StateB$, and sends one agent to an input state of each of the subcomputers. This is all we need to define the potential function:

\begin{definition}
Let $\Prot$ be the computer for a given predicate $\varphi$ constructed as in Section~\ref{sec:construction}, and let $Q$ be its set of states. The function $\Phi \colon Q \rightarrow \N$ is defined as follows, with $q\in Q$:
\begin{itemize}
\item If $q=\StateB$, then $\Phi(q):=0$.
\item If $q$ is an initial state and $\multiset{q, (k-1)\cdot \StateB} \mapsto r$ is its corresponding distribution transition, then
$\Phi(q) := k-1 + \sum_{q' \in r} \Phi(q')$.
\item If $q$ is a state of a subcomputer for a threshold predicate, then $\Phi(q):=1$.
\item If $q$  is a state of a subcomputer for a remainder predicate, then $\Phi(q)$ is defined as follows. 
As described in Section~\ref{sec:construction}, apart from the shared state $\StateB$ the states of this subcomputer are $\{2^0,2^1,...,2^d\}$ for some $d \geq 0$.
Let $d':=d-\lceil\log_26d\rceil$. For every $i\in0,...,d'-1$ we set $\Phi(2^i):=2$, and for every $j\in\{0,...,d-d'\}$ we set
$\Phi(2^{d'+j}):=2^j+1$.
\end{itemize}
\end{definition}

We extend $\Phi$ to configurations $\N^Q$ by choosing the unique linear function with weights $\Phi$.
We show that $\Phi$ is a potential function for $\Prot$.

\begin{proposition}\label{prop:CMTpotential}
The function $\Phi$ is a potential function for $\Prot$. Additionally, $\Phi(\StateB)=0$ and $\Phi(q)\in\O(\Abs{\varphi})$ for all $q\in Q$.
\end{proposition}
\begin{proof}
It follows immediately from the definition of $\Phi$ and from Section \ref{sec:construction} that distribution transitions and transitions of threshold subcomputers decrease the potential. It remains to prove the same for transitions of remainder subcomputers. For \TraRef{mod:combine:main}, we need to show $\Phi(2^i)+\Phi(2^i)>\Phi(2^{i+1})$, which, depending on $i$, reduces to either $2+2>2$ or $2^j+1+2^j+1>2^{j+1}+1$ for some $j\ge0$. In the case of \TraRef{mod:modulo:main}, we note that $2\Modulus\ge 2^d$ (as $d=\lceil\log_2 \Modulus\rceil$), so $2^d-\Modulus\le2^0+2^1+...+2^{d-2}$. It thus suffices to show $\Phi(2^d)\ge\sum_{i=0}^{d-2}\Phi(2^i)+(d-2)$, and we get
\[\sum_{i=0}^{d-2}\Phi(2^i)=2d'+(d-d'-2)+2^{d-d'-1}-1\le2d+2^{d-d'-1}\]
So $\Phi(2^d)-\sum_{i=0}^{d-2}\Phi(2^i)\ge2^{d-d'}+1-(2d+2^{d-d'-1})$ which is at least $d-2$ if $2^{d-d'-1}\ge3d-3$. The latter then follows from our choice of $d'$. Finally, \TraRef{mod:fast-modulo:main} obviously reduces the potential as well. 
\end{proof}

\subparagraph{The potential function $\Phi$ is $\O(\Abs{\varphi})^3$-rapidly decreasing.}
Let $\Prot$ be the computer of Section \ref{sec:construction} for a predicate $\varphi$. 
Recall that it consists of $s$ subcomputers, each of which corresponds to either a remainder or a threshold predicate $\varphi_j$. Subcomputer $j$ has a degree $d_j$, for $j=1,...,s$, corresponding to the bits of the representation it encodes.
Further $\Prot$ has a helper state $\StateB$, shared by all subcomputers. We introduce some notations. 

\begin{itemize}
\item 
We need to reference specifically the highest bits of threshold subcomputers, and so we define $\Qhh:=\{(2^{\h_j})_j, (-2^{\h_j})_j:\varphi_j\text{ is threshold predicate}\}$.
\item Furthermore, we are interested in the states $q$ for which a transition using only agents in $q$ exists. The set of these states is $\Qother:=Q\setminus(\Qinit\cup\Qhh\cup\{\StateB\})$.
\item We denote by $\Value_j(q)$ the value of state $q$ for the $j$-th subcomputer. Formally, for $q\in Q$ and a subcomputer $j\in\{1,...,s\}$ we define  $\Value_j((q)_j):=q$ for each $q\in Q_j$, $\Value_j(0):=0$ and $\Value_j(x_i):=a_i^j$ for input $i\in\{1,..,v\}$, where $a_i^j$ is the coefficient of the variable $x_i$ in the predicate $\varphi_j$.
\end{itemize}

Note that $\sum_q\Value_j(q)C(q)$ is invariant for configurations $C$ of a run, if $j$ is a threshold predicate. (For remainder predicates it would be invariant modulo $\Modulus_j$, but that is not relevant for this section.) Also, the sum $\sum_q\Abs{\Value_j(q)}C(q)$ is non-increasing, for all $j$.

We show that $\Phi$ is $\O(\Abs{\varphi})^3$-rapid in all well-initialised configurations.
First, we prove a lemma bounding the number of agents in the states of $Q_d$ in reachable configurations. 

\begin{lemma}\label{lem:highvalue}
Let $\Prot$ be the computer of Section \ref{sec:construction} for a predicate $\varphi$, and let $C$ be a reachable configuration of $\Prot$. We have:
\begin{enumerate}
\item $C(\Qhh)\le\Abs{C}/8$, and
\item Let $t_j:=((2^{\h_j})_j,(-2^{\h_j})_j\mapsto0,0)\in\textnormal{\TraRef{thr:cancel:app}}$, where $j$ is the index of a threshold subcomputer, and let $C\rightarrow C'$. Then $C(\Qhh)-C'(\Qhh)\le2\sum_j\Tmin{t_j}{C}+C(\Qinit)+C(\Qother)$.
\end{enumerate}
\end{lemma}
\begin{proof}
For part 1., let $j$ denote the index of the threshold predicate, $q=(\pm2^{\h_j})_j\in\Qhh$ one of its largest states, and $p\in I$ an input state. We then have
\[\Abs{\Value_j(p)}\le a^j_\mathrm{max}\le \frac{2^{d_j}}{16s}=\frac{\Value_j(q)}{16s}\]
due to the choice of $d_j$. As $C$ is reachable from some initial configuration $C_0$, and $\sum_q\Abs{\Value_j(q)}C(q)$ cannot increase, we sum over $p$ to get
\[\Abs{\Value_j(q)}\,C(q)
\le\sum_{p\in\Qinit}\Abs{\Value_j(p)}\,C_0(p)
\le\sum_{p\in\Qinit}\Abs{\Value_j(q)}\,C_0(p)/16s
=\Abs{\Value_j(q)}\,C_0(\Qinit)/16s\]
So we have $C(q)\le C_0(\Qinit)/16s$ and summing over $q$ yields the desired statement.

For part 2., let $D\in\N^{\Qhh}$ with $D(q):=\max\{C(q)-C'(q),0\}$ for $q\in\Qhh$. Note that $D(q)$ is a lower bound on how many agents leave state $q$ in any run from $C$ to $C'$, and that $C(\Qhh)-C'(\Qhh)\le D(\Qhh)$.
Let $j$ be the index of a threshold subcomputer. As the only ways to leave \(\Qhh\) are \TraRef{thr:cancel:app} and \TraRef{thr:cancel 2nd highest:app}, we know that
\[\sum_{q\in Q_{j+}}\Abs{\Value_j(q)}\,C(q)\ge2^{\h_j-1}\Big(D((2^{\h_j})_j)+D((-2^{\h_j})_j)\Big)\]
where $Q_{j+}:=\{q\in Q:\Value_j(q)>0\}$. The same inequality holds when replacing $Q_{j+}$ with $Q_{j-}:=\{q\in Q:\Value_j(q)<0\}$. From these, we derive 
\[2\,C((\mathord{\sim}2^{\h_j})_j)+2\sum_{q\in Q\setminus\Qhh}2^{-\h_j}\Abs{\Value_j(q)}\,C(q)\ge D((2^{\h_j})_j)+D((-2^{\h_j})_j)\]
for $\mathord{\sim}\in\{+,-\}$, as $Q\setminus\Qhh \cup\{(\mathord{\sim}2^{\h_j})_j\}\supseteq Q_{j\sim}$. We can combine the two inequalities into
\[2\Tmin{t_j}{C}+2\sum_{q\in Q\setminus\Qhh}2^{-\h_j}\Abs{\Value_j(q)}\,C(q)\ge D((2^{\h_j})_j)+D((-2^{\h_j})_j)\]
Summing over $j$ then yields
\[2\sum_{t\in T_\h}\Tmin{t}{C}+2\sum_{q\in Q\setminus\Qhh}\left [ C(q)\sum_j 2^{-\h_j}\Abs{\Value_j(q)} \right ]\ge D(\Qhh)\]
If $\sum_j2^{-\h_j}\Abs{\Value_j(q)}\le\tfrac{1}{2}$ were to hold for all $q\in Q\setminus\Qhh$, then we would get the desired statement (noting $\Value(\StateB)=0$), so it remains to show this claim. For $q\in Q_j\setminus\Qhh$ for some $j$ we have $\Abs{\Value_j(q)}\le2^{\h_j-1}$ and $\Value_{j'}(q)=0$ for $j'\ne j$, and for $q\in I$ it follows from $\Abs{\Value_j(q)}\le a^j_\mathrm{max}\le 2^{\h_j}/16s$ for all $j\in\{1,...,s\}$.
\end{proof}

We  now show that $\Phi$ is $\O(\Abs{\varphi}^3)$-rapidly decreasing. The proof uses the following inequality, which follows immediately from the 
Cauchy-Bunyakovsky-Schwarz inequality.

\begin{lemma}\label{lem:sumofsquares}
	Let $x_1,...,x_n\in\mathbb{R}$. Then
	$\Big(\sum_{i=1}^n x_i\Big)^2\le n\sum_ix_i^2$.
\end{lemma}

\begin{proposition}\label{prop:rapidpot}
The function $\Phi$ is a $\O(\Abs{\varphi}^3)$-rapidly decreasing potential function for $\Prot$ in all well-initialised configurations. 
\end{proposition}
\begin{proof}
Let $C$ be a well-initialised configuration and let $\Cterm$ be a terminal configuration such that 
$C \rightarrow \Cterm$. Let $C':=C-\Cterm$ and let $W:=\max_{q\in Q}\Phi(q)$ denote the largest weight of $\Phi$. Then
\[\Phi(C)-\Phi(\Cterm)=\Phi(C')\le W \cdot C'(Q\setminus\{\StateB\})\]
Applying Lemma~\ref{lem:highvalue}(2), as well as $C'\le C$, yields
\begin{equation}\label{eqn:1627}
C'(Q\setminus\{\StateB\})=C'(\Qhh)+C'(\Qinit)+C'(\Qother)\le2\sum_{t\in T_\h}\Tmin{t}{C}+2C(\Qinit)+2C(\Qother)\tag{$*$}
\end{equation}
where $T_\h:=\{t_j:\varphi_j\text{ threshold predicate}\}$ with $t_j$ as for Lemma~\ref{lem:highvalue}(2). Since $C$ is well-initialised we have $C(\Qinit)\le2(C(\Qother)+C(\StateB)+C(\Qhh))$, and Lemma~\ref{lem:highvalue}(1) implies $C(\Qhh)\le C(Q\setminus\Qhh)/7$. We use both to derive
\begin{align*}
&C(\Qinit)\le 2C(\Qother)+2C(\StateB)+\tfrac{2}{7}\big(C(\Qinit)+C(\Qother)+C(\StateB)\big)\\
\Rightarrow\;&\tfrac{5}{7}C(\Qinit)\le\tfrac{16}{7}(C(\Qother)+C(\StateB))\\
\Rightarrow\;&C(\Qinit)\le\tfrac{16}{5}\big(C(\Qother)+C(\StateB)\big)\le4\big(C(\Qother)+C(\StateB)\big) 
\end{align*}
It follows that $C(\Qinit)\le4C(\Qother)+4\min\{C(\StateB),C(\Qinit)\}$ holds. Writing $T_I:=\TraRef{distribute}$, and using $\min\{C(0),C(I)\} \leq \sum_{q \in I} \min\{C(0),C(q)\}$ we get
\[C(\Qinit)\le4C(\Qother)+4\sum_{t\in T_I}\Tmin{t}{C} \]
Every state $q\in\Qother$ has a (unique) transition using only agents in $q$; we use $T_o$ to denote this set and set $T:=T_I\cup T_o\cup T_\h$. We can now insert the previous inequality into (\ref{eqn:1627}) to get $C'(Q\setminus\{\StateB\})\le8\sum_{t\in T}\Tmin{t}{C}$. Noting $\Abs{T}\le2\Abs{Q}$ and applying Lemma~\ref{lem:sumofsquares} we get:
\[\Phi(C')^2\le \Big(8\,W\sum_{t\in T}\Tmin{t}{C}\Big)^2\le 128\,W^2\Abs{Q}\Speed(C)\]
The desired statement then follows from $\Abs{Q},W\in\O(\Abs{\varphi})$.
\end{proof}

\subparagraph{Proof of Theorem \ref{thm:mainCa}} We are now ready to prove:

\thmmainCa*
\begin{proof}
We show that the computers satisfy the conditions of the definition of $\alpha$-rapid protocols (Definition~\ref{def:rapid}) for
$\alpha \in \O(\Abs{\varphi}^3)$.

Condition 1 follows immediately from Proposition \ref{prop:rapidpot}.
Conditions 2-4  are easy to check: For condition 2, we note that only the reservoir state $0$ (shared by all subcomputers) has more than two outgoing transitions. Condition~3 is ensured by transition~\TraRef{distribute} always taking at least one agent from the reservoir $0\notin I$. Similarly, this transition is the only transition affecting the input states $I$, so Condition~4 is met.
\end{proof}

\subsection{Removing multiway transitions preserves speed}\label{ssec:rapid-conv-1}

We show that, loosely speaking, the $\binarise$ conversion of Section~\ref{ssec:kwayconversion} preserves the speed of configurations. Formally, we prove that if the input to the conversion has a rapidly decreasing potential function with parameter $\alpha$, then the output also has a rapidly decreasing potential function, and its parameter is not much larger than $\alpha$.

\begin{proposition}
\label{prop:12345}
Let \(\Prot=(Q,\delta,I,O,H)\) denote a bounded population computer satisfying conditions 2 and 3 of the output specification of $\binarise$ (page~\pageref{spec:binarise}). If some potential function for $\Prot$ is $\alpha$-rapidly decreasing in all well-initialised configurations, then some potential function for $\binarise(\Prot)$ is $\O(\Abs{Q}^2 k^2 \alpha)$-rapidly decreasing in all well-initialised configurations, where $k$ is the maximal arity of the transitions of $\Prot$.
\end{proposition}
\begin{proof}
Let $\Phi$ be a potential function for $\Prot$ that is $\alpha$-rapidly decreasing in all well-initialised configurations,
and let $\Prot':=\binarise(\Prot)$. We first construct a potential function $\Phi'$ for $\Prot'$. Intuitively, the potential of a state  corresponds directly to the original potential of states it owns, with some additional accounting to pay for overhead of executing a multiway transition. At this point it becomes important that our definition of potential function requires a transition of arity $k$ to reduce the potential by $k-1$, as this means that we have to increase the total potential by only a constant factor.

We first adjust $\Phi$ by multiplying it with $5$, so that $\Phi(r)\ge\Phi(s)+5(\Abs{s}-1)$ for all transitions $(r\mapsto s)\in \delta $. Now, let $q,p\in Q$, $t=(r\mapsto s)\in \delta $ a transition where 
$\Support{r}=\{q,p\}$ and $q$ is primary, and let $s_1,...,s_\kwayL$ denote the enumeration of $s$ from Section~\ref{app:multiway_construction}. 
	We define $\Phi'$ as
	\begin{align*}
		\Phi'((q,0))&:=0&&\\
		\Phi'((q,i))&:=i\Phi(q)+1&&\text{for }i\in\{1,...,m(q)-1\}\\
		\Phi'((q,m(q)))&:=m(q)\Phi(q)&&\\
		\Phi'((q,i,t))&:=i\Phi(q)+\Phi(s)+2\kwayL+2&&\text{for }i\in\{1,...,m(q)\}\\
		\Phi'((t,i))&:=\sum_{j=i}^{\kwayL}(\Phi(s_j)+2)&&\text{for }i\in\{1,...,\kwayL-1\}
	\end{align*}

\medskip\noindent\textbf{Claim 1}. $\Phi'$ is a potential function for $\Prot'$. \\
For most transitions, it is easy to see that $\Phi'$ decreases. However, we need to verify that \TraRef{commit} does so as well. If $q\ne p$, we have to prove the inequality
	\[i\Phi(q)+j\Phi(p)+2\ge (i-r(q))\Phi(q)+\Phi(s)+2l+2+(j-r(p))\Phi(p)+1\]
	which boils down to $\Phi(r)\ge\Phi(s)+2l+1$. This then follows from $\Phi(r)\ge\Phi(s)+5(l-1)$. The case $q=p$ is shown analogously. This concludes the proof of the claim.

\medskip The next three claims show technical properties that are needed for the proof that $\Phi'$ is rapidly decreasing. The first gives a relation between the potential of $\Prot'$ and of the refined computer $\Prot$ (recall Section~\ref{app:multiway_correctness}). Let \(\pi: \N^{Q'} \to \N^Q\) be the mapping relating configurations of \(\Prot'\) and \(\Prot\) defined in Section~\ref{app:multiway_construction}.

\medskip\noindent\textbf{Claim 2}. $\Phi(\pi(C))\le\Phi'(C)\le\Phi(\pi(C))+2C(S)$ for $S:=Q'\setminus\{(q,m(q)), (q,0):q\in Q\}$. \\
	For the first inequality we simply observe $\Phi(\pi(q'))\le\Phi'(q')$ for each $q'\in Q'$. Each state $q'=(q,m(q))$, for $q\in Q$, satisfies $\Phi'(q')=\Phi(\pi(q'))$. For each other state $q'\in S$ we show $\Phi'(q')-\Phi(\pi(q'))\le 2c$, where $c$ is the amount of agents “owned” by an agent in state $q'$. E.g.\ $q'=(q,i)$ for $q\in Q$ and $0<i<m(q)$ has $\Phi'(q,i)-\Phi(\pi(q,i))=1$ and $c=i$. For $(q,i,t)$, we have $c=\Abs{s}+i$, and require \(2c \geq 2 \Abs{s}+2\). The respective inequalities follow immediately, which concludes the proof of the claim.

In Section~\ref{ssec:ours-are-rapid} we have seen that states which can initiate a transition by themselves are useful to show speed bounds. More precisely, states $q$ such that there exists a transition $t=(r\mapsto s)$ with $\Support{r}=\{q\}$, implying $\Tmin{t}{C}=C(q)$ for all $C$. The next claim shows that most states \(q\) can be assigned a transition with the same useful property on reachable configurations (even if not all of them use only a single state).

\medskip\noindent\textbf{Claim 3}. Let $S:=Q'\setminus\{(q,m(q)), (q,0):q\in Q\}$. There is an injection $g:S\rightarrow \delta '$, s.t.\ $C(q')=\Tmin{g(q')}{C}$ for any reachable configuration $C$ and $q'\in S$. \\
Let $q\in Q$, and $t=(r\mapsto s)\in \delta $.
	
	If $q'=(q,i)$ for $i<m(q)$, then there is a \TraRef{stack} transition using only agents in $q$, which we use as $g(q')$. If $q'=(q,i,t)$, then we know that $C(q')\le C((q,0))$, as $q'$ owns at least $\Abs{r}\ge 1$ agents, so we can pick the \TraRef{transfer} transitions for $g(q')$. Finally, if $q'=(t,i)$, for $i<\Abs{s}$, then $q'$ owns $\Abs{s}-i\ge1$ agents other than itself, and we choose the corresponding \TraRef{execute:multiway} transition. This proves the claim.

In the end, we want to show that $\Phi'$ is rapidly decreasing in all well-initialised configurations if $\Phi$ is. For this, we need to argue briefly that being a well-initialised configuration corresponds.

\medskip\noindent\textbf{Claim 4}. If $C$ is well-initialised, then so is $\pi(C)$. \\
Due to Condition~3 in the specification of $\binarise$, $m(q)=1$ for each $q\in I$. In combination with Condition~2 we get $C((q,1))=\pi(C)(q)$ for all $q\in I$. This implies $C(I')=\pi(C)(I)$. Noting $\Abs{H}=\Abs{H'}$, the statement follows immediately.

\bigskip\noindent  Finally, we prove that $\Phi'$ is rapidly decreasing.
Let $k$ be the maximal arity of the transitions of $\Prot$. We show that $\Phi'$ is $\O(\alpha k^2\Abs{Q}^2)$-rapidly decreasing in $C$ if $\Phi$ is $\alpha$-rapidly decreasing in $\pi(C)$, for all reachable configurations $C$ and $\alpha\ge1$.
Let $\Cterm$ denote a terminal configuration reachable from $C$. Using Claim~2 we get $\Phi'(C)-\Phi'(\Cterm)\le\Phi(\pi(C))-\Phi(\pi(\Cterm))+2C(S)$. We know that $\pi(\Cterm)$ is reachable from $\pi(C)$, and, as shown in the proof of \cref{thm:kwaycorrect}, Claim 1, it is terminal as well. Hence we can use that $\Prot$ is $\alpha$-rapidly decreasing in $\pi(C)$ to get $\big(\Phi(\pi(C))-\Phi(\pi(\Cterm))\big)^2\le\alpha\Speed(\pi(C))$.

Let us now estimate $\Speed(\pi(C))$. Let  $t^*$ be the \TraRef{commit} transition corresponding to $t\in \delta $ using agents in states $(q,m(q)),(p,m(p))$ for $q,p\in Q$. Additionally, let $H_q:=\sum_{q'\in S}C(q')\pi(q')(q)$ for $q\in Q$ denote the contribution of agents in states $S$ to $\pi(C)(q)$, for $q\in Q$. We have 
$$\pi(C)(q)=H_q + m(q)C((q,m(q)))\le H_q+kC((q,m(q))) \ . $$

Now, let $t=(r\mapsto s)\in \delta $ and $q,p\in Q$ with $\Support{r}=\{q,p\}$. Then the above (together with the well-known inequality stating that $\min\{x_1+x_2,y_1+y_2\}\le\min\{x_1,y_1\}+x_2+y_2$ holds for non-negative $x_1,x_2,y_1,y_2$) yields
\[\Tmin{t}{\pi(C)}=\min\{\pi(C)(q),\pi(C)(p)\}\le k\Tmin{t^*}{C}+H_q+H_p\]
Squaring the right-hand side gives at most $3(k^2\Tmin{t^*}{C}^2+H_q^2+H_p^2)$, and so summing over $t$ and applying Lemma \ref{lem:sumofsquares} we get a first bound for $\Speed(\pi(C))$:
\[\Speed(\pi(C))\le 3k^2\sum_{t\in \delta }\Tmin{t^*}{C}^2+6\Abs{Q}\sum_{q\in Q}H_q^2 \ . \]

Let us bound $\sum_{q\in Q}H_q^2$. We have  $\sum_{q\in Q}H_q=\Abs{\pi(C_S)}$, where $C_S(q):=C(q)$ for $q\in S$ and $0$ otherwise. Moreover, the definition of $\pi$ yields $\Abs{\pi(q)}\le 2k$, and so $\Abs{\pi(C_S)}\le 2k\Abs{C_S}$. Finally, by Claim 3 we obtain $\Abs{C_S}\le\sum_{t'\in g(S)}\Tmin{t'}{C}$. Putting this together we get $\sum_{q\in Q}H_q^2 \le  \big(\sum_{q\in Q}H_q\big)^2 \leq 4k^2 \big(  \sum_{t'\in g(S)}\Tmin{t'}{C} \big)^2$. Applying Lemma \ref{lem:sumofsquares},  we finally get 
\[\Speed(\pi(C))\le 3k^2\sum_{t\in \delta }\Tmin{t^*}{C}^2+24k^2\Abs{Q}^2\sum_{t'\in g(S)}\Tmin{t'}{C}^2\le 24k^2\Abs{Q}^2\Speed(C)\]
\noindent  We are now ready to complete the proof (note $C(S)=\Abs{C_S}$):
\begin{align*}
\big(\Phi'(C)-\Phi'(\Cterm)\big)^2 & \le 2\big(\Phi(\pi(C))-\Phi(\pi(\Cterm))\big)^2+8C(S)^2\\
& \le2\alpha\Speed(\pi(C))+8\Abs{Q}\Speed(C) \\
& \le49\alpha k^2\Abs{Q}^2\Speed(C)
\end{align*}
\end{proof}

\subsection{Generalised potential function analysis}\label{ssec:speeddefinitions}

We generalise the notion of potential function (described in Section~\ref{ssec:fastoutputbroadcast}) to better analyse the conversions $\focalise$ and $\autarkify$. We start by briefly recalling the main definitions of Sections~\ref{ssec:fastoutputbroadcast} and~\ref{sec:rapidly decreasing}.

\defPotentialFunction*
\defSpeed*
\defRapidlyDecreasing*
\defWellInitialised*

While the above definitions can be applied to all of our conversions, they lead to large constants in the final speed. These are merely the result of a loose analysis – they do not reflect the actual speed of our protocols. Mainly, this is due to a single potential function being unable to model computations that consist of multiple phases efficiently. A concrete explanation of this problem in the context of $\focalise$ is given in Section~\ref{ssec:rapid-conv-2}. In this section we introduce the formal machinery necessary to better adapt our technique to those constructions, leading to better constants and easier proofs.

We start by extending the definition of rapidly decreasing to handle linear functions which are not potentials. Here, we do not need to deal with multiway transitions, so let $\Prot=(Q,\delta,I,H,O)$ denote a binary population computer.

\begin{definition}
Let $\Phi:\N^Q\rightarrow\N$ be linear, let $\delta_>:=\{(r\mapsto s)\in\delta:\Phi(r)>\Phi(s)\}$ be the transitions decreasing $\Phi$, and let $\alpha>0$. If $\Speed_{\delta_>}(C)\ge(\Phi(C)-\Phi(\Cterm))^2/\alpha$ for a configuration $C$ and all terminal configurations $\Cterm$ with $C\rightarrow\Cterm$, we say that $\Phi$ is \emph{$\alpha$-rapidly decreasing in $C$}.
\end{definition}
The only change compared to Definition~\ref{def:rapidlydecreasing} is that the speed considers only transitions which reduce the given linear function. If $\Phi$ is a potential function, we have $\delta_>=\delta$ and $\Phi(C)>\Phi(\Cterm)$ for all non-terminal configurations $C$, making this definition coincide with Definition~\ref{def:rapidlydecreasing}.

To model phases, the general idea is that we construct a family of linear functions $\Phi_1,...,\Phi_e$. For each configuration $C$, one of these will be rapidly decreasing (we refer to it as “active”). That alone would not be enough to guarantee a quadratic number of interactions (or any time bound at all), as it would not prevent the other functions from increasing their value. So we require the stronger property that a $\Phi_i$ cannot increase once it has been active. We also need that $\Phi_i$ can decrease at $C$, which certifies that some progress can be made. Otherwise, $\Phi_i$ might be “rapidly decreasing” but already at its lowest point.

\begin{definition}
A tuple $\Phi=(\Phi_1,...,\Phi_e)$, where $\Phi_1,...,\Phi_e:\N^Q\rightarrow\N$ denote linear maps, is a \emph{potential group (of size $e$)}.
A potential group $\Phi$ is \emph{$\alpha$-rapidly decreasing in a configuration $C$}, for $\alpha\ge1$,  if $C$ is terminal or there is some $i\in\{1,...,e\}$ s.t.\ $\Phi_i$ is $\alpha$-rapidly decreasing in $C$, some transition reducing $\Phi_i$ is enabled at $C$, and no transition increasing $\Phi_i$ can be executed at any configuration reachable from $C$. We then call $\Phi_i$ \emph{active} at $C$.
\end{definition}

The definition places no restrictions on the order in which the $\Phi_i$ are listed. However, in our proofs we will generally order them in the same fashion as they would become active in a run. Further, our potential groups have the additional property that they decrease lexicographically with each transition.

To close out the section, we show that the above notion does actually lead to a strong speed bound when applied to population protocols.

\def\P{\mathbb{P}}
\begin{restatable}{proposition}{propmultipotentialspeed}\label{prop:multipotential_speed}
Let $\Prot$ denote a population protocol and $\Phi$ a potential group for $\Prot$ of size $e$ which is \(\alpha\)-rapidly decreasing in all reachable configurations with at least $m$ agents. Then $\Prot$ reaches a terminal configuration after $\O(e(\sqrt{\alpha}\Abs{Q}+\alpha)\,n^2)$ random interactions in expectation for all initial configurations with at least \(m\) agents.
\end{restatable}
\begin{proof}
\newcommand{\Len}{l}
Let $\sigma=C_0C_1...$ denote a fair run of $\Prot$, and pick the smallest $\Len$ s.t.\ $C_\Len$ is terminal. We define
\[X_i^c:=\Abs{\{j:\text{ $\Phi_i$ is active at $C_j$ and $\Phi_i(C_j)-\Phi_i(C_\Len)=c$}\}}\]
We observe that $\Len\le\sum_{i,c}X_i^c$ holds and will now proceed to prove a bound on the expected value $\E(X_i^c)$, for all $i,c$, if $\sigma$ is generated via random interactions.

Consider $\P(X_i^c\ge k+1\mid X_i^c\ge k)$, for $k\ge1$. We note that $\sigma$ is generated by a (homogeneous) Markov chain and the index $\tau$ of the $k$-th configuration counting towards $X_i^c$ is a stopping time. By the strong Markov property, the above probability is equal to the probability that $\Prot$ reaches some configuration $C$ counting towards $X_i^c$ when started in the configuration $C_\tau$. This is at most $1-\gamma$, where $\gamma$ is the probability that $\Prot$ executes a transition reducing $\Phi_i$ at $C_\tau$, as an active $\Phi_i$ cannot increase at any later point.

First, we know that $\Phi_i$ is active at $C_\tau$, so some transition reducing $\Phi_i$ is enabled at $C_\tau$ and $\gamma\ge1/n(n-1)$. However, if $c$ is large enough we can get a better bound due to the fact that $\Phi_i$ is rapidly decreasing at $C_\tau$.

Let $\delta_>\subseteq\delta$ denote the transitions reducing $\Phi_i$. Let $\xi:=\Tmin{t}{C_\tau}$ for some $t=(q,p\mapsto q',p')\in\delta_>$. By definition, we have $C_\tau(q),C_\tau(p)\ge\xi$ and thus the probability of executing $t$ at $C_\tau$ is at least $\xi(\xi-1)/n(n-1)$ (note that $q=p$ is possible). As $\xi(\xi-1)\ge\xi^2/2-1$, we find $n(n-1)\gamma\ge\Speed_{\delta_>}(C_\tau)-\Abs{\delta_>}$ by summing over $t$. Since $\Abs{\delta_>}\le\Abs{Q}^2$, and rapidly decreasing implies $\Speed_{\delta_>}(C_\tau)\ge c^2/\alpha$, we get $\gamma\ge\max\{1,c^2/\alpha-\Abs{Q}^2\}/n(n-1)$.

From $\P(X_i^c\ge k+1\mid X_i^c\ge k)\le1-\gamma$ for all $k\ge1$ we get $\E(X_i^c)\le1/\gamma$ (similar to the geometric distribution). Summing over $i$ and $c$ and using $c^2/\alpha-\Abs{Q}^2\ge c^2/2\alpha$ for $c\ge\sqrt{2\alpha}\Abs{Q}$,
we get:
\begin{align*}
\E(\Len)& \le\sum_{i=1}^e\sum_{c=0}^\infty\E(X_i^c) \\
& \le en(n-1)\sum_{c=0}^\infty\frac{1}{\max\{1,c^2/\alpha-\Abs{Q}^2\}} \\
& \le en(n-1)\Big(\sqrt{2\alpha}\Abs{Q}+\frac{\alpha\pi^2}{3}\Big)
\end{align*}
\end{proof}

\subsection{Converting to marked consensus preserves speed}\label{ssec:rapid-conv-2}

{\def\TraNs{tra:output}

We prove that \(\focalise\), the conversion of Section~\ref{ssec:outputfunction1} is fast
(we use the notion of potential groups introduced in Section~\ref{ssec:speeddefinitions}):

\begin{proposition}\label{prop:focalise-speed}
Let \(\Prot=(Q,\delta,I,O,H)\) be a bounded binary population computer fulfilling 
the specification of $\focalise$ (page~\pageref{spec:focalise}), and let $\Phi$ denote a potential function for $\Prot$ which is \(\alpha\)-rapidly decreasing in all well-initialised configurations.

Then $\Prot':=\focalise(\Prot)$ has a potential group of size \(5\) which is \(\O(\alpha+\size_2(\Prot)^2)\)-rapidly decreasing in all well-initialised configurations.
\end{proposition}
\newcommand{\Tset}{T}
\begin{proof}
We construct a potential group $\Phi'=(\Phi'_1,...,\Phi'_5)$ and show that it is rapidly decreasing in all well-initialised configurations. So let $C$ denote such a configuration, and let $\Cterm$ denote a terminal configuration reachable from $C$.

For the sake of readability we defer the definition of the $\Phi'_i$ until they are used. However, note that the definition will be independent of $C$.

The proof proceeds via case distinction based on the properties of $C$. For the $i$-th case we show that $\Phi'_i$ is active. We implicitly assume that prior cases are excluded, so the proof for case $i$ assumes that the conditions for cases $1,...,i-1$ are not being met.

\smallskip
\textbf{Case 1.} $C(\Qsupp\cup\Qgate\cup\Qreset)>\Abs{Q}+\beta+1$.
For $\Phi'_1$, the goal is to show that the “leader elections” for each state happen quickly. We set $\Phi'_1(q):=2$ for $q\in\Qsupp\cup\Qgate$, $\Phi'_1(q):=1$ for $q\in\Qreset$, and $\Phi'_1(q):=0$ for $q\in\Qorig$. Clearly, the only transitions that affect $\Phi'_1$ are \TraRef{leader} and the third part of \TraRef{init-reset}, both of these reducing the potential by $1$. It is thus not possible for $\Phi'_1$ to increase. One of these transitions is enabled, so $\Phi'_1$ can decrease at $C$.

In particular, note that for each $q\in S$, where $S:=\{q\in Q':\Phi'_1(q)>0\}$ are the states with positive potential, there is a transition reducing $\Phi'_1$ using two agents in $q$. Using $\Tset$ to denote these transitions, we get $\Phi(C)\le2C(S)\le2\sum_{t\in \Tset}\Tmin{t}{C}$ and thus (via Lemma~\ref{lem:sumofsquares}), $(\Phi(C)-\Phi(\Cterm))^2\le\Phi(C)^2\le4\Abs{\Tset}\Speed_{\Tset}(C)$. Finally, we note $\Abs{\Tset}=\Abs{Q}+\beta+1$.

\smallskip
\textbf{Case 2.} $\pi(C)$ is not terminal. In this case, we will argue that the refined transitions of $\Prot$ are likely to occur. We define $\Phi'_2((q,\pm)):=\Phi(q)$ for $q\in Q$ (recall that $\Phi$ is the potential function of $\Prot$), and set $\Phi'_2$ to $0$ elsewhere. $\Phi'_2$ is reduced precisely by the \TraRef{execute} transitions, and increased only by the third case of \TraRef{init-reset}.

As we exclude Case~1, we have $C(\Qsupp\cup\Qgate\cup\Qreset)=\Abs{Q}+\beta+1$. This implies $\Phi(\pi(C))=\Phi'_2(C)$ (we even get $\pi(C)(q)=C((q,\pm))$ for $q\in Q$) and ensures that the third case of \TraRef{init-reset} cannot be executed by any configuration reachable from $C$. Having $\pi(C)(q)=C((q,\pm))$ for $q\in Q$ then ensures that for a transition of $\Prot$ enabled at $\pi(C)$, there is a corresponding \TraRef{execute} transition enabled at $C$.

We now want to show that $\pi(C)$ is well-initialised.
\[\pi(C)(I)
\stackrel{(1)}{=}C(I')
\stackrel{(2)}{\le}\tfrac{2}{3}\Abs{C}-\Abs{H'}
=\tfrac{2}{3}\Abs{C}-\Abs{H}-\Abs{Q}-\beta-1
\stackrel{(3)}{\le}\tfrac{2}{3}C(\Qorig)-\Abs{H}\]
At (1), we use that states in $I$ have no incoming transitions in $\Prot$, so $I\times\{+\}$ has no incoming transitions in $\Prot'$ and is always empty. (2) follows from $C$ being well-initialised. For (3) we use $C(Q'\setminus\Qorig)=C(\Qsupp\cup\Qgate\cup\Qreset)=\Abs{Q}+\beta+1$. Finally, due to $C(\Qorig)=\Abs{\pi(C)}$ we derive that $\pi(C)$ is well-initialised.

This allows us to use that $\Phi$ is $\alpha$-rapidly decreasing:
\[(\Phi'_2(C)-\Phi'_2(\Cterm))^2=(\Phi(\pi(C))-\Phi(\pi(\Cterm)))^2\le\alpha\Speed(\pi(C))\]
It remains to show $\Speed(\pi(C))\le\Speed_{\Tset}(C)$, where $\Tset$ are the \TraRef{execute} transitions. For each transition $t\in T$ we have four corresponding transitions $t_1,...,t_4\in \Tset$, one for each choice of $\pm$. The bound $\Tmin{t}{\pi(C)}\le\sum_i\Tmin{t_i}{C}$ then follows from the well-known inequality  stating that $\min\{\sum_ix_i,\sum_iy_i\}\le\sum_{i,j}\min\{x_i,y_j\}$ holds for any non-negative numbers $x_1,...,x_k,y_1,...,y_k$.

\smallskip
\textbf{Case 3.} $C(Q\times\{+\})>0$. Here, we show that all “$+$” flags are eliminated quickly. We set $\Phi'_3((q,+))=1$ for $q\in Q$ and $0$ elsewhere. We know that $\pi(C)$ is terminal (else we would be in Case~2), and it must remain so. Hence \TraRef{execute} is disabled and no transition increases $\Phi'_3$. Also, the first case of \TraRef{init-reset} is enabled and can reduce the potential.

For every $q\in Q'$ with $\Phi'_3(q)>0$ we have a \TraRef{denotify} transition which decreases $\Phi'_3$ and uses only agents in $q$. Similarly to $\Phi'_1$, we use $\Tset$ to denote the set of these transitions, and find $(\Phi(C)-\Phi(\Cterm))^2\le\Abs{\Tset}\Speed_{\Tset}(C)$, noting $\Abs{\Tset}\le\Abs{Q}$.

\smallskip
\textbf{Case 4.} $C(\{0,...,\Abs{Q}-1\})>0$ or $C((q,1))\ne1$ for some $q\in\Support{\pi(C)}$. In this case, we show that the agents in $\Qsupp$ stabilise quickly. We use the potential
\begin{align*}
\Phi'_4((q,0))&:=\Phi'_4((q,!))+1:=2&&\text{ for }q\in Q\\
\Phi'_4(i)&:=3(\Abs{Q}-i) &&\text{ for }i=0,...,\Abs{Q}
\end{align*}
Again, $\Phi'_4$ is $0$ elsewhere. Due to the conditions on Cases~1 and~3, the only transition producing a state $\{0,...,\Abs{Q}-1\}$ is the first part of \TraRef{reset}, which decreases $\Phi'_4$. Otherwise, state $Q\times\{0\}$ cannot be produced. The only other transitions affecting the potential are \TraRef{detect} and the second case of \TraRef{init-reset}, which both decrease $\Phi'_4$. One of the above transitions, which we again denote by $\Tset$, is always enabled, so $\Speed_{\Tset}(C)\ge 1$. Additionally, we have $C(\Qsupp)=\Abs{Q}$ and $C(\Qreset)=1$, so $\Phi'_4(C)\le5\Abs{Q}\le5\Abs{Q}\Speed_{\Tset}(C)$.

\smallskip
\textbf{Case 5.} $C$ is not terminal. Finally, we consider the speed at which gates stabilise and the computer terminates.
\newcommand{\IsBot}{\mathbf{1}_\bot}%
\begin{align*}
\Phi'_5((g,b_1,b_2,b_3))&:=\IsBot(b_2)+\IsBot(b_3)&&\text{ for }g\in G\\
\Phi'_5(\Abs{Q}+i)&:=3(\beta-i) &&\text{ for }i=0,...,\beta
\end{align*}
where $\IsBot(\bot):=1$ and $\IsBot(0):=\IsBot(1):=0$. At this point, only transitions \TraRef{gate} and the second part of \TraRef{reset} are active, and both reduce $\Phi'_5$. We denote them by $\Tset$ and, analogous to Phase~4, we get the estimate $\Phi'_5(C)\le5\beta\Speed_{\Tset}(C)$ and find that one of these transitions is always enabled.
\end{proof}

\begin{remark}
 While it is possible to provide a potential function for $\Prot'$ based on a potential function for $\Prot$, this results in large constants for the speed of the protocol. 
 The reason lies in the nature of our computation, which proceeds in multiple phases. As an example, take transition \TraRef{execute}. One of the resulting agents has its flag set to $+$, which may initiate a reset of every agent in $\Qsupp\cup\Qgate$. To pay for this work, every transition of $\Prot$ would have to reduce the potential by $\Abs{Q}+\beta$. However, most of this cost would be wasted; only the last reset needs to be executed fully, and the other resets are likely to be interrupted before completion.
 \end{remark}

}

\subsection{Removing helpers preserves speed}\label{ssec:rapid-conv-3}
We now show that \(\autarkify\), the conversion of Section~\ref{ssec:simulatehelpers}, is fast.

\begin{proposition}\label{prop:autarkify-speed}
Let \(\Prot=(Q,\delta,I,O,H)\) denote a bounded binary population computer with marked consensus output, and let $\Phi$ denote a potential group of size $e$ for $\Prot$ which is \(\alpha\)-rapidly decreasing in all well-initialised configurations.

Then $\Prot':=\autarkify(\Prot)$ (see Section~\ref{ssec:simulatehelpers}) has a potential group of size \(e+1\) which is \(\O(\alpha+\size_2(\Prot)^3)\)-rapidly decreasing in all reachable configurations of size at least \(6\Abs{I}+10\Abs{H}\).
\end{proposition}

\begin{proof}
Let $(\Phi_1,...,\Phi_e):=\Phi$. We define $\Phi':=(\Phi'_1,\Phi_1,\Phi_2,...,\Phi_e)$, where:
\begin{alignat*}{2}
\Phi'_1(q)&:=2&\qquad&\text{ for }q\in I\\
\Phi'_1(\Up_i)&:=i+1&&\text{ for }i\in\{1,...,\Abs{H}-1\}\\
\Phi'_1(\Down_i)&:=i&&\text{ for }i\in\{2,...,\Abs{H}\}
\end{alignat*}
As usual, other states have potential $0$, and we extend $\Phi_i$ to $\N^{Q'}$ by setting the weight of states in $Q'\setminus Q$ to $0$.

Clearly, transitions Double and Helper decrease $\Phi'_1$, while transitions in $\delta$ cannot increase it due to the input specification of $\autarkify$ (page~\pageref{spec:autarkify}).

Now, let $C$ denote a configuration reachable from an input of size at least $2\Abs{H}+\Abs{I}$. To show that $\Phi'$ is rapidly decreasing in $C$, we differentiate between two cases.

\smallskip\textbf{Case 1.} If either Double or Helper is enabled at $C$, we show that $\Phi'_1$ is active. It has already been shown that $\Phi'_1$ cannot decrease. To show that $\Phi'_1$ is rapidly decreasing, let $S:=\{q\in Q':\Phi'_1(q)>0\}$ denote the states with positive potential. For each state $q\in I\cup\{\Up_1,...,\Up_{\Abs{H}-1}\}\subseteq S$ we have a transition using only agents in $q$. For each other state, i.e.\ $q=\Down_i\in S$ for some $i$, we observe $C(\Up_0)\ge C(\Down_i)$, as the construction guarantees that enough agents in $\Up_0$ exist. So in total we have $\Phi(C)\le\Abs{H}C(S)$ and $C(S)^2\le \Abs{S}\Speed_{\delta_>}(C)$ by Lemma~\ref{lem:sumofsquares}. Using $\Abs{S}\le \Abs{I}+2\Abs{H}$, $\Phi'_1$ is $(\Abs{I}+2\Abs{H})\Abs{H}^2$-rapidly decreasing in $C$. 

\smallskip\textbf{Case 2.} Otherwise, $C(I)\le\Abs{I}$, $C(\Qhelper)<\Abs{H}$ and, due to our construction, $C(I')\le \Abs{C}/2$. From the second, we derive $C(Q)>\Abs{C}-\Abs{H}$, which we combine with the other two to get $C(I\cup I')\le \Abs{I}+\Abs{C}/2<\Abs{I}+(C(Q)+\Abs{H})/2$. Rearranging terms yields $C(I\cup I')+\Abs{H}\le C(Q)/2+\Abs{I}+\tfrac{3}{2}\Abs{H}$. 

Now, we use $\Abs{C}\ge6\Abs{I}+10\Abs{H}$ to get $C(Q)>\Abs{C}-\Abs{H}\ge 6\Abs{I}+9\Abs{H}$, so $C(I\cup I')+\Abs{H}\le\Abs{I}+\tfrac{3}{2}\Abs{H}+C(Q)/2\le\tfrac{2}{3}C(Q)$. Noting $C(Q)=\Abs{\pi(C)}$, we find that $\pi(C)$ is well-initialised, so $\Phi$ is $\alpha$-rapidly decreasing in $\pi(C)$. This extends directly to $\Phi'$.
\end{proof}

\subsection{Converting to consensus output preserves speed}\label{ssec:rapid-conv-4}
Finally, we prove that speed is preserved by $\distribute$, the conversion of Section~\ref{ssec:converttoconsensusoutput}.

\begin{proposition}\label{prop:distribute-speed}
Let \(\Prot=(Q,\delta,I,O,\multiset{})\) denote a bounded binary population computer with marked consensus output, and let $\Phi$ denote a potential group of size $e$ for $\Prot$ which is \(\alpha\)-rapidly decreasing in all reachable configurations of size at least $k$.

Then on inputs of size at least \(k\), $\Prot':=\distribute(\Prot)$ stabilises in \(\O(e (\sqrt{\alpha}\Abs{Q}+\alpha) n^2)\) interactions in expectation.
\end{proposition}
\begin{proof}
By Proposition~\ref{prop:multipotential_speed}, $\Prot$ reaches a terminal configuration after $\O(e(\sqrt{\alpha}\Abs{Q}+\alpha)\,n^2)$ interactions in expectation.
So it suffices to show that $\Prot'$ can broadcast the result to all agents also in $\O(e(\sqrt{\alpha}\Abs{Q}+\alpha)\,n^2)$ interactions. We prove a more
general result: If $\Prot'$ reaches a configuration $C$ with $\pi(C)$ terminal within $f(\Prot',n)$ random interactions in expectation, them $\Prot'$ stabilises after $\O(f(\Prot',n)+n^2)$ random interactions in expectation. This was already shown in Section \ref{app:consensus_correctness}.
\end{proof}

\subsection{Proof of Theorem \ref{thm:mainCb}}

We collect the results of the previous sections to prove:

\thmmainCb*

\begin{proof}
Let $\Prot=(Q,\delta,I,O,H)$ be an $\alpha$-rapid population computer of size $m$ deciding $\Double(\varphi)$. 
Since $\binarise$ satisfies its specification (page~\pageref{spec:binarise}) and $\Prot$ satisfies the input specification, 
the computer $\Prot_1:= \binarise(\Prot)$ satisfies the postcondition. In particular,
$\Prot_1$ is binary. Further, the postcondition contains a conjunction of three implications, stating that if $\Prot$ satisfies
additional conditions, then $\Prot'$ enjoys  additional properties. By the definition of
$\alpha$-rapid computers (Definition \ref{def:rapid}),  $\Prot$ satisfies the premises of these
three implications, and so $\Prot_1$ satisfies their consequences. This shows that:
$\size(\Prot_1) \in \O(\size(\Prot))$ (because we have $\beta \leq 3$); no initial state of $\Prot_1$ has incoming transitions;
and all configurations that only populate the initial states of $\Prot_1$ are terminal. 

Now, let $\Prot_2 := \focalise(\Prot_1)$,
$\Prot_3:=  \autarkify(\Prot_2)$, and $\Prot_4:=  \distribute(\Prot_3)$. Proceeding exactly as in 
the proof in Section~\ref{ssec:puttogether}, we obtain that $\Prot_4$ is a population protocol of adjusted size $\O(m)$, 
and so with $\O(m)$ states, that decides $\varphi$ for all inputs of size \(\Omega(m)\). It remains to prove that $\Prot_4$ stabilises within $\O(\alpha m^4 n^2)$ expected interactions, which we achieve in several steps:
\begin{itemize}
\item By Proposition \ref{prop:multipotential_speed},
there is a potential function for $\Prot_1$ that is $\O(\Abs{Q}^2 k^2 \alpha)$-rapidly decreasing 
for every well-initialised configuration, where $k$ is the maximum arity of the transitions of $\Prot$.
So, in particular, $\Prot_1$ has an $\alpha_1\in\O(m^4 \alpha)$-rapidly decreasing potential function.
\item By Proposition \ref{prop:focalise-speed}, there is a potential group of size 5 for $\Prot_2$ that is
$\O(\alpha_1 + m^2)$-rapidly decreasing, and so $\alpha_2\in\O(m^4 \alpha)$-rapidly decreasing, in all well-initialised configurations. 
\item By Proposition \ref{prop:autarkify-speed}, there is a potential group of size $5+1=6$ for $\Prot_3$ that is
$\O(\alpha_2+ m^3)$-rapidly decreasing, i.e.\ \(\alpha_3\in\O(m^4 \alpha)\)-rapidly decreasing in all reachable configurations
of size $\Omega(m)$.
\item By Proposition \ref{prop:distribute-speed}, $\Prot_4$ stabilises in 
$\O((\sqrt{m^4 \alpha} \, m + m^4 \alpha) n^2) = \O(\alpha m^4 n^2)$ interactions in expectation. 
\end{itemize}
\end{proof}

%

\section{Conclusions} \label{sec:conclusions}
We have shown that every predicate $\varphi$ of quantifier-free Presburger arithmetic has a population protocol with 
$\O(\Abs{\varphi})$ states and $\O(\Abs{\varphi}^7 \cdot n^2)$ expected interactions to stabilisation for all inputs of size $\Omega(\Abs{\varphi})$. Therefore, every Presburger predicate has a protocol that is at the same time fast and succinct.
Our construction is close to optimal. Indeed, for every construction there is an infinite family of predicates for which it yields protocols with $\Omega(\Abs{\varphi}^{1/4})$ states \cite{BlondinEJ18}; further, it is known that every protocol for the majority predicate requires $\Omega(n^2/\Polylog n)$ interactions.  

Our construction is very modular. We have introduced population computers, a model that extends population protocols with three very useful features: interactions of arbitrary arity, helpers, and generalised output functions. We have designed conversions that, loosely speaking, allow us to transform an arbitrary computer into an equivalent protocol by eliminating each of these features. The conversions are independent of each other. Further, we have proved a powerful theorem showing that  in order to prove 
quantitative properties about the speed of the protocol it suffices to prove qualitative properties of the computer. 

In future work we plan to study the existence of succinct protocols with $\O(\Polylog(n))$ convergence time, either in the presence of leaders, as in ~\cite{AAE08}, or using the construction of ~\cite{KosowskiU18}. As mentioned in the introduction, our work can be considered a first step in this direction.

\smallskip\noindent \textbf{Acknowledgements.} We thank the anonymous reviewers for many helpful remarks. In particular, one remark led to Lemma~\ref{lem:boundedpotential}, which in turn led to a nicer formulation of Theorem~\ref{thm:mainB1}, one of our main results.

\bibliography{lipics}


\end{document}